\documentclass[prl,showpacs,twocolumn,
amsmath,amssymb,superscriptaddress,floatfix,nofootinbib]{revtex4-1}
\usepackage[utf8]{inputenc}
\usepackage[sort&compress]{natbib}
\usepackage[normalem]{ulem}
\usepackage{lipsum} 
\usepackage{bm}
\usepackage{times}
\usepackage{amssymb,amsbsy,amsmath,amsfonts}
\usepackage{graphicx}
\usepackage{float}
\usepackage{caption}
\usepackage{wrapfig}
\usepackage{subcaption}
\usepackage{color}
\usepackage{morefloats}
\usepackage{rotating}
\usepackage{srcltx}
\usepackage{slashed}
\usepackage{multirow}
\usepackage{verbatim}
\usepackage{hyperref}
\usepackage{tabularx}
\usepackage{adjustbox}
\usepackage[dvipsnames]{xcolor}
\usepackage{nicefrac}

\def\l{\left}
\def\r{\right}

\usepackage{hyperref}
\hypersetup{
	colorlinks	=true,
	urlcolor	=blue,
	linkcolor	=blue,
	citecolor	=blue,
	pdftitle	={},
	pdfauthor	={ Jun-Xu Lu, Yang Xiao, Zhi-Wei Liu, Li-Sheng Geng},
	pdfsubject	={the Fifth Force}
	}

\begin{document}

\title{
Relativistic chiral nuclear forces: status and prospects
}

\author{Jun-Xu Lu}
\affiliation{School of Physics, Beihang University, Beijing 102206, China}

\author{Yang Xiao}
\affiliation{School of Space, Beihang University, Beijing 102206, China}
\affiliation{School of Physics, Beihang University, Beijing 102206, China}

\author{Zhi-Wei Liu}
\affiliation{School of Physics, Beihang University, Beijing 102206, China}

\author{Li-Sheng Geng}
\email[Corresponding author: ]{lisheng.geng@buaa.edu.cn}
\affiliation{School of Physics, Beihang University, Beijing 102206, China}
 \affiliation{Sino-French Carbon Neutrality Research Center, \'Ecole Centrale de P\'ekin/School of General Engineering, Beihang University, Beijing 100191, China}
\affiliation{Peng Huanwu Collaborative Center for Research and Education, Beihang University, Beijing 100191, China}
\affiliation{Beijing Key Laboratory of Advanced Nuclear Materials and Physics, Beihang University, Beijing 102206, China }
\affiliation{Southern Center for Nuclear-Science Theory (SCNT), Institute of Modern Physics, Chinese Academy of Sciences, Huizhou 516000, China}

\begin{abstract}

Understanding nuclear structure, reactions, and the properties of neutron stars from  \textit{ab initio} calculations from the nucleon degrees of freedom has always been a primary goal of nuclear physics, in which the microscopic nuclear force serves as the fundamental input. So far, the Weinberg chiral nuclear force, first proposed by the Nobel laureate Weinberg, has become the \textit{de facto} standard input for nuclear \textit{ab initio} studies. However, compared to their non-relativistic counterparts, relativistic \textit{ab initio} calculations, which describe better nuclear observables, have only begun. The lack of modern relativistic nucleon-nucleon interactions is an important issue restricting their development. In this work, we briefly review the development and status of the Weinberg chiral nuclear force, as well as its limitations. We further present a concise introduction to the relativistic chiral nuclear force, show its description of the scattering phase shifts and observables such as differential cross sections, and demonstrate its unique features. Additionally, we show that the relativistic framework could be naturally extended to the antinucleon-nucleon interaction.

\end{abstract}


\maketitle


\section{Introduction}

The atomic nucleus is a vital microscopic level of material structure, lying between atoms and nucleons. Based on the degree of freedom of nucleons, the nuclear force is the foundation for understanding the rich and intricate nuclear physics phenomena related to nuclear astrophysics, exotic hadron states, and new physics beyond the standard model.

The attempts to understand the nuclear force can be traced back to the 1930s when Yukawa proposed the pion exchange theory~\cite{Yukawa:1935xg}, initiating the microscopic study of nuclear forces. Later in the 1950s and 1960s, with the experimental discoveries of the pion and other heavier mesons (such as $\rho$ and $\omega$), the one-boson exchange model was developed~\cite{DeTourreil:1975gz, Holinde:1976mkn, Nagels:1977ze}, which could qualitatively describe the properties of the nuclear force and deuteron properties. In the 1970s, the fundamental theory describing the strong interaction - Quantum Chromodynamics (QCD), was established, whose basic degrees of freedom are quarks and gluons. Then, physicists attempted to explain the nuclear force at the quark level. Various types of quark models~\cite{Oka:1980ax,Faessler:1982ik,Shimizu:1989ye, Valcarce:2005em,Straub:1988gj,Zhang:1994pp,Zhang:1997ny,Dai:2003dz,Wang:1992wi,Wu:1996fm,Ping:1998si,Wu:1998wu,Pang:2001xx,Fujiwara:1995td,Fujiwara:1995fx,Fujiwara:1996qj,Fujita:1998sg} were proposed. However, these models based on quark degrees of freedom are phenomenological, and their connection with QCD is unclear.

A breakthrough in the study of the nuclear force was achieved in the 1990s. The Nobel laureate Weinberg proposed that the chiral effective field theory (ChEFT) developed from QCD could be used to describe the nuclear force. These chiral nuclear forces started the era of modern microscopic nuclear forces~\cite{Weinberg:1990rz, Weinberg:1991um, Weinberg:1992yk}. For the early history and literature on the development of nuclear forces, we refer to the reviews ~\cite{Epelbaum:2008ga,Machleidt:2011zz,Machleidt:2024bwl}. After more than 30 years of development, Weinberg's chiral nuclear force has proven to be a great success and become the standard input for nuclear ab initio calculations, pioneering the study of non-perturbative strong interactions based on effective field theories. 

It is worth noting that the Weinberg chiral nuclear force is non-relativistic. While it has achieved great success, it has also encountered many difficulties (both fundamental and empirical), such as slow convergence, failure to satisfy renormalization group invariance, inconsistency in treating three-body forces, etc. To solve these problems, various attempts have been made, for example, developing chiral nuclear forces that do not explicitly contain pions~\cite{Bedaque:2002mn}, considering the contribution of the excited state $\Delta(1232)$~\cite{Ordonez:1993tn,Ordonez:1995rz,Kaiser:1998wa,Krebs:2007rh,Epelbaum:2008td,Piarulli:2014bda,Ekstrom:2017koy,Strohmeier:2020dkb,Nosyk:2021pxb,vanKolck:1994yi,Pandharipande:2005sx,Epelbaum:2007sq,Kaiser:2015yca,Krebs:2018jkc}, utilizing renormalization group invariance to modify the Weinberg power counting~\cite{PhysRevC.72.054006,Birse:2005um,PhysRevLett.114.082502,PhysRevC.95.024001}, developing the relativistic chiral nuclear force~\cite{Ren:2016jna, Xiao:2018jot, Xiao:2020ozd, Wang:2021kos, Lu:2021gsb}, etc. This article will mainly introduce the efforts and progress in the recent development of the relativistic chiral nuclear force.

The present literature is organized as follows. First, we introduce the Weinberg chiral nuclear force's status and shortcomings. Then, we show how to construct a nuclear force in the relativistic framework, the reproduction of the scattering phase shifts and observables, and introduce the recent efforts in describing the antinucleon-nucleon interaction. Finally, we summarize and present the prospects for future developments.

\subsection{Weiberg's chiral nuclear force}

\begin{figure*}
  \centering
  \includegraphics[width=\textwidth]{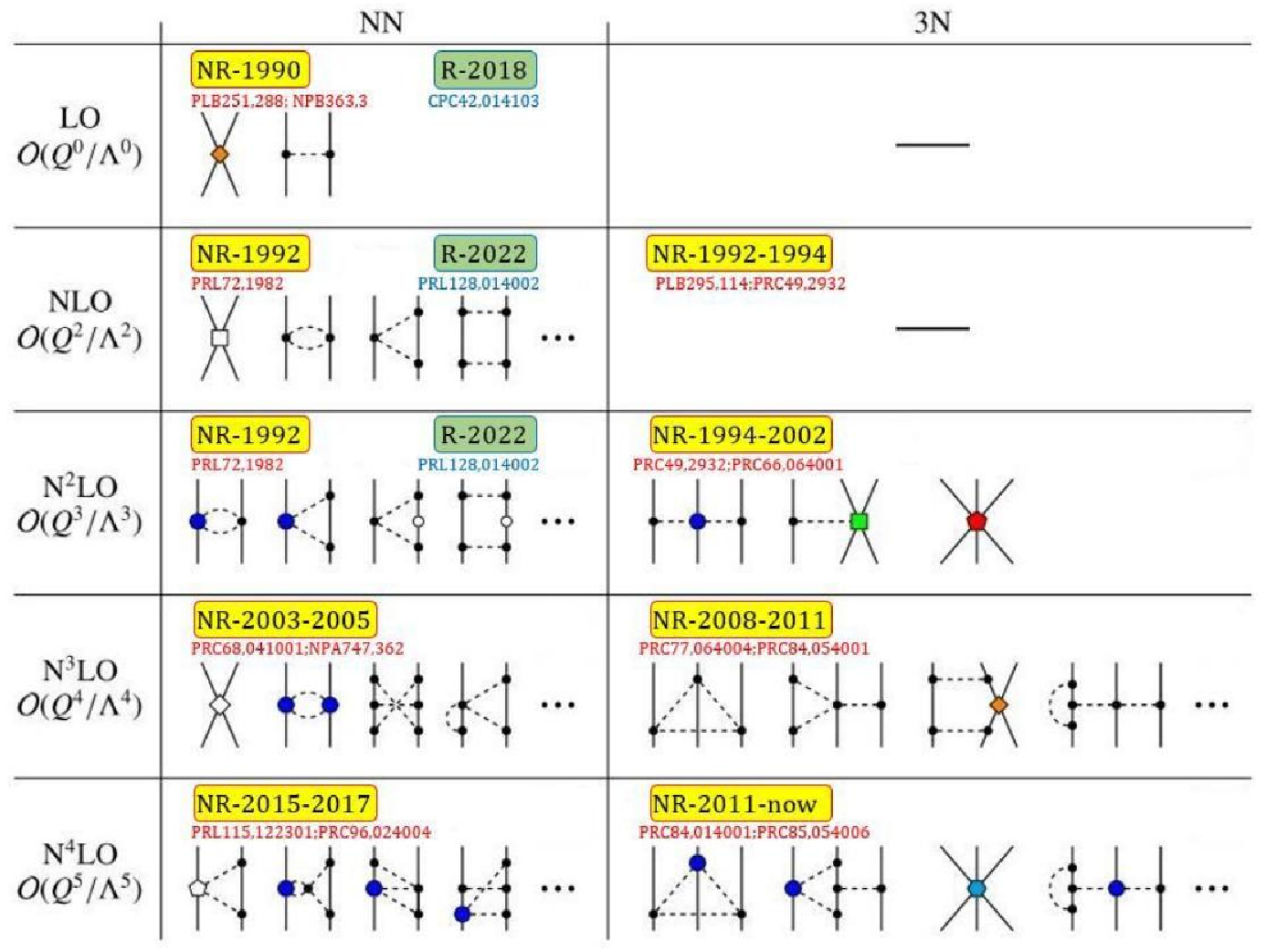}
  \caption{Hierarchy of the Weinberg chiral nuclear force up to the fifth order (N$^4$LO). This figure also shows the status of the relativistic chiral nuclear force developed by the Beihang group and their collaborators. In the figure, ``NR" represents the Weinberg chiral nuclear force, and ``R" represents the relativistic chiral nuclear force.}
  \label{fig:NRChNF}
\end{figure*}

In 1990, Weinberg proposed that one could first calculate the nuclear potential perturbatively according to the chiral order and then substitute it into the Lippmann-Schwinger (LS) equation to consider the non-perturbative effects~\cite{Weinberg:1990rz,Weinberg:1991um,Weinberg:1992yk}. The resulting nuclear force is referred to as the Weinberg chiral nuclear force. Over the past 30 years, the non-relativistic chiral nuclear force has achieved remarkable progress through the unremitting efforts of many physicists. Figure~\ref{fig:NRChNF} shows the hierarchy of the chiral nuclear force up to the fifth order (N$^4$LO) based on the Weinberg power counting from the naive dimensional analysis.

Weinberg constructed the leading-order (LO) chiral nuclear force~\cite{Weinberg:1990rz,Weinberg:1991um}. In this order, the chiral nuclear force contains the long-range one-pion-exchange (OPE) term, namely the well-known Yukawa interaction~\cite{Yukawa:1935xg} and the short-range contact terms. Subsequently, van Kolck et al. advanced the Weinberg chiral nuclear force to the next-to-next-to-leading order (NNLO) based on the time-ordered chiral effective field theory~\cite{Ordonez:1993tn,Ordonez:1995rz}. In the next-to-leading order (NLO) and NNLO, in addition to new short-range  contact terms, there are two-pion exchanges (TPE) and relativistic correction terms. The studies by Epelbaum et al. pointed out~\cite{Epelbaum:1998ka,Epelbaum:1999dj} that the nuclear force based on the time-ordered chiral effective field theory depends on the energy of the incident nucleons and proposed a new NNLO nuclear force scheme based on unitary transformations. In 2003, Epelbaum et al. further improved the convergence of this chiral nuclear force scheme using the spectrum function regularization technique~\cite{Epelbaum:2003gr,Epelbaum:2003xx}. 

Almost simultaneously, the two groups, led by Machleidt and Epelbaum, advanced the chiral nuclear force to the next-to-next-to-next-to-leading order (N$^3$LO)~\cite{Entem:2003ft,Epelbaum:2004fk}. New short-range contact terms, two-loop TPEs, three-pion-exchange terms, and the corresponding relativistic correction terms emerged in this order. Also, in this order, the description of the nucleon-nucleon scattering observables reached the level of the most sophisticated phenomenological nuclear forces. The chiral nuclear force enters the era of ``high precision". In 2014, Epelbaum et al. further improved the chiral nuclear force at N$^3$LO~\cite{Epelbaum:2014efa}, proposing to use the local regularization factor in coordinate space to replace the regularization factor in momentum space to improve the short-range contribution and convergence, and proposed a theoretical uncertainty estimation method based on the convergence properties of chiral effective field theories. Recently, Machleidt et al. completed the construction of the local N$^3$LO chiral nuclear force in coordinate space~\cite{Saha:2022oep}. 

The chiral nuclear force at the fifth order (N$^4$LO) was constructed by Machleidt and Epelbaum separately in 2015-2017~\cite{Entem:2014msa,Entem:2017gor,Epelbaum:2014sza,Reinert:2017usi}. Up to the sixth order (N$^5$LO), the chiral nuclear force has become extremely complicated. Parts of the long-range contributions were studied by Machleidt et al. ~\cite{Entem:2015xwa}, but a complete N$^5$LO chiral nuclear force is not yet available.

Contrary to the phenomenological nuclear forces (such as AV18~\cite{Wiringa:1994wb}, Reid93~\cite{Stoks:1994wp}, and CD-Bonn~\cite{Machleidt:2000ge}), the Weinberg chiral nuclear force is intimately related to the fundamental theory of the strong interaction, QCD. In the chiral effective field theory, through the power counting, the relative magnitude of a certain part of the nuclear force can be estimated a priori, thereby achieving a systematic order-by-order calculation, which provides a method to systematically improve the description according to the chiral orders and estimate the theoretical uncertainties~\cite{Epelbaum:2008ga,Meissner:2014lgi,Hammer:2019poc}. 

The power counting of the chiral effective field theory can consistently incorporate three-body and even four-body interactions. Notably, these so-called ``many-body nuclear forces" reflect that nucleons are not point-like elementary particles but composite particles with internal structures~\cite{Hebeler:2020ocj}. According to the Weinberg power counting rule, the NLO three-body interaction is vanishing~\cite{Weinberg:1992yk}, so the non-zero lowest-order three-body nuclear force appears at the next-to-next-to-leading order (NNLO)~\cite{vanKolck:1994yi,Epelbaum:2002vt}. At the NNLO, the three-body interaction includes short-range contact terms, medium-range one-pion-exchange terms, and long-range two-pion-exchange terms. In 2007,  the three-body nuclear force was calculated up to the N$^3$LO~\cite{Bernard:2007sp,Bernard:2011zr,Drischler:2017wtt}, and around 2011,  the long-range, medium-range and some short-range contributions at the N$^4$LO were calculated~\cite{Krebs:2012yv,Krebs:2013kha,Girlanda:2011fh}, but it is currently incomplete.

The lowest-order four-body interaction appears at the N$^3$LO~\cite{Epelbaum:2005bjv,Epelbaum:2007us}, constructed by Epelbaum, and the role of the four-body interaction in $^4$He was discussed~\cite{Rozpedzik:2006yi}. Currently, there is no study on the N$^4$LO four-body interaction.

\subsection{Issues encountered in the Weinberg chiral nuclear force }

The Weinberg chiral nuclear force has been well developed to date. However, several issues have been encountered in its development and application in first-principles calculations. The primary issue lies in the power counting or the convergence. The Weinberg chiral nuclear force is based on the heavy baryon chiral effective field theory. According to the naive dimensional analysis, this non-relativistic expansion of the baryon propagator and Dirac spinors results in a slow convergence. Epelbaum pointed out that accurately describing the neutron-deuteron scattering process requires at least the N$^4$LO chiral nuclear force~\cite{Epelbaum:2019zqc}.

The discussions on the power counting are usually associated with the renormalization of the LS equation and renormalization group invariance. The essence of the problem is that when the LS equation is truncated up to a certain order of the chiral expansion, it is non-renormalizable. The ultraviolet divergence of the loop diagrams in the scattering amplitude cannot be completely absorbed by the LECs truncated to the corresponding order~\cite{Epelbaum:2018zli}. Relevant discussions can be found in the review article~\cite{Hammer:2019poc}. Kaplan et al. proposed~\cite{Kaplan:2019znu} that renormalization can be achieved by considering the pion-exchange contribution only perturbatively, but this scheme has convergence issues. Epelbaum and Gegelia pointed out~\cite{Baru:2019ndr} that in the covariant framework, the non-perturbative OPE is renormalizable, but higher-order contributions still require perturbative treatments. Currently, there is no satisfactory solution to this problem.

The Weinberg power counting accounts that the contribution of the three-body force is much smaller than that of the two-body nuclear force since the lowest-order three-body force appears at N$^2$LO~\cite{Epelbaum:2008ga,Machleidt:2011zz}. However, realistic calculations show that the three-body force is key in describing finite nuclei and nuclear matter~\cite{Hebeler:2015hla,Hebeler:2020ocj}. Therefore, current ab initio calculations usually consider the three-body nuclear force. However, the current three-body nuclear force based on dimensional regularization (the loop contribution of N$^3$LO) breaks chiral symmetry when applied to the scattering equation (Faddeev equation) regularized with a certain momentum cutoff. This issue has been emphasized by Epelbaum and Machleidt~\cite{Epelbaum:2019zqc,Epelbaum:2019kcf,Epelbaum:2022cyo,Machleidt:2023jws}. Therefore, before the regularization problem of the three-body nuclear force is solved self-consistently, reliable predictions can only be based on LO, NLO, and N$^2$LO chiral nuclear forces.

The non-relativistic chiral nuclear force has become the standard microscopic input for ab initio calculations in nuclear physics. It has been widely applied in the studies of nuclear structure, nuclear reactions, nuclear matter, and nuclear astrophysics, achieving great success~\cite{Hebeler:2020ocj}. However, the results of ab initio calculations using the current chiral nuclear force as input are highly sensitive to the observables described, the nuclei of interest, and the details of the nuclear force. Currently, there is no nuclear force that, while exhibiting systematic chiral expansion convergence, can accurately predict within uncertainties the few-body and many-body observables of all nuclei across the entire nuclear chart. There is an urgent need to further develop the chiral nuclear force to correctly reproduce the properties of known atomic nuclei and nuclear matter and predict the properties of exotic nuclei far away from the beta stability line in a reliable and controlled manner. The two-body force and its application in \textit{ab initio} studies are clear, but the three-body force and its applications face significant challenges.

It is worth noting that to adapt to different many-body methods based on the chiral nuclear force, various optimized or softened nuclear forces have been developed, including NNLO$_{\text{sat}}$\cite{Ekstrom:2015rta}, NNLO$_{\text{opt}}$\cite{Ekstrom:2013kea}, NNLO$_{\text{sim}}$\cite{Carlsson:2015vda}. The differences lie in the methods used in determining the LECs of NN, 3N, and $\pi$N interactions and whether to introduce many-body observables for constraints. The technique of softening the nuclear force is based on the similarity renormalization group method~\cite{Jurgenson:2009qs,Roth:2011ar,Roth:2011vt}. The basic idea is to decouple the high-energy and low-energy parts of the Hamiltonian through a series of unitary transformations, thereby softening the nuclear force. It should be noted that this will introduce equivalent three-body nuclear forces.

\section{Relativistic chiral nuclear force}

The relativistic chiral effective field theory has been widely applied in the one-baryon system (baryon mass \cite{Ren:2012aj,Ren:2013oaa,Ren:2013wxa,Ren:2013dzt, Ren:2016aeo}, sigma term \cite{Ren:2014vea,Ren:2017fbv}, magnetic moment \cite{Geng:2008mf,Liu:2018euh,Shi:2018rhk,Xiao:2018rvd}, 
$\pi$N scattering \cite{Siemens:2016hdi}, hyperon weak radiative decay \cite{Shi:2022dhw}, etc.) and shows a faster convergence than its non-relativistic counterpart. Moreover, current studies based on the Dirac-Bruckener-Hatree-Fock (DBHF) framework indicate that the nuclear many-body calculations in the relativistic framework can describe observables better than the non-relativistic Bruckner-Hartree-Fock (BHF) method \cite{Shen:2019dls}, showing the promise of a simultaneous description of nuclear matter and finite nuclei.

Considering the problems encountered by the non-relativistic chiral nuclear force, especially the slow convergence and the failure to satisfy the renormalization group invariance, we proposed to develop the relativistic chiral nuclear force based on the covariant baryon chiral effective field theory \cite{Ren:2016jna}. This method follows the EOMS scheme to restore the power counting in the one-baryon system. Unlike the heavy baryon chiral effective field theory, it retains the complete Dirac spinor and Clifford algebra in the covariant chiral effective Lagrangian. Additionally, the non-perturbative effects are considered by solving the relativistic scattering equation instead of the Lippmann-Schwinger equation. 

\subsection{Covariant power counting and Lagrangians}

The contact Lagrangian describing the short-range nucleon-nucleon interaction has been constructed up to the $\mathcal{O}(p^4)$~\cite{Xiao:2018jot}. In the relativistic framework, the construction of the chiral effective Lagrangian follows the following principles: The chiral effective Lagrangian should be a Lorentz scalar and satisfy chiral symmetry, parity transformation, charge conjugation transformation, Hermitian conjugation transformation, and time-reversal invariance. Its construction should satisfy certain proper power counting so that the relative magnitude of each term can be estimated in advance and guarantee that the number of terms of the chiral effective Lagrangian up to a certain order of the chiral expansion is finite.

Next, one should define the chiral order of each component of the chiral effective Lagrangian to construct a self-consistent power counting. In the relativistic framework, the chiral orders of the nucleon bilinear structures, differential operators, Dirac matrices, and Levi-Civita tensors, as well as their properties under parity ($\mathcal{P}$), charge conjugation ($\mathcal{C}$), and Hermitian conjugation (h.c.) transformations, are shown in Table~\ref{Nucleon bilinears}, with which the relativistic chiral effective Lagrangians can be constructed up to $\mathcal{O}(p^0)$, $\mathcal{O}(p^2)$, and $\mathcal{O}(p^4)$ as given in Table~\ref{tb:NNLO NN Lagrangian}.

\begin{table}[h]
\caption{Chiral orders of the nucleon bilinear structures, differential operators, Dirac matrices, and Levi-Civita tensors, as well as their properties under parity ($\mathcal{P}$), charge conjugation ($\mathcal{C}$), and Hermitian conjugation (h.c.) transformations}
\label{Nucleon bilinears}
\centering
\begin{tabular}{ccccccccc}
 \hline\hline 
 Building Blocks&  $\bm{1}$    &  $\gamma_5$    &    $\gamma_\mu$     &   $\gamma_5\gamma_\mu$   &   $\sigma_{\mu\nu}$  &   $\epsilon_{\mu\nu\rho\sigma}$ &   $\overleftrightarrow \partial_{\mu}$ & $\partial_{\mu}$\\
 \hline 
  $\mathcal{P}$   & $+$ & $-$ & $+$ & $-$ & $+$ & $-$ & $+$ & $+$\\
  $\mathcal{C}$   & $+$ & $+$ & $-$ & $+$ & $-$ & $+$ & $-$ & $+$\\
  h.c.            & $+$ & $-$ & $+$ & $+$ & $+$ & $+$ & $-$ & $+$\\
  \hline
  Chiral order   & $0$ & $1$ & $0$ & $0$ & $0$ & $-$ & $0$ & $1$\\
 \hline 
\end{tabular}
\end{table}

\begin{table*}[!h]
\caption{Chiral effective Lagrangians up to $\mathcal{O}(p^4)$. } \label{tb:NNLO NN Lagrangian}
\centering
\begin{tabular}{c|c|c|c}
\hline\hline 
  $\widetilde{{O}}_{1}$          &$\left(\bar{\psi} \psi\right)\left(\bar{\psi} \psi\right)$   &$\widetilde{{O}}_{21}$       &$\frac{1}{16m^4}\left(\bar{\psi} i\overleftrightarrow{\partial}^{\mu} \psi\right) \partial^{2} \partial^{\nu}\left(\bar{\psi} \sigma_{\mu\nu} \psi\right)$    \\
  $\widetilde{{O}}_{2}$          &$\left(\bar{\psi} \gamma^{\mu} \psi\right)\left(\bar{\psi} \gamma_{\mu} \psi\right)$  & $\widetilde{{O}}_{22}$       &$\frac{1}{16m^4}\left(\bar{\psi} \sigma^{\mu \alpha} \psi\right) \partial^{2} \partial_{\alpha}\partial^{\nu}\left(\bar{\psi} \sigma_{\mu\nu} \psi\right)$  \\
  $\widetilde{{O}}_{3}$          &  $\left(\bar{\psi} \gamma_5 \gamma^{\mu} \psi\right)\left(\bar{\psi} \gamma_5 \gamma_{\mu} \psi\right)$    & $\widetilde{{O}}_{23}$&  $\frac{1}{16m^4}\left(\bar{\psi} \sigma^{\mu\nu} i\overleftrightarrow{\partial}^{\alpha} \psi\right) \partial^{\beta} \partial_{\nu}\left(\bar{\psi} \sigma_{\alpha \beta} i\overleftrightarrow{\partial}_{\mu} \psi\right)$ \\
  $\widetilde{{O}}_{4}$          & $\left(\bar{\psi} \sigma^{\mu\nu} \psi\right)\left(\bar{\psi} \sigma_{\mu\nu} \psi\right)$ & $\widetilde{{O}}_{24}$       &$\frac{1}{16m^4}\left(\bar{\psi} \psi\right) \partial^{4}\left(\bar{\psi} \psi\right)$\\
  \cline{1-2}
  $\widetilde{{O}}_{5}$          & $\left(\bar{\psi} \gamma_5 \psi\right)\left(\bar{\psi} \gamma_5 \psi\right)$ & $\widetilde{{O}}_{25}$       &$\frac{1}{16m^4}\left(\bar{\psi} \gamma^{\mu} \psi\right) \partial^{4} \left(\bar{\psi} \gamma_{\mu} \psi\right)$    \\
  $\widetilde{{O}}_{6}$          &$\frac{1}{4m^2}\left(\bar{\psi} \gamma_5 \gamma^{\mu} i\overleftrightarrow{\partial}^{\alpha} \psi\right)\left(\bar{\psi} \gamma_5 \gamma_{\alpha} i \overleftrightarrow{\partial}_{\mu} \psi\right)$   & $\widetilde{{O}}_{26}$       &$\frac{1}{16m^4}\left(\bar{\psi} \gamma_5 \gamma^{\mu} \psi\right) \partial^{4} \left(\bar{\psi} \gamma_5 \gamma_{\mu} \psi\right)$  \\
  $\widetilde{{O}}_{7}$          &$\frac{1}{4m^2}\left(\bar{\psi} \sigma^{\mu\nu} i\overleftrightarrow{\partial}^{\alpha} \psi\right)\left(\bar{\psi} \sigma_{\mu\alpha} i\overleftrightarrow{\partial}_{\nu} \psi\right)$ & $\widetilde{{O}}_{27}$       &$\frac{1}{16m^4}\left(\bar{\psi} \sigma^{\mu\nu} \psi\right) \partial^{4} \left(\bar{\psi} \sigma_{\mu\nu} \psi\right)$ \\
  $\widetilde{{O}}_{8}$          &$\frac{1}{4m^2}\left(\bar{\psi}  i\overleftrightarrow{\partial}^{\mu} \psi\right)\partial^{\nu}\left(\bar{\psi} \sigma_{\mu\nu} \psi\right)$ & $\widetilde{{O}}_{28}$  &   $\frac{1}{4m^2}\left(\bar{\psi} \gamma_5 i\overleftrightarrow{\partial}^{\alpha} \psi\right)\left(\bar{\psi} \gamma_5 i\overleftrightarrow{\partial}_{\alpha} \psi\right)-\widetilde{O}_{5}$\\
  $\widetilde{{O}}_{9}$          &$\frac{1}{4m^2}\left(\bar{\psi}  \sigma^{\mu \alpha} \psi\right)\partial_{\alpha}\partial^{\nu}\left(\bar{\psi} \sigma_{\mu\nu} \psi\right)$   & $\widetilde{{O}}_{29}$  & $\frac{1}{16m^4}\left(\bar{\psi} \gamma_5\gamma^{\mu} i\overleftrightarrow{\partial}^{\alpha} i\overleftrightarrow{\partial}^{\beta} \psi\right)\left(\bar{\psi} \gamma_5\gamma_{\alpha} i\overleftrightarrow{\partial}_{\mu} i\overleftrightarrow{\partial}_{\beta} \psi\right)-\widetilde{O}_{6}$\\
  $\widetilde{{O}}_{10}$  &$\frac{1}{4m^2}\left(\bar{\psi} \psi\right) \partial^{2}\left(\bar{\psi} \psi\right)$  & $\widetilde{{O}}_{30}$  & $\frac{1}{16m^4}\left(\bar{\psi} \sigma^{\mu\nu} i\overleftrightarrow{\partial}^{\alpha} i\overleftrightarrow{\partial}^{\beta} \psi\right)\left(\bar{\psi} \sigma_{\mu\alpha} i\overleftrightarrow{\partial}_{\nu} i\overleftrightarrow{\partial}_{\beta} \psi\right)-\widetilde{O}_{7}$ \\
  $\widetilde{{O}}_{11}$  &$\frac{1}{4m^2}\left(\bar{\psi} \gamma^{\mu} \psi\right) \partial^{2} \left(\bar{\psi} \gamma_{\mu} \psi\right)$  & $\widetilde{{O}}_{31}$  & $\frac{1}{16m^4}\left(\bar{\psi}  i\overleftrightarrow{\partial}^{\mu} i\overleftrightarrow{\partial}^{\beta} \psi\right) \partial^{\alpha}\left(\bar{\psi} \sigma_{\mu\alpha}  i\overleftrightarrow{\partial}_{\beta} \psi\right)-\widetilde{O}_{8}$\\
  $\widetilde{{O}}_{12}$       &$\frac{1}{4m^2}\left(\bar{\psi} \gamma_5 \gamma^{\mu} \psi\right) \partial^{2} \left(\bar{\psi} \gamma_5 \gamma_{\mu} \psi\right)$   & $\widetilde{{O}}_{32}$  & $\frac{1}{16m^4}\left(\bar{\psi}  \sigma^{\mu \alpha} i\overleftrightarrow{\partial}^{\beta} \psi\right)\partial_{\alpha}  \partial^{\nu}\left(\bar{\psi} \sigma_{\mu\nu}  i\overleftrightarrow{\partial}_{\beta} \psi\right)-\widetilde{O}_{9}$ \\
  $\widetilde{{O}}_{13}$       &$\frac{1}{4m^2}\left(\bar{\psi} \sigma^{\mu\nu} \psi\right) \partial^{2} \left(\bar{\psi} \sigma_{\mu\nu} \psi\right)$  &  $\widetilde{{O}}_{33}$ &  $\frac{1}{16m^4}\left(\bar{\psi} i\overleftrightarrow{\partial}^{\alpha} \psi\right)\partial^{2}\left(\bar{\psi} i\overleftrightarrow{\partial}_{\alpha} \psi\right)-\widetilde{O}_{10}$ \\
  $\widetilde{{O}}_{14}$  & $\frac{1}{4m^2}\left(\bar{\psi} i\overleftrightarrow{\partial}^{\alpha} \psi\right)\left(\bar{\psi} i\overleftrightarrow{\partial}_{\alpha} \psi\right)-\widetilde{O}_{1}$    & $\widetilde{{O}}_{34}$ &  $\frac{1}{16m^4}\left(\bar{\psi} \gamma^{\mu} i\overleftrightarrow{\partial}^{\alpha} \psi\right)\partial^{2}\left(\bar{\psi} \gamma_{\mu} i\overleftrightarrow{\partial}_{\alpha} \psi\right)-\widetilde{O}_{11}$ \\
  $\widetilde{{O}}_{15}$  & $\frac{1}{4m^2}\left(\bar{\psi} \gamma^{\mu} i\overleftrightarrow{\partial}^{\alpha} \psi\right)\left(\bar{\psi} \gamma_{\mu} i\overleftrightarrow{\partial}_{\alpha} \psi\right)-\widetilde{O}_{2}$   &$\widetilde{{O}}_{35}$ &  $\frac{1}{16m^4}\left(\bar{\psi} \gamma_5\gamma^{\mu} i\overleftrightarrow{\partial}^{\alpha} \psi\right)\partial^{2}\left(\bar{\psi} \gamma_5\gamma_{\mu} i\overleftrightarrow{\partial}_{\alpha} \psi\right)-\widetilde{O}_{12}$  \\
   $\widetilde{{O}}_{16}$  &   $\frac{1}{4m^2}\left(\bar{\psi} \gamma_5\gamma^{\mu} i\overleftrightarrow{\partial}^{\alpha} \psi\right)\left(\bar{\psi} \gamma_5 \gamma_{\mu} i\overleftrightarrow{\partial}_{\alpha} \psi\right)-\widetilde{O}_{3}$    &  $\widetilde{{O}}_{36}$ &  $\frac{1}{16m^4}\left(\bar{\psi} \sigma^{\mu\nu} i\overleftrightarrow{\partial}^{\alpha} \psi\right)\partial^{2}\left(\bar{\psi} \sigma_{\mu\nu} i\overleftrightarrow{\partial}_{\alpha} \psi\right)-\widetilde{O}_{13}$ \\
  $\widetilde{{O}}_{17}$  &   $\frac{1}{4m^2}\left(\bar{\psi} \sigma^{\mu\nu} i\overleftrightarrow{\partial}^{\alpha} \psi\right)\left(\bar{\psi} \sigma_{\mu\nu} i\overleftrightarrow{\partial}_{\alpha} \psi\right)-\widetilde{O}_{4}$  & $\widetilde{{O}}_{37}$ & $\frac{1}{16m^4}\left(\bar{\psi} i\overleftrightarrow{\partial}^{\alpha} i\overleftrightarrow{\partial}^{\beta} \psi\right)\left(\bar{\psi} i\overleftrightarrow{\partial}_{\alpha} i\overleftrightarrow{\partial}_{\beta} \psi\right)-2\widetilde{O}_{14}-\widetilde{O}_{1}$ \\
  \cline{1-2}
  $\widetilde{{O}}_{18}$       &$\frac{1}{4m^2}\left(\bar{\psi} \gamma_5 \psi\right) \partial^{2}\left(\bar{\psi} \gamma_5 \psi\right)$  &$\widetilde{{O}}_{38}$ & $\frac{1}{16m^4}\left(\bar{\psi} \gamma^{\mu} i\overleftrightarrow{\partial}^{\alpha} i\overleftrightarrow{\partial}^{\beta} \psi\right)\left(\bar{\psi} \gamma_{\mu} i\overleftrightarrow{\partial}_{\alpha} i\overleftrightarrow{\partial}_{\beta} \psi\right)-2\widetilde{O}_{15}-\widetilde{O}_{2}$ \\
  $\widetilde{{O}}_{19}$       &$\frac{1}{16m^4}\left(\bar{\psi} \gamma_5 \gamma^{\mu} i\overleftrightarrow{\partial}^{\nu} \psi\right) \partial^{2} \left(\bar{\psi} \gamma_5 \gamma_{\nu} i\overleftrightarrow{\partial}_{\mu}  \psi\right)$ &   $\widetilde{{O}}_{39}$ & $\frac{1}{16m^4}\left(\bar{\psi} \gamma_5 \gamma^{\mu} i\overleftrightarrow{\partial}^{\alpha} i\overleftrightarrow{\partial}^{\beta} \psi\right)\left(\bar{\psi} \gamma_5 \gamma_{\mu} i\overleftrightarrow{\partial}_{\alpha} i\overleftrightarrow{\partial}_{\beta} \psi\right)-2\widetilde{O}_{16}-\widetilde{O}_{3}$ \\
  $\widetilde{{O}}_{20}$       &$\frac{1}{16m^4}\left(\bar{\psi} \sigma^{\mu\nu} i\overleftrightarrow{\partial}^{\alpha} \psi\right) \partial^{2} \left(\bar{\psi} \sigma_{\mu\alpha} i\overleftrightarrow{\partial}_{\nu}  \psi\right)$ &  $\widetilde{{O}}_{40}$ & $\frac{1}{16m^4}\left(\bar{\psi} \sigma^{\mu\nu} i\overleftrightarrow{\partial}^{\alpha} i\overleftrightarrow{\partial}^{\beta} \psi\right)\left(\bar{\psi} \sigma_{\mu\nu} i\overleftrightarrow{\partial}_{\alpha} i\overleftrightarrow{\partial}_{\beta} \psi\right)-2\widetilde{O}_{17}-\widetilde{O}_{4}$  \\
  \hline 
    \end{tabular}
\end{table*}

\subsection{Pion-exchange contributions}

To describe the long-range one-pion exchange and medium-range two-pion exchanges, we need the Lagrangian for the $\pi$N interaction, which takes the form as ~\cite{Chen:2012nx},
\begin{equation}
  \mathcal{L}^{(1)}_{\pi N} = \bar{\psi}(i\slashed{D}-m+\frac{g_A}{2}\slashed{u}\gamma_5)\psi ,
\end{equation}
\begin{align}
\label{eq:Lag1}
  \mathcal{L}^{(2)}_{\pi N} = &c_1\langle\chi_+\rangle\bar{\psi}\psi-\frac{c_2}{4m^2}\langle\ u^\mu u^\nu \rangle(\bar{\psi}D_\mu D_\nu \psi+h.c.) + \nonumber\\
  & \frac{c_3}{2}\langle\ u^2 \rangle \bar{\psi}\psi - \frac{c_4}{4} \bar{\psi} \gamma^\mu\gamma^\nu[u_\mu,u_\nu] \psi ,
\end{align}
where $\bar{\psi}$ and $\psi$ are the nucleon fields. $D_\mu$ is the covariant derivative defined as $D_\mu=\partial_\mu+\Gamma_\mu$ with $\Gamma_\mu=\frac{1}{2}(u^\dag\partial_\mu u+u \partial_\mu u^\dag)$, $u=\text{exp}(\frac{i \Phi}{2f})$. $\Phi$ is the pion field. The axial current $u_\mu=i(u^\dag\partial_\mu u-u \partial_\mu u^\dag)$ and the chiral symmetry breaking term $\chi_+=u\dag \chi u+u \chi u\dag$ with $\chi=\mathrm{diag}(m^2_\pi,m^2_\pi)$. $g_A$ is the axial coupling constant and $c_{1,2,3,4}$ are the NLO LECs.

According to the covariant power counting, the pion exchange contributions at each order are shown in Figure~\ref{fig:NRChNF}. In the leading order, the pion-exchange contribution only contains the one-pion-exchange term. At the next-to-leading order, the medium-range two-pion exchanges begin to contribute. In addition, it also includes the one-loop correction of the OPE. However, this contribution can be absorbed by renormalizing the leading-order OPE. At the next-to-next-to-leading order, the second-order TPEs contribute, including the low-energy constant of the NLO $\pi$N scattering. Relevant details can be found in Refs.~\cite{Xiao:2020ozd,Wang:2021kos}.

\subsection{Nucleon-nucleon scattering in the relativistic framework}

\begin{figure}[h]
\centering
\includegraphics[width=0.46 \textwidth]{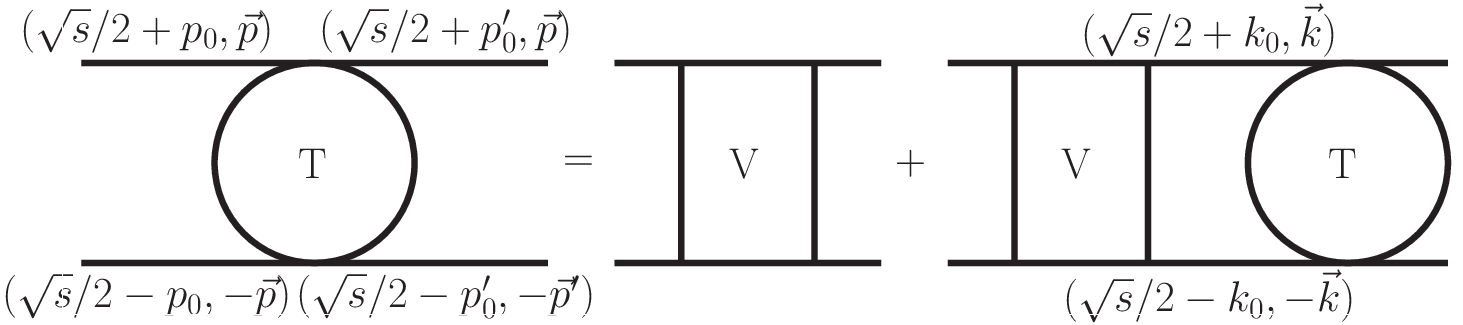}
\caption{Schematic diagram for the kinematics of nucleon-nucleon scattering.}
\label{kinematics}
\end{figure}

The existence of the deuteron indicates that the nucleon-nucleon interaction is nonperturbative. To account for this, we solve the following relativistic Blankenbecler-Sugar (BbS) equation~\cite{Blankenbecler:1965gx},
\begin{equation}\label{BbSE}
\begin{split}
  T(\bm{p}',\bm{p},s)&=V(\bm{p}',\bm{p},s)\\
  +\int\frac{\mathrm{d}^3\bm{k}}{(2\pi)^3}&V(\bm{p}',\bm{k},s)\frac{m^2}{E_k}\frac{1}{\bm{q}_{cm}^2-\bm{k}^2-i\epsilon}T(\bm{k},\bm{p},s),
\end{split}
\end{equation}
where $|\bm{q}_{\rm cm}|=\sqrt{s/4-m^2}$ is the nucleon momentum in the c.m. frame with $\sqrt{s}$ the total energy and $m$ the nucleon mass. $V$ is the chiral nuclear force as input. The notations for the incoming and outgoing nucleons are shown in Fig.~\ref{kinematics}.

To remove the ultraviolet divergence, the potential is regularized as follows,
\begin{equation}
V_{l'l}^{sj}({p}',{p}|\sqrt{s})=f_{R}({p})V({p}',{p}|\sqrt{s})f_{R}({p'}),
\end{equation}
where the regulator is taken to be
\begin{equation}
f_{R}({p})=f_{R}^{\textrm{sharp}}({p})=\theta(\Lambda^2-p^2).
\end{equation}

The S-matrix for each partial wave can be expressed as 
\begin{align}
S_{l'l}^{sj}({p}_{\text{cm}}) = \delta_{l'l}^{sj} + 2\pi i \rho T_{l'l}^{sj}({p}_{\text{cm}}),\quad \rho=-\frac{|{p}_{\text{cm}}|m^2}{16\pi^2E_{\text{cm}}}
\end{align}
in which {${p}_{\text{cm}}=\sqrt{T_{\text{lab}}m/2}$}. 
The phase shifts are then calculated as
\begin{align}
S_{jj}^{0j}=\textrm{exp}{\left(2i\delta_{j}^{0j}\right)}, \quad S_{jj}^{1j}=\textrm{exp}{\left(2i\delta_{j}^{1j}\right)},
\end{align}
where $\delta^{0j}_j$ and $\delta^{1j}_j$ are the phase shifts of the corresponding partial waves. Similarly, for coupled channels we adopt the Stapp parametrization~\cite{Stapp:1956mz},
\begin{align}
S=&\begin{pmatrix} S_{--}^{1j} & S_{-+}^{1j} \\ S_{+-}^{1j} & S_{++}^{1j} \end{pmatrix}\nonumber\\
=&\begin{pmatrix} \textrm{exp}{\left(i\delta_{-}^{1j}\right)} & 0 \\ 0 & \textrm{exp}{\left(i\delta_{+}^{1j}\right)} \end{pmatrix}\begin{pmatrix} \textrm{cos}(2\varepsilon) & i\textrm{sin}(2\varepsilon) \\ i\textrm{sin}(2\varepsilon) & \textrm{cos}(2\varepsilon) \end{pmatrix} \nonumber \\
&\quad \quad \quad \quad \cdot \begin{pmatrix} \textrm{exp}{\left(i\delta_{-}^{1j}\right)} & 0 \\ 0 & \textrm{exp}{\left(i\delta_{+}^{1j}\right)} \end{pmatrix}\nonumber\\
=&\begin{pmatrix} \textrm{cos}(2\varepsilon)\textrm{exp}{\left(2i\delta_{-}^{1j}\right)} & i\textrm{sin}(2\varepsilon)\textrm{exp}{\left(i\delta_{-}^{1j}+i\delta_{+}^{1j}\right)} \\
i\textrm{sin}(2\varepsilon)\textrm{exp}{\left(i\delta_{-}^{1j}+i\delta_{+}^{1j}\right)}  &
\textrm{cos}(2\varepsilon)\textrm{exp}{\left(2i\delta_{+}^{1j}\right)}
\end{pmatrix}
\end{align}
The subscripts``$+$'' denotes $j+1$, while``$-$''denotes $j-1$.

The LECs are determined by fitting the partial-wave scattering phase shifts. We adopted the same fitting strategy as in the studies of the Weinberg chiral nuclear force~\cite{Epelbaum:2014sza,Ordonez:1993tn}. Specifically, we fitted all the proton-neutron scattering phase shifts with angular momentum $J\leq 2$ from the Nijmegen collaboration~\cite{Stoks:1993tb}. Note that at LO, since the contact terms do not contribute to the partial waves with $J=2$, only the partial wave phase shifts with $J=0,1$ are considered in the fitting. We selected eight experimental points for each partial wave corresponding to the laboratory kinetic energies $T_{\rm{lab}}=1,5,10,25,50,100,150,200$ MeV, respectively. The $\tilde{\chi}^2$ between the theoretical results and the experimental data is defined as
\begin{equation}
    \tilde{\chi}^2=\sum_i(\delta^i-\delta^i_{\rm{PWA93}})^2,
\end{equation}
In the above equation, $\delta^i$ is the theoretical partial-wave phase shift or mixing angle, and $\delta^i_\mathrm{PWA93}$ is the corresponding Nijmegen result~\cite{Stoks:1993tb}.

\subsection{LO relativistic chiral nuclear force}

\begin{figure}
  \centering
  \includegraphics[width=\linewidth]{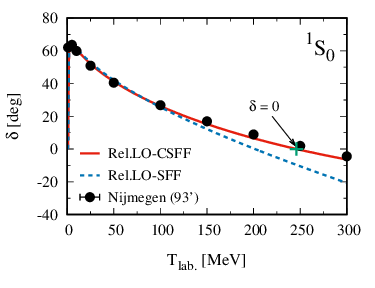}
  \caption{The $^1S_0$ partial-wave phase shift as a function of the laboratory energy $T_{\text{lab}}$. The red solid and blue dashed lines represent the leading-order relativistic chiral nuclear force results with the separable covariant and non-covariant regulators, respectively. The crosses in the figure mark the positions where the phase shift is zero. The black solid dots represent the results of the Nijmegen partial wave analysis~\cite{Stoks:1993tb}.}
  \label{fig:single-column}
\end{figure}

Like the non-relativistic chiral nuclear force, the LO relativistic chiral nuclear force includes short-range contact interactions and long-range OPE contributions. However, the covariant power counting dictates that the LO relativistic chiral nuclear force already contains some NLO operators of the Weinberg chiral nuclear force, thereby significantly accelerating the convergence of the chiral expansion. At LO, the description of the scattering phase shifts with total angular momentum $J\leq 1$ by the relativistic chiral nuclear force is already comparable to the non-relativistic NLO \cite{Ren:2016jna}. Furthermore, with the leading-order relativistic chiral nuclear force, we have conducted a detailed study on the proton-neutron scattering data of the $^1S_0$ partial wave \cite{Ren:2017yvw}, including the scattering phase shift of the $^1S_0$ partial wave, the position of the virtual state, the scattering length, the effective range, etc., and discussed the influence of different regularization factors. The description of the scattering phase shift of the $^1S_0$ partial wave by the LO relativistic chiral nuclear force is shown in Figure~\ref{fig:single-column}. Compared with the Weinberg chiral nuclear force that requires calculations up to NLO, the relativistic chiral nuclear force can describe most details of the scattering process at the leading order, obviously conforming more to the expectations for an effective field theory.

\begin{figure}
  \centering
  \includegraphics[width=\linewidth]{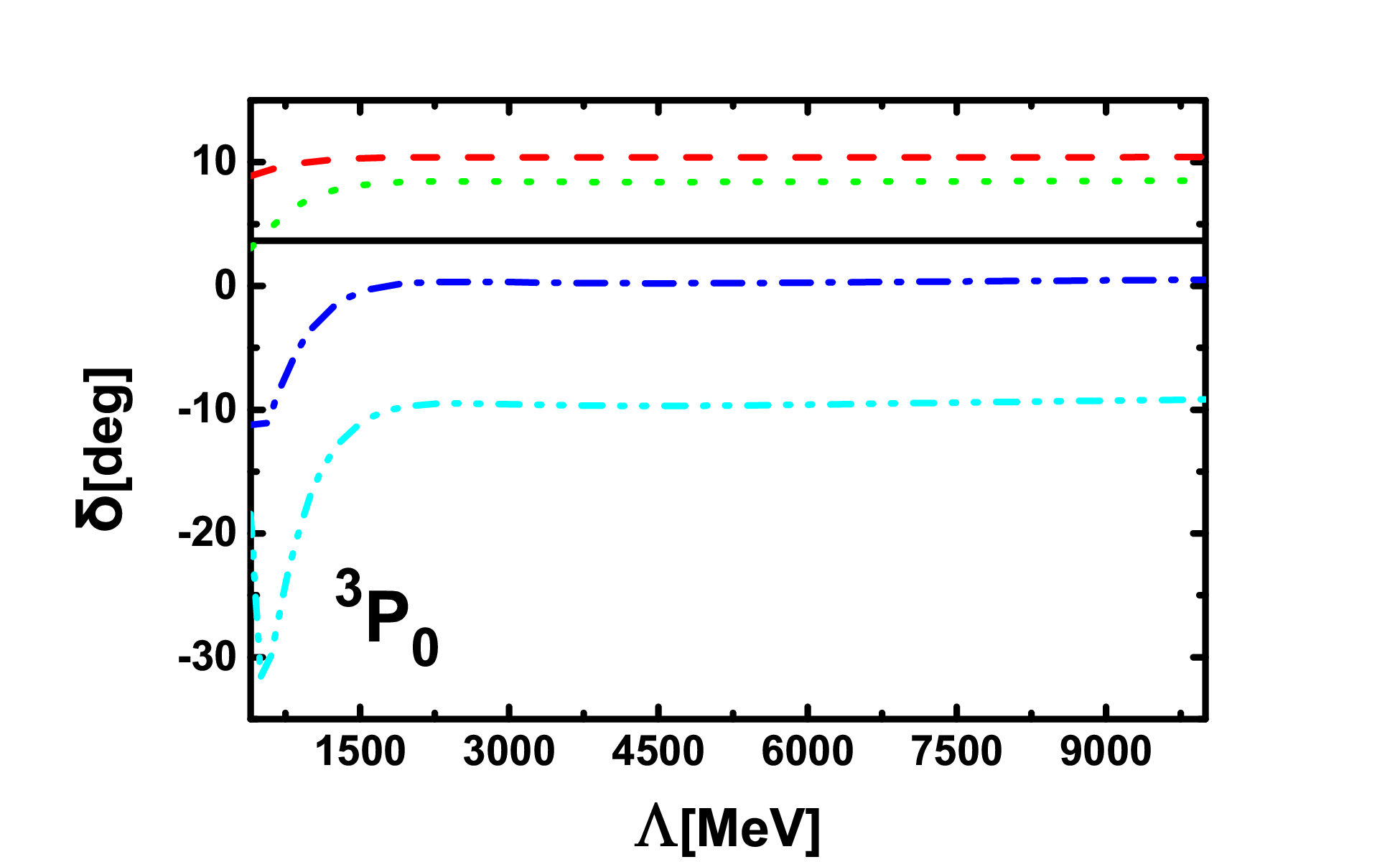}\\
  \includegraphics[width=\linewidth]{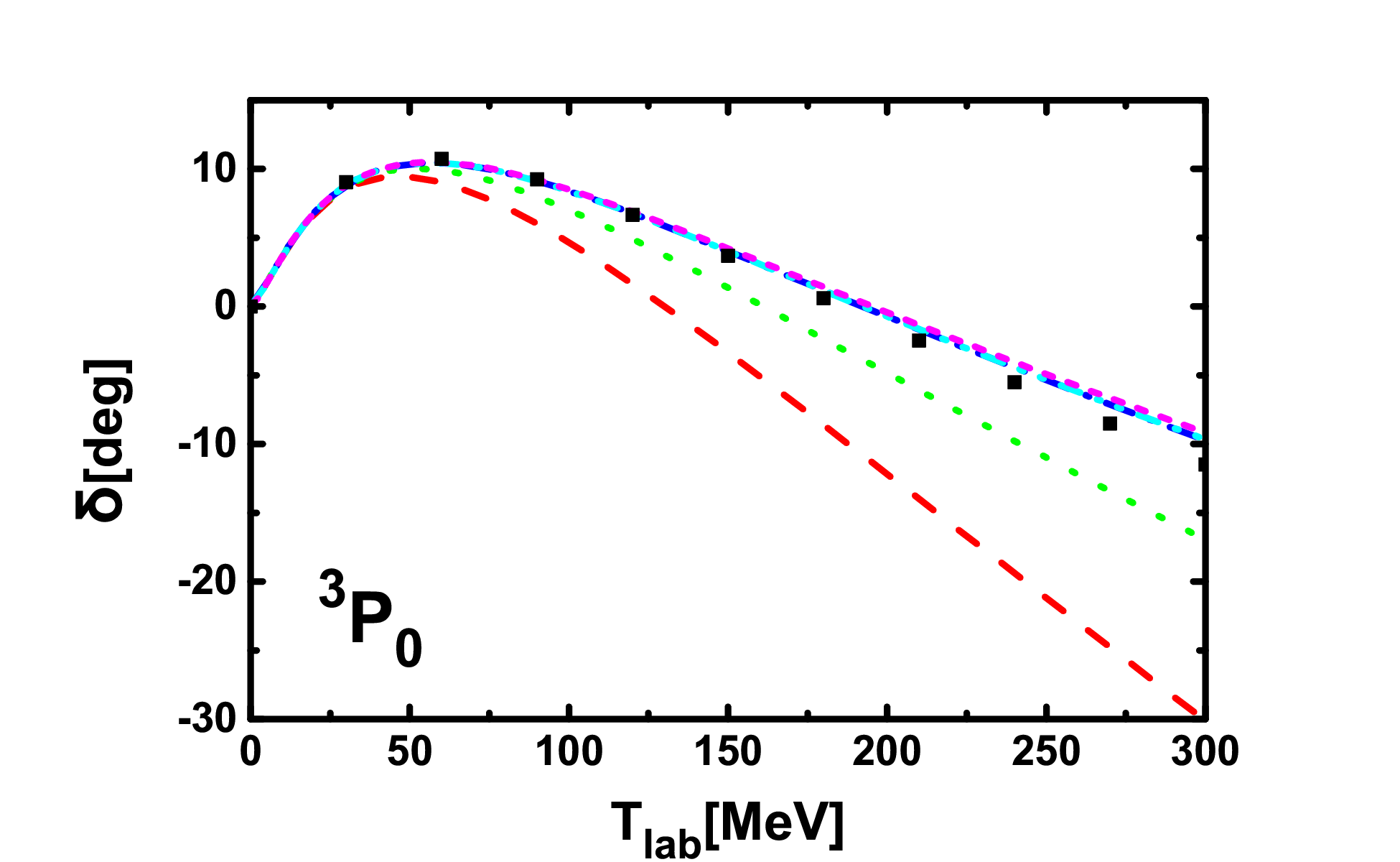}\\
  \caption{Upper panel: the $^3P_0$ partial wave phase shift as a function of the momentum cutoff $\Lambda$ for $T_{\text{lab}}$ equals 10 MeV (black solid line), 50 MeV (red dashed line), 100 MeV (green dotted line), 190 MeV (blue dot-dashed line), and 300 MeV (cyan double-dot-dashed line), respectively. Lower panel: the $^3P_0$ partial wave phase shift for momentum cutoffs of 600 MeV (red dashed line), 1000 MeV (green dotted line), 2000 MeV (blue dot-dashed line), 5000 MeV (cyan double-dot-dashed line), and 10000 MeV (purple short-dashed line), respectively.}
  \label{fig:3P0-RGI}
\end{figure}

An effective field theory with proper power counting should satisfy the renormalization group invariance. In other words, its description of observables should not (significantly) depend on the momentum cutoff introduced when solving the scattering equation. However, as mentioned earlier, the description of the phase shifts by the Weinberg chiral nuclear force does not satisfy the renormalization group invariance. We note that in the relativistic framework, the renormalization group evolution behavior of observables is improved ~\cite{Wang:2020myr}. We find that under the covariant power counting, the partial-wave phase shifts of $^3S_1$, $^3D_1$, $\epsilon_1$, $^3P_0$, $^1P_1$ satisfy the renormalization group invariance. In the $^3P_0$ partial wave, the Weinberg chiral nuclear force does not satisfy the renormalization group invariance, as shown in Fig.~\ref{fig:3P0-RGI}. Furthermore, we have further studied the chiral evolution of the relativistic chiral nuclear force~\cite{Bai:2020yml, Bai:2021uim}, which is in good agreement with the results of the HALQCD collaboration~\cite{Inoue:2011ai}.

\begin{figure*}[htbp]
\centering
\includegraphics[width=0.36\textwidth]{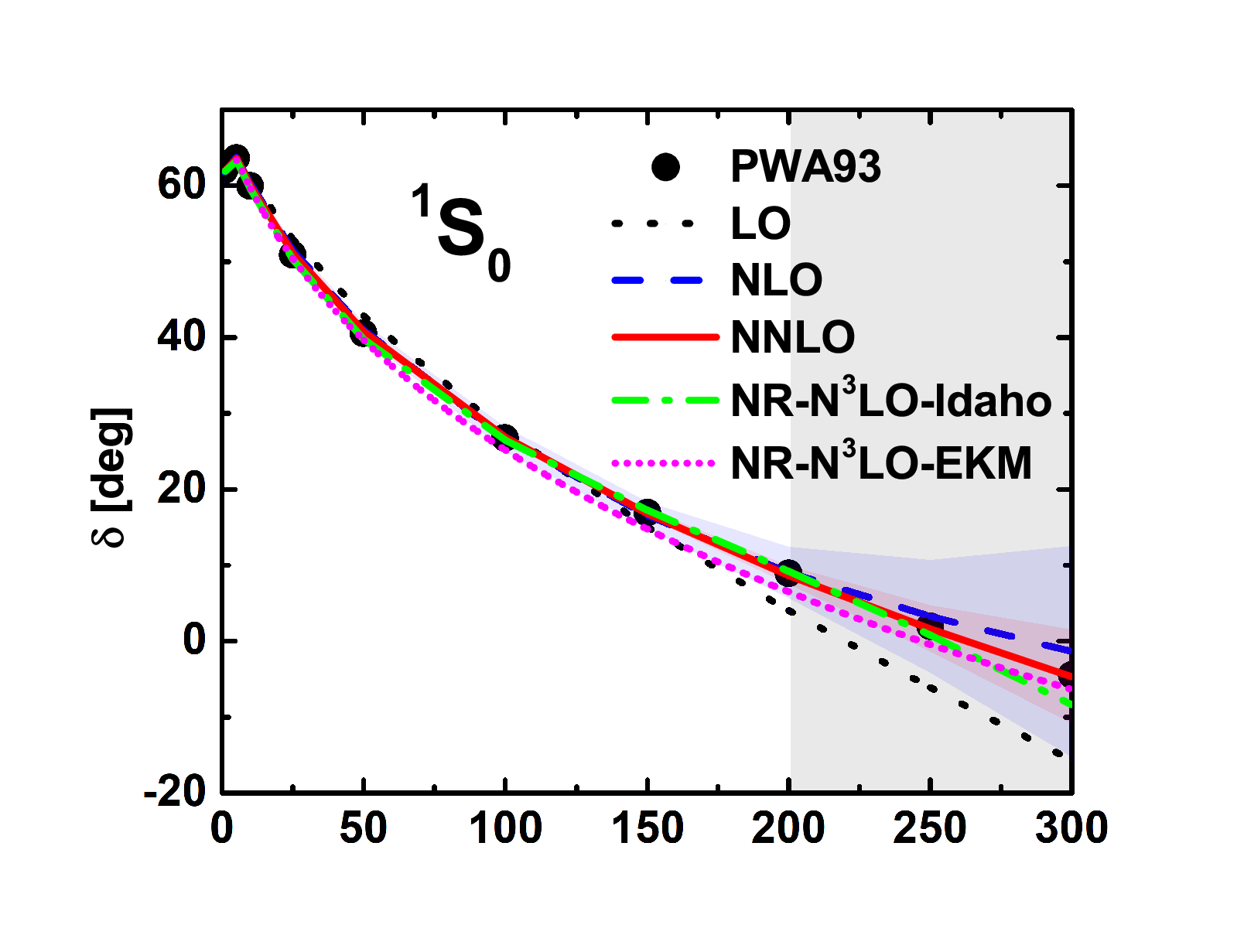}\hspace{-13mm}
\includegraphics[width=0.36\textwidth]{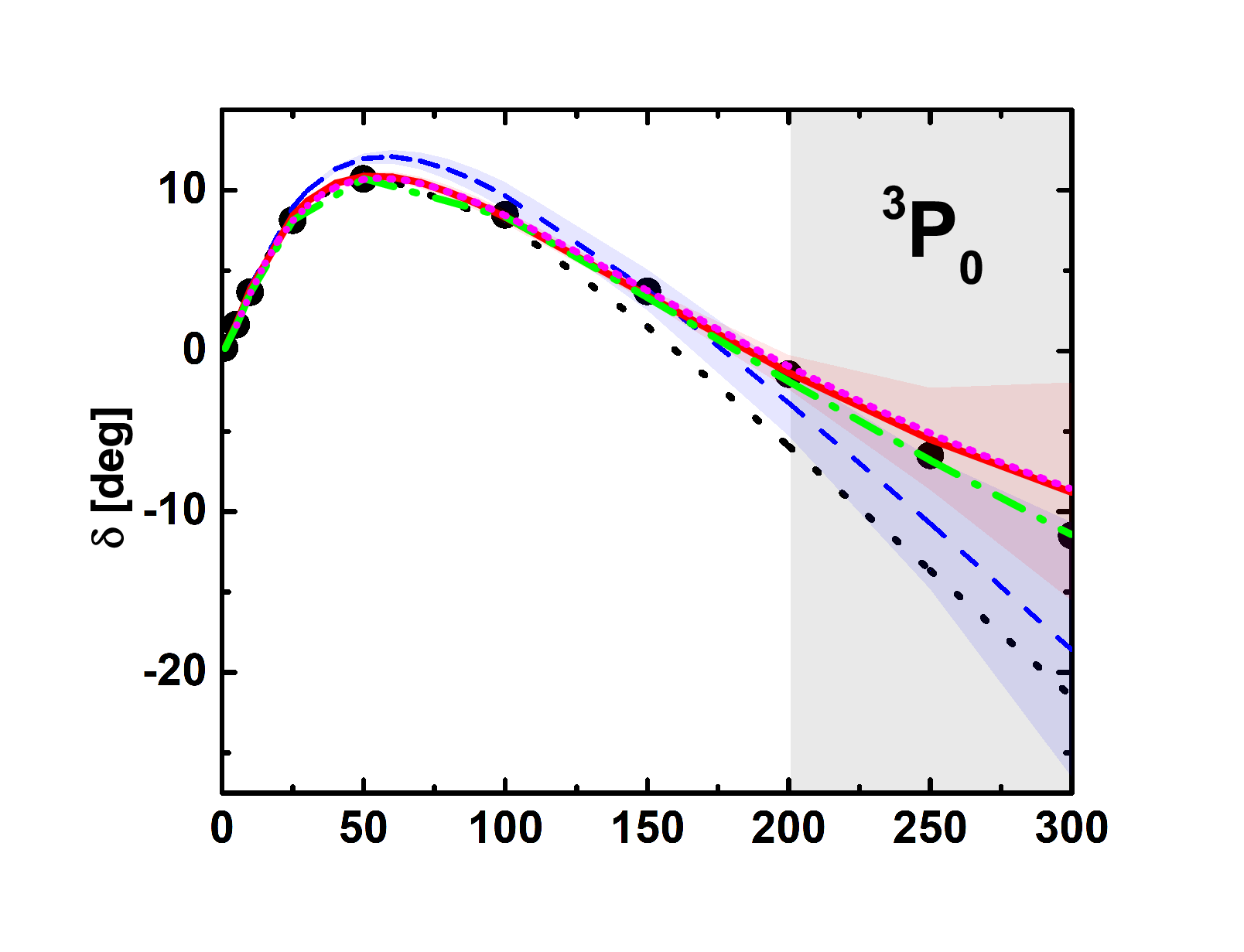}\hspace{-13mm}
\includegraphics[width=0.36\textwidth]{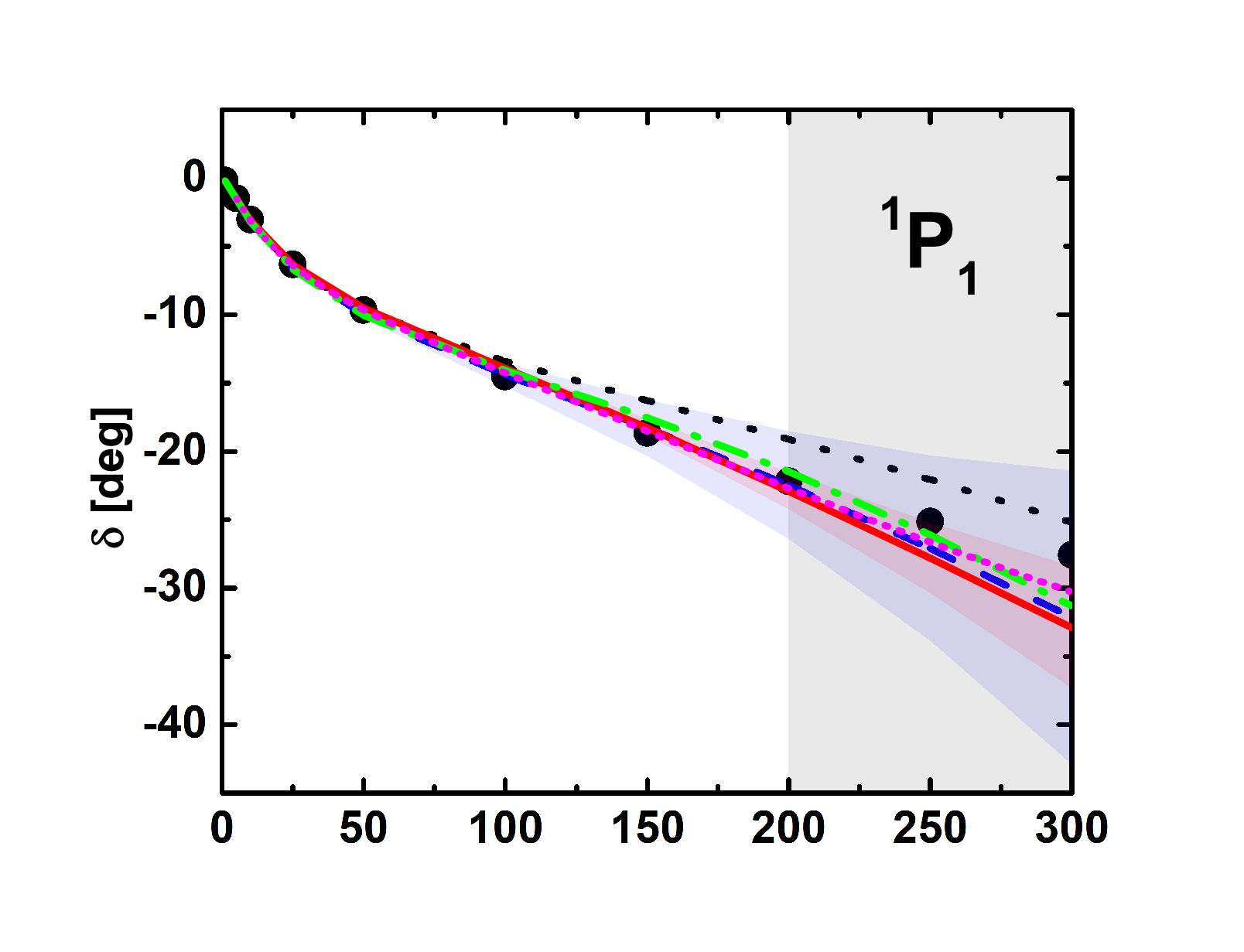}\\ \vspace{-9mm}
\includegraphics[width=0.36\textwidth]{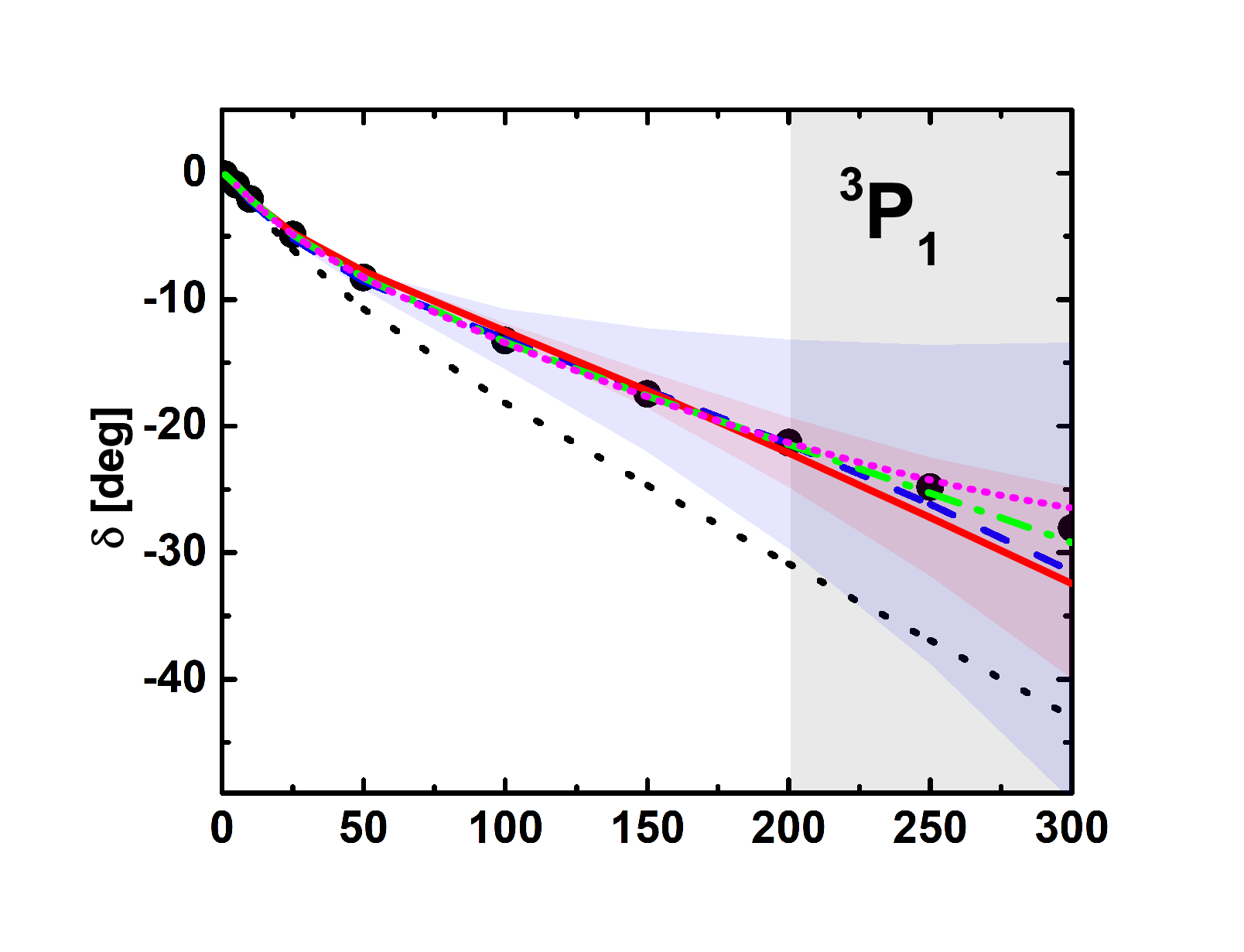}\hspace{-13mm}
\includegraphics[width=0.36\textwidth]{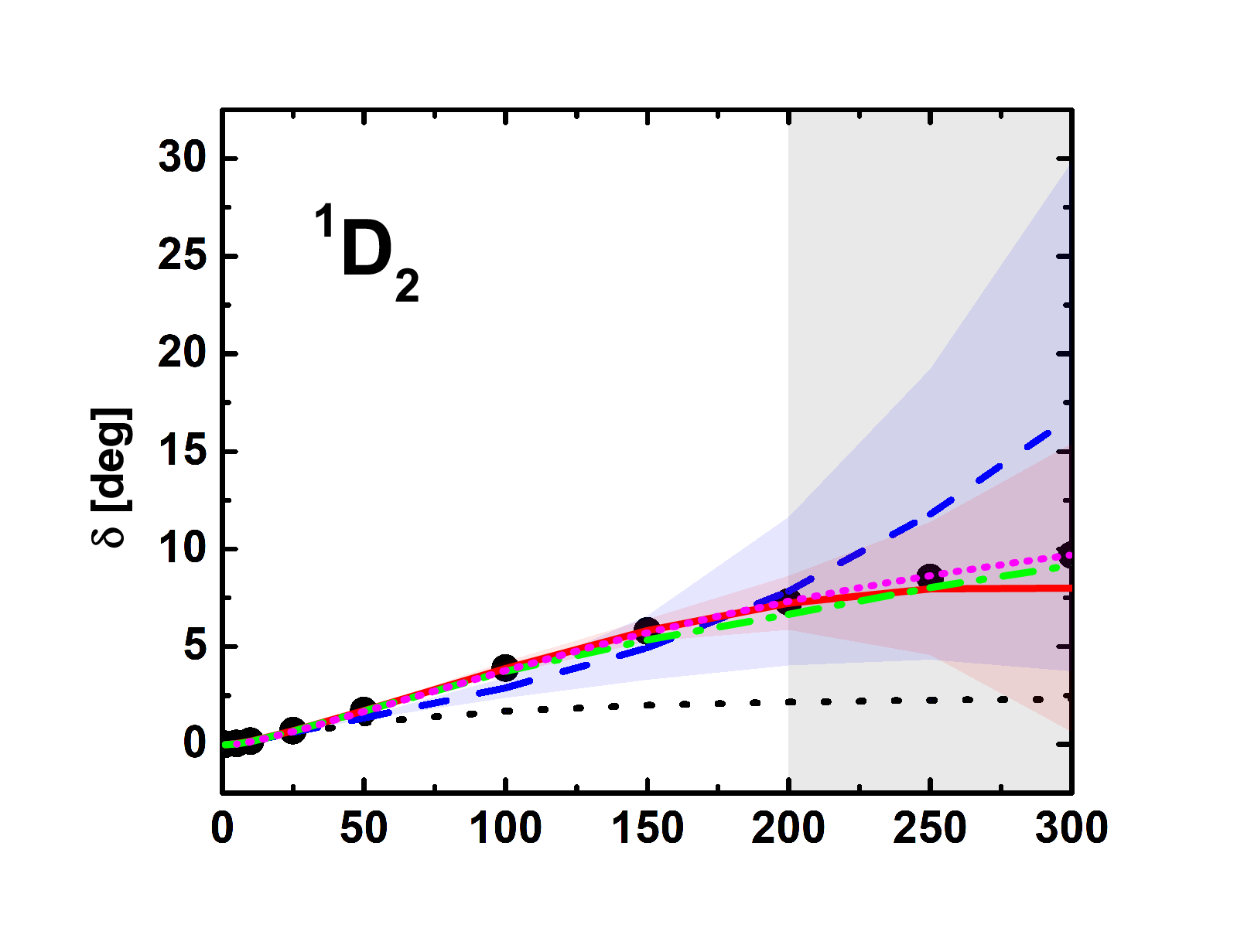}\hspace{-13mm}
\includegraphics[width=0.36\textwidth]{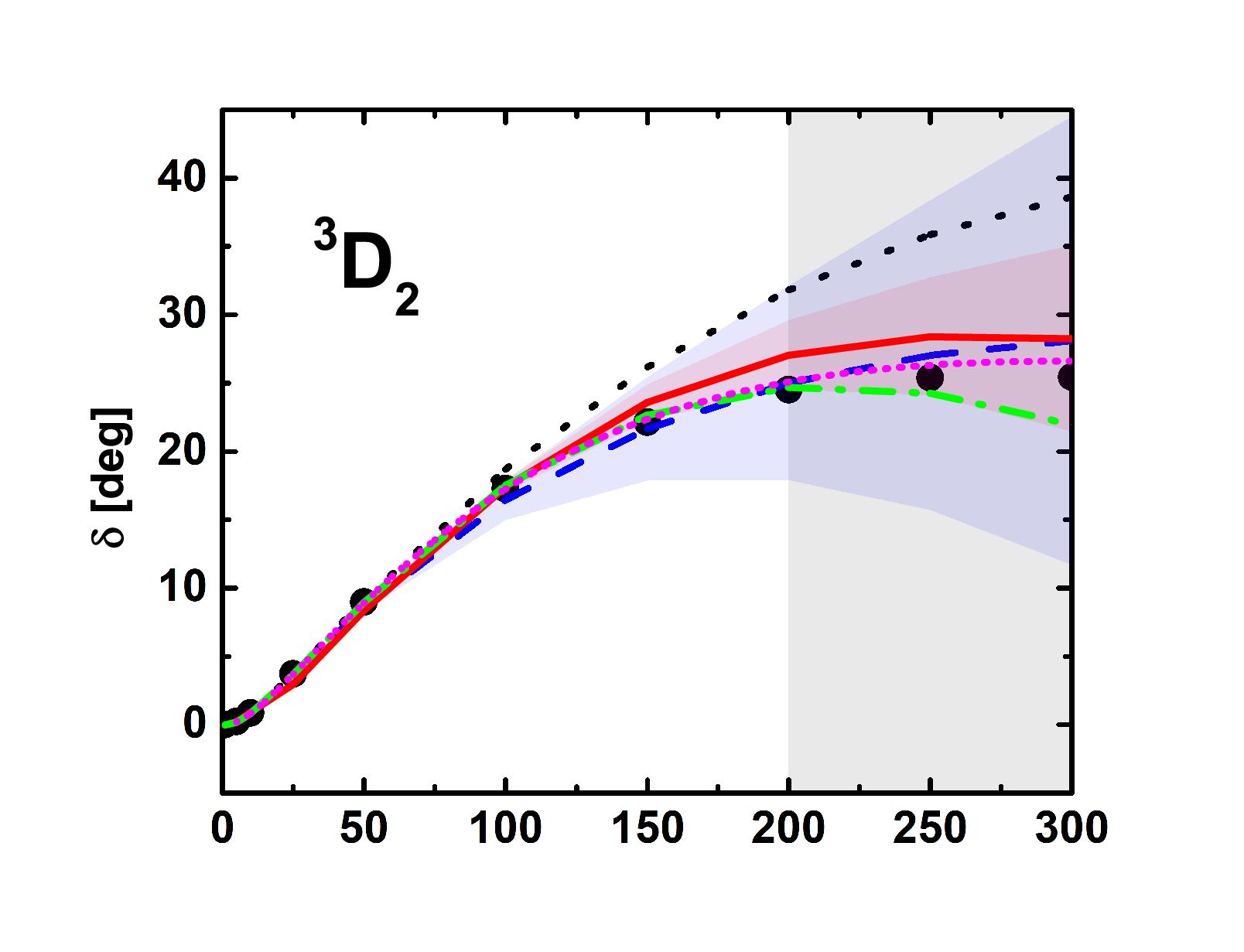}\\ \vspace{-9mm}
\includegraphics[width=0.36\textwidth]{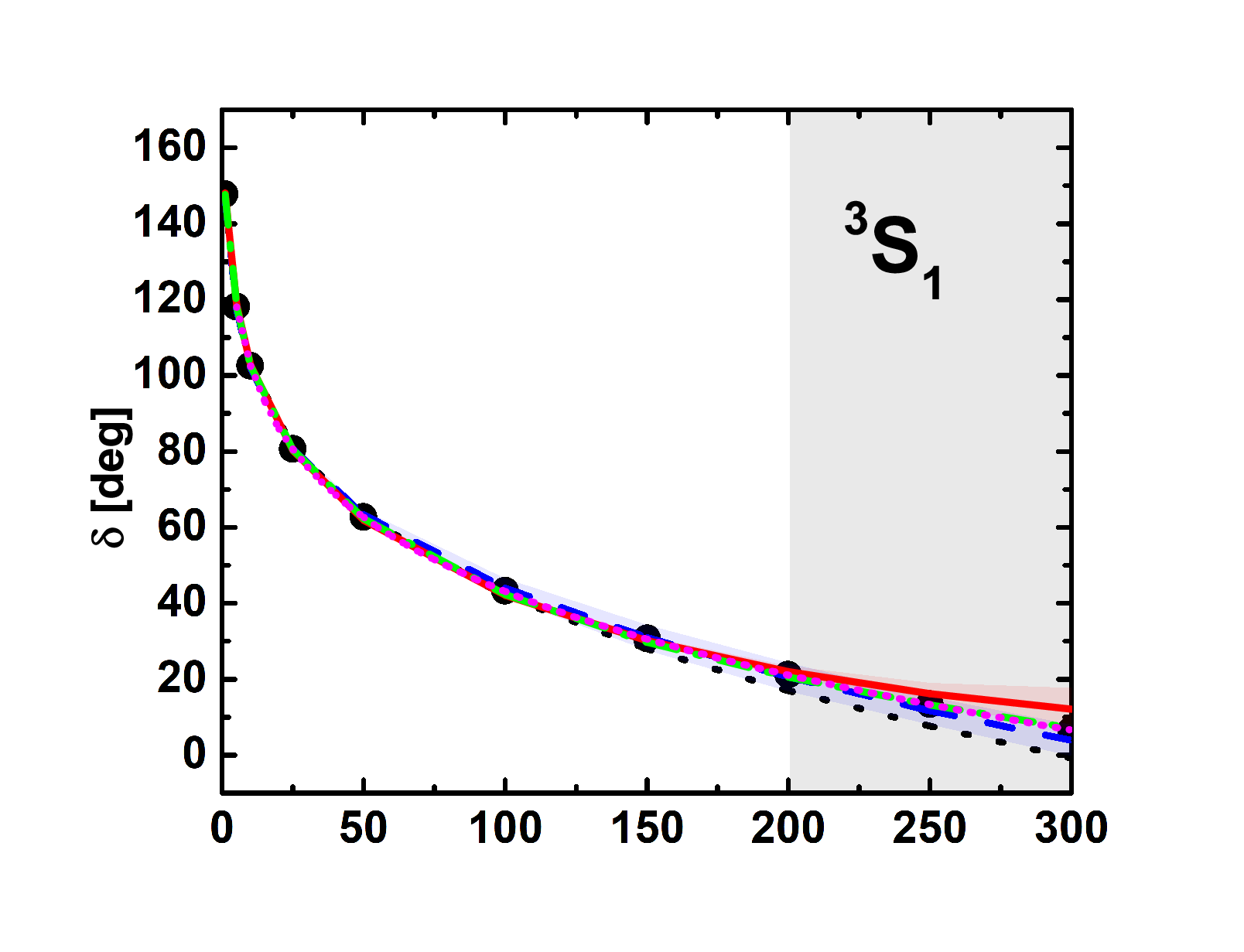}\hspace{-13mm}
\includegraphics[width=0.36\textwidth]{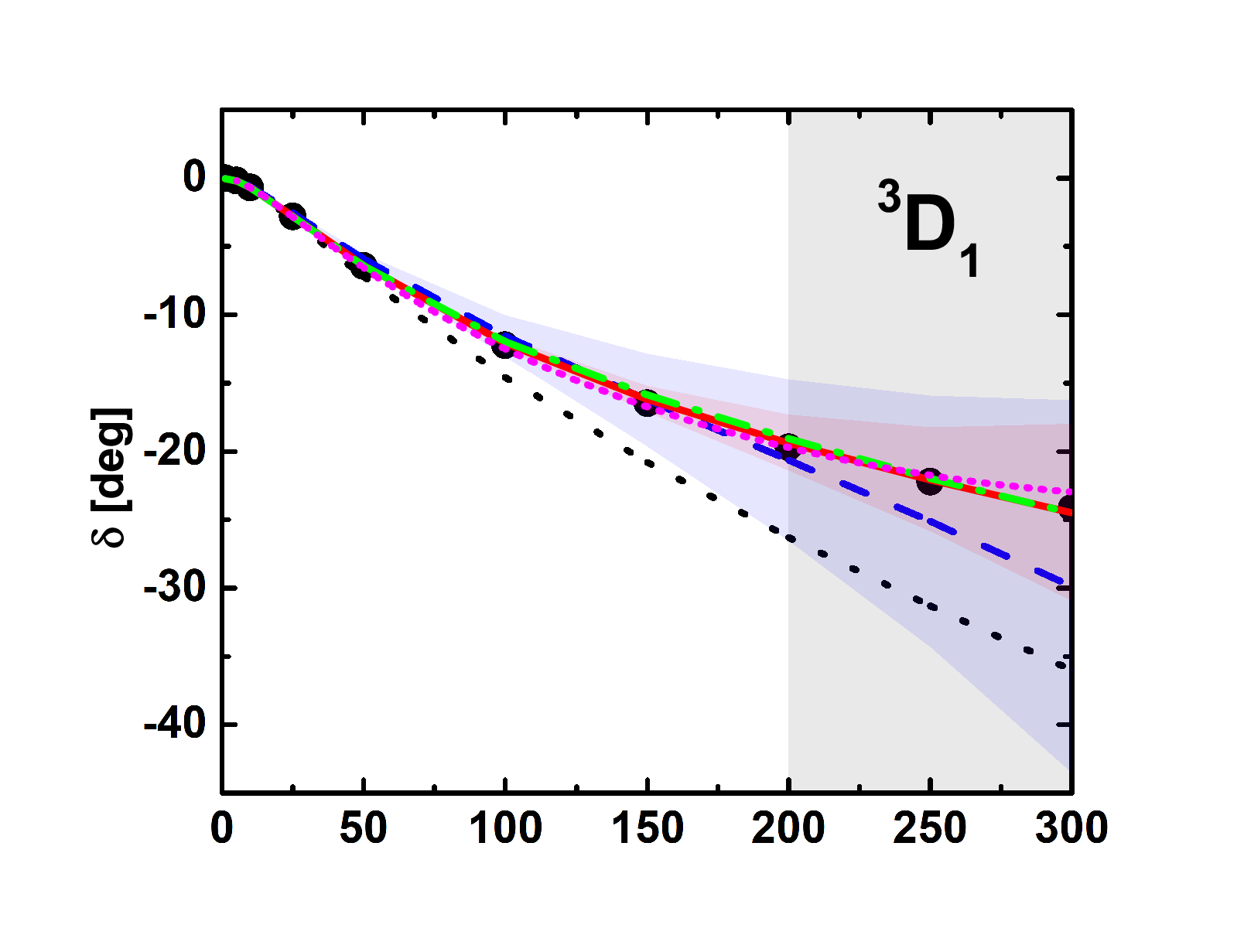}\hspace{-13mm}
\includegraphics[width=0.36\textwidth]{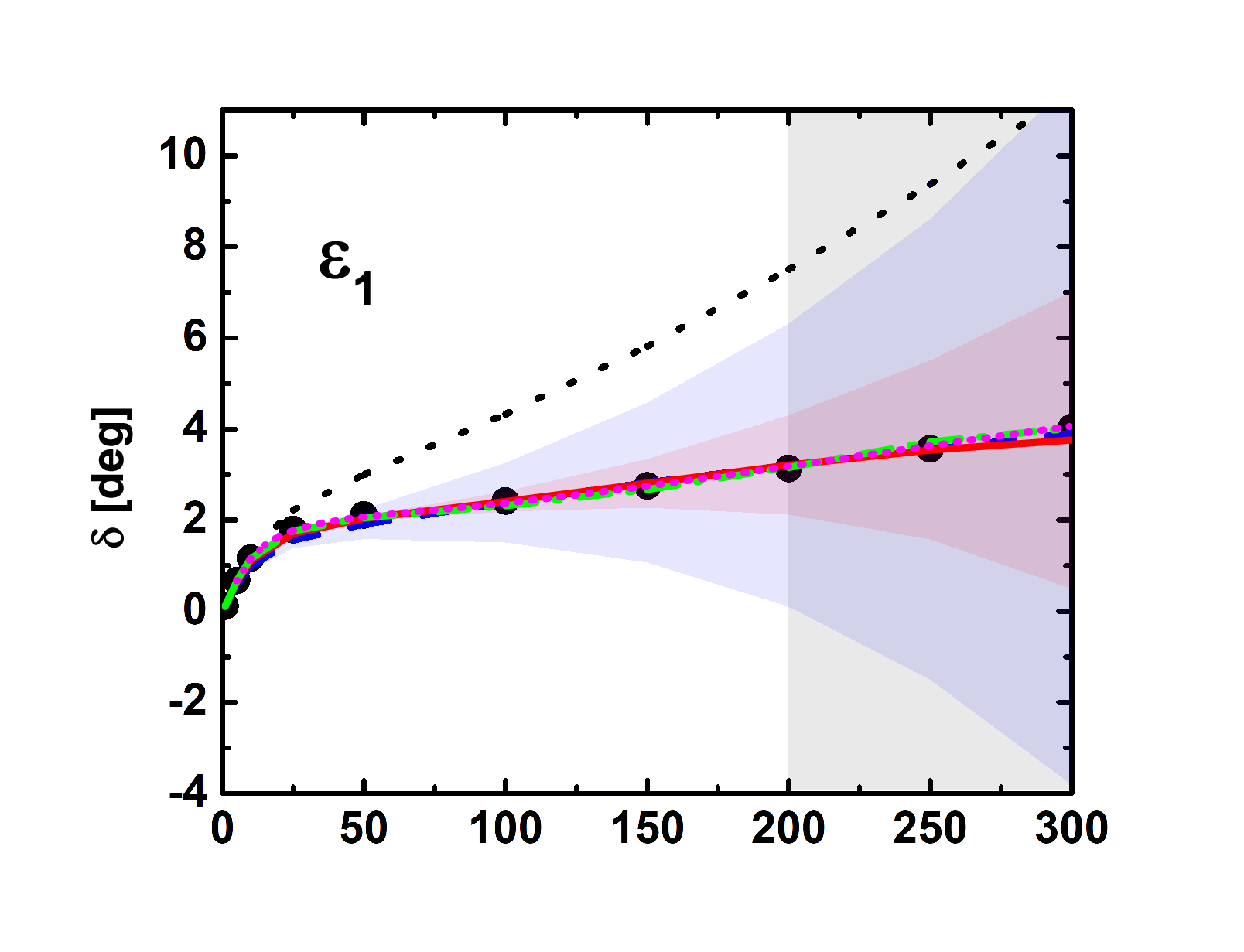}\\ \vspace{-9mm}
\includegraphics[width=0.36\textwidth]{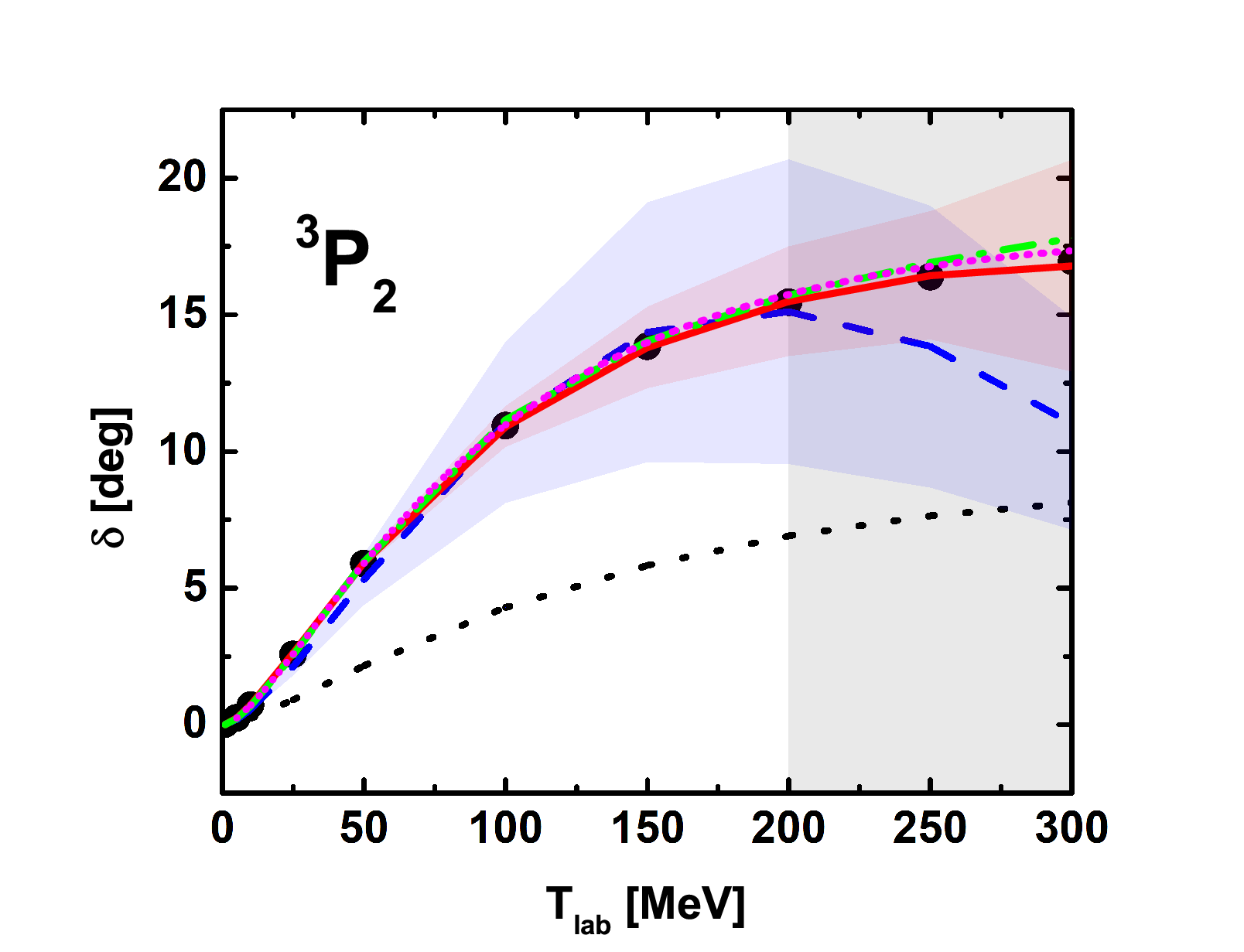}\hspace{-13mm}
\includegraphics[width=0.36\textwidth]{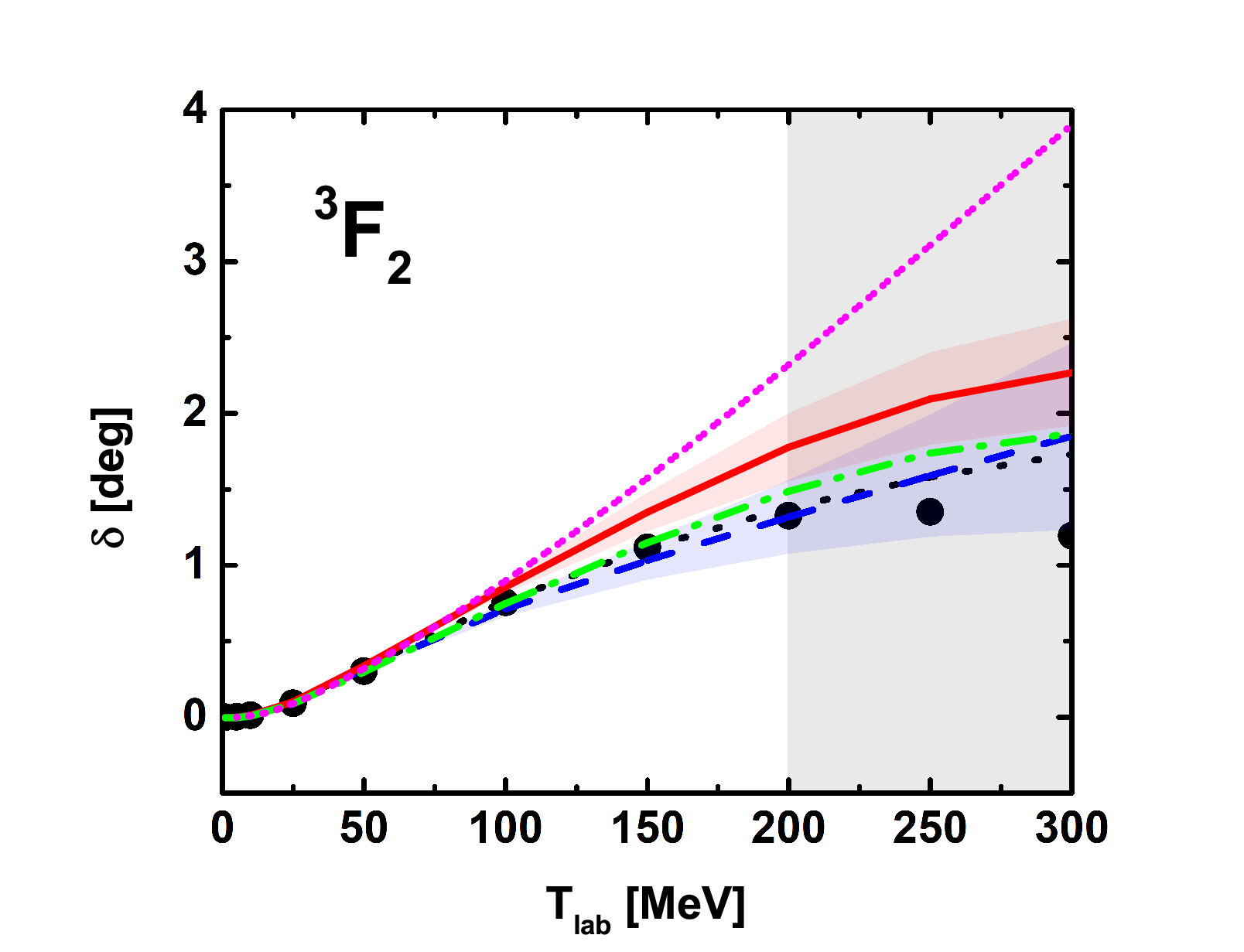}\hspace{-13mm}
\includegraphics[width=0.36\textwidth]{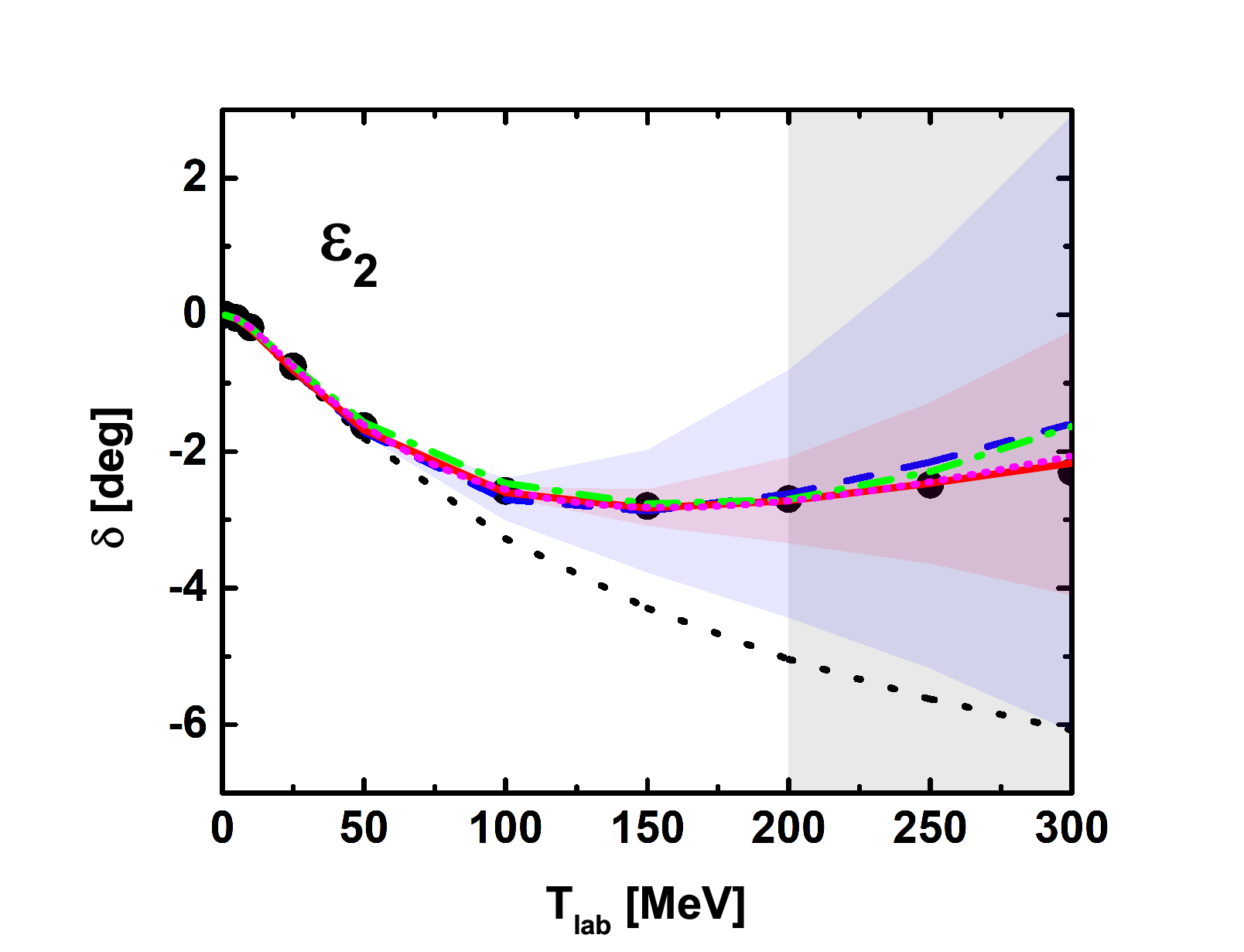}
\caption{Scattering phase shifts of the $J\leq 2$ partial waves provided by different nuclear forces. The red solid lines in the figure result from our NNLO relativistic chiral nuclear force and the momentum cutoff is set at $\Lambda=0.9$ GeV. The blue dashed lines result from the NLO relativistic chiral nuclear force, and the momentum cutoff is set at $\Lambda=0.6$ GeV. The corresponding shaded intervals are the theoretical uncertainties at the 68$\%$ confidence level. For comparison, we also present the results of the relativistic leading order (black dotted line, momentum cutoff $\Lambda=0.6$ GeV) and two non-relativistic N$^3$LO chiral nuclear forces, namely NR-N$^3$LO-Idaho ($\Lambda=0.5$ GeV, green dot-dashed line)~\cite{Entem:2003ft,Machleidt:2011zz}, and NR-N$^3$LO-EKM ($\Lambda=0.9$ fm, purple short dot-dashed line)~\cite{Epelbaum:2014efa,Epelbaum:2014sza}. The black solid points are the Nijmegen partial-wave analysis results~\cite{Stoks:1993tb}.}
\label{fig:EX-uncertainties}
\end{figure*}

\begin{figure*}[htpb]
\centering
\includegraphics[scale=0.24]{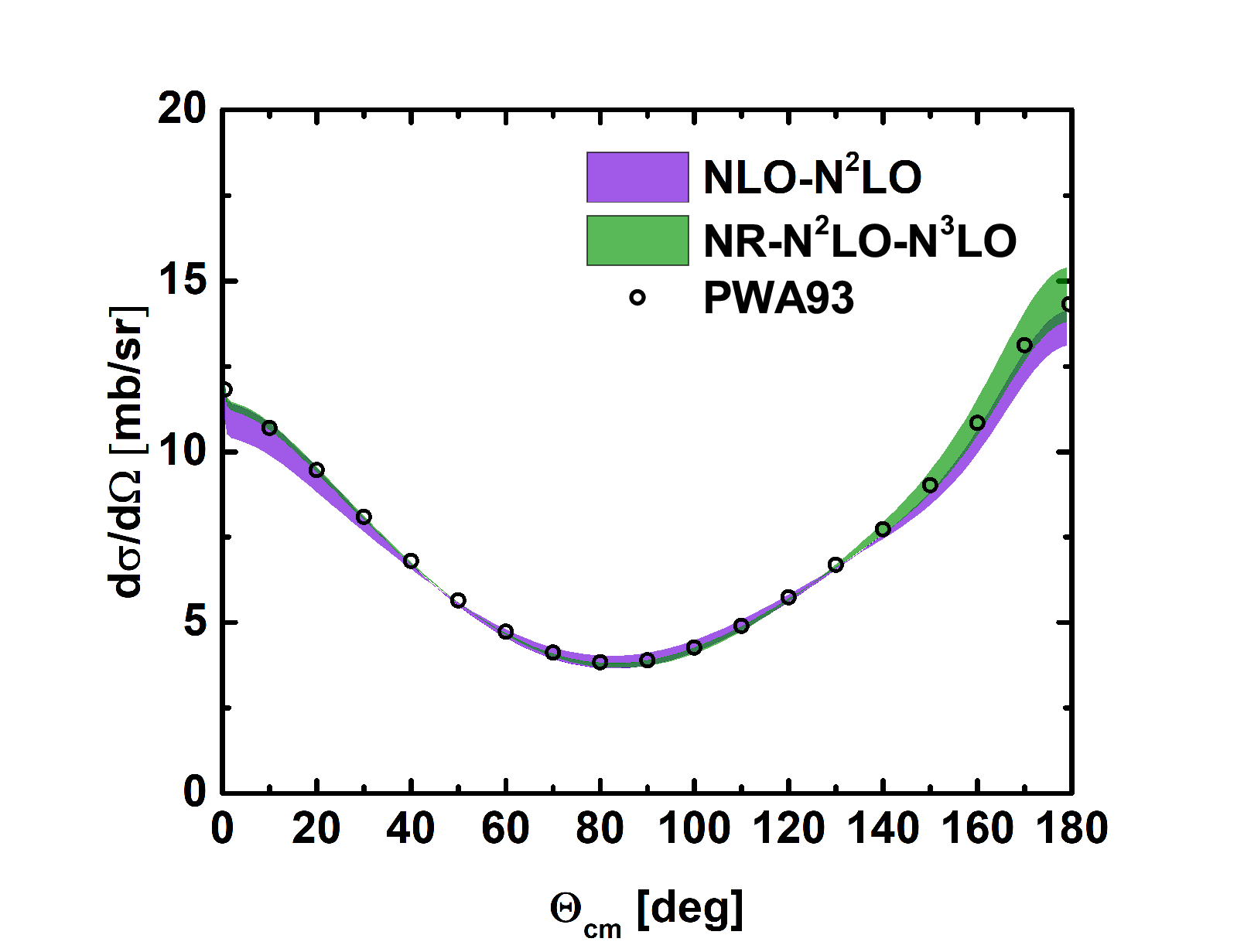}\hspace{-13mm}
\includegraphics[scale=0.24]{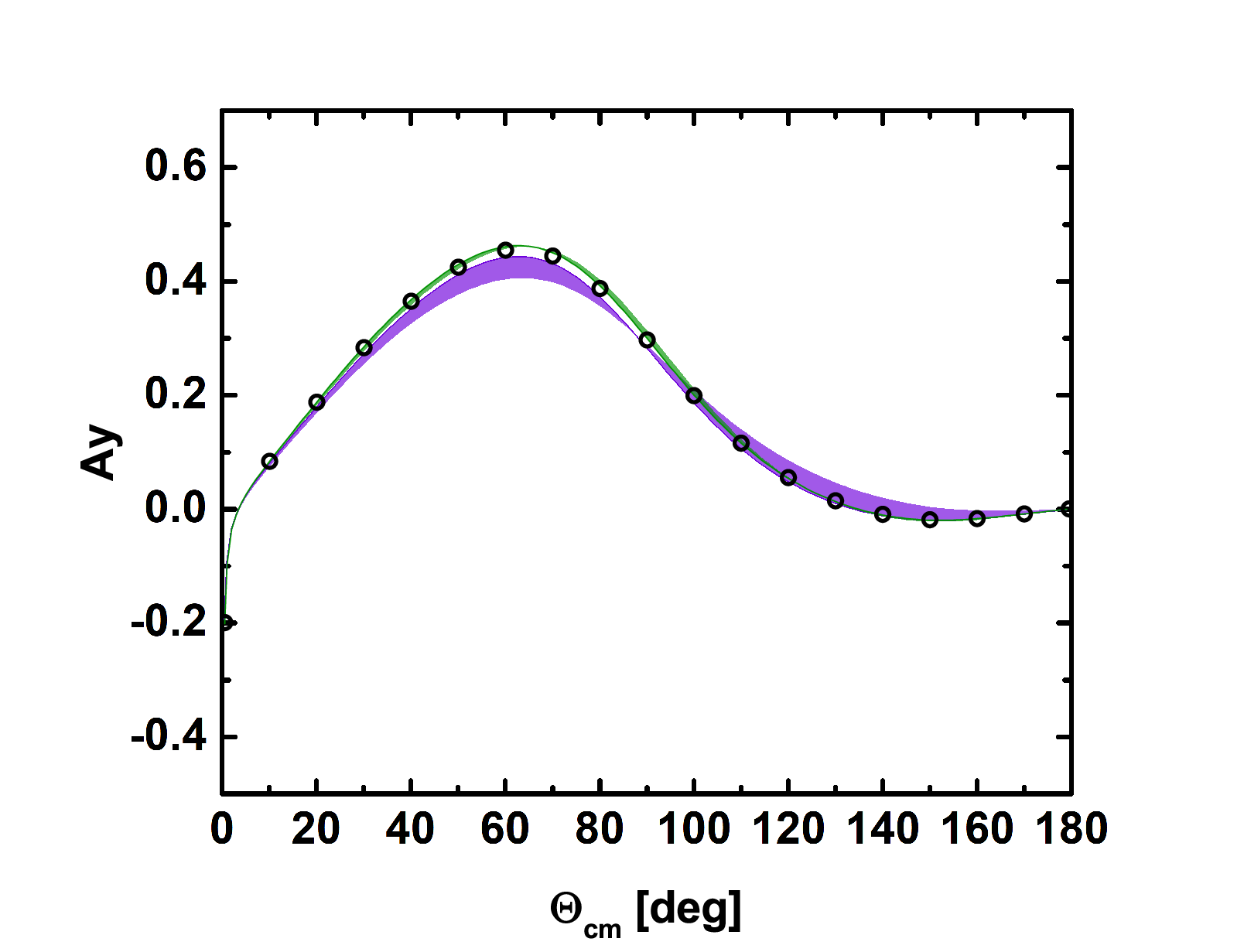}\hspace{-13mm}
\includegraphics[scale=0.24]{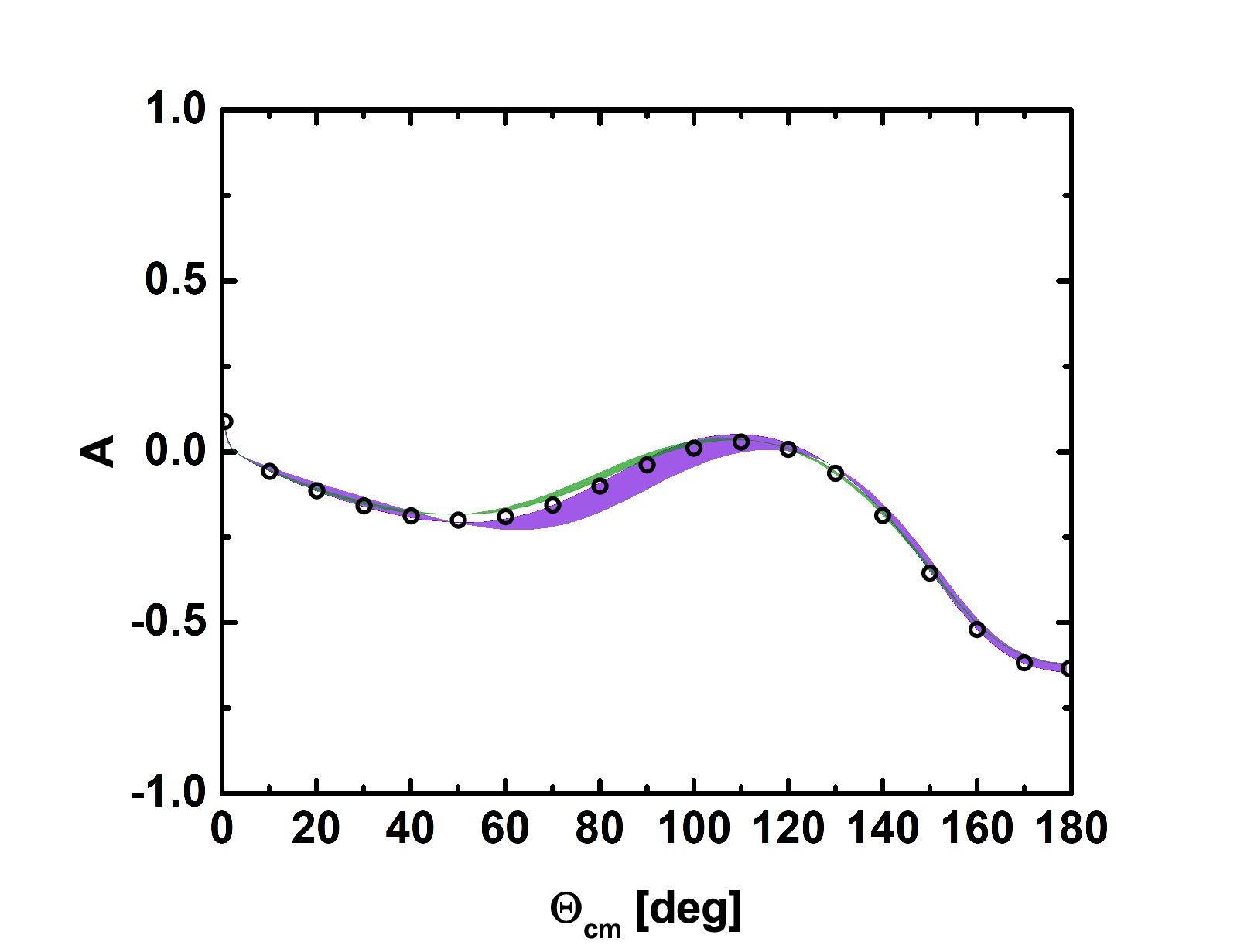}\\
\includegraphics[scale=0.24]{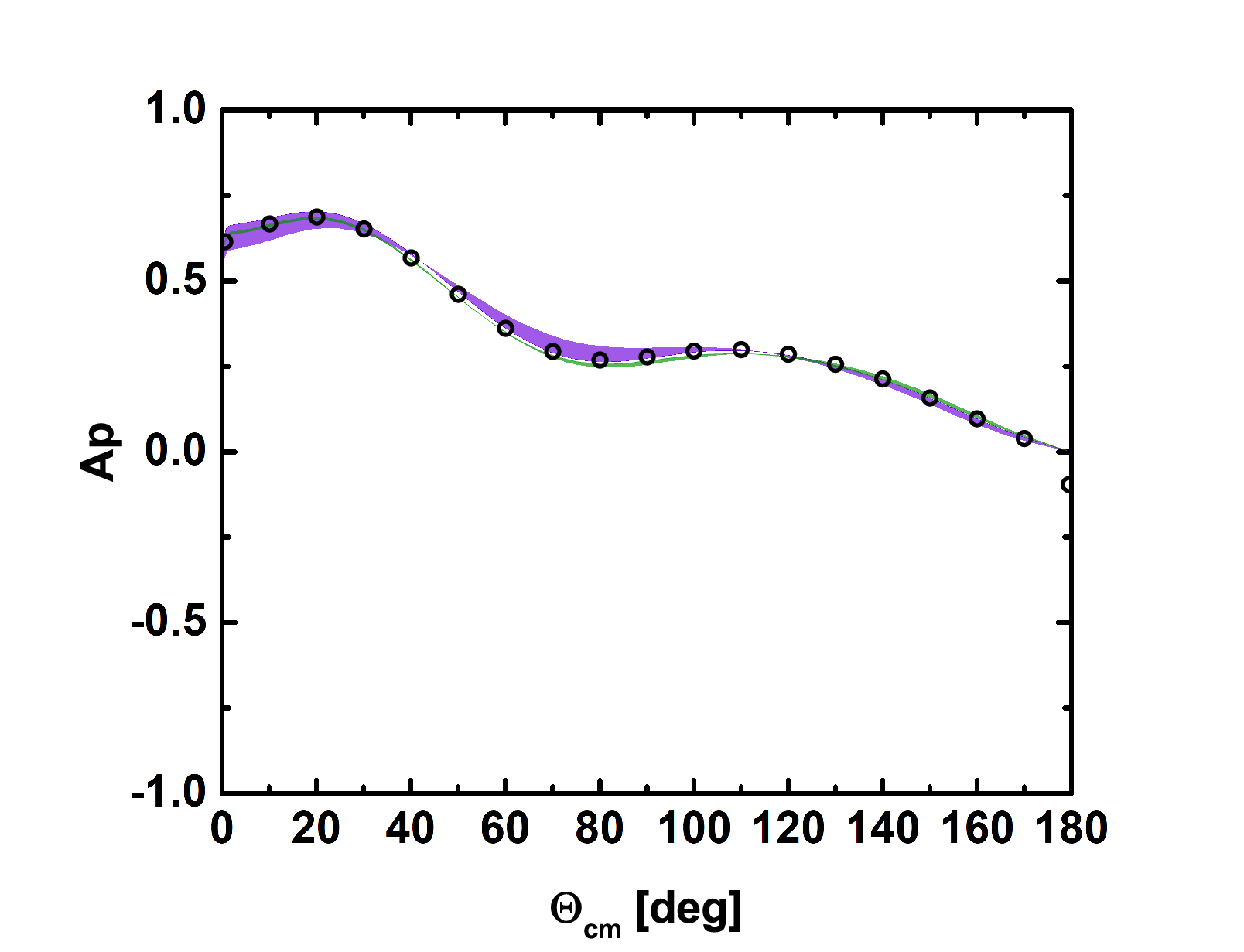}\hspace{-13mm}
\includegraphics[scale=0.24]{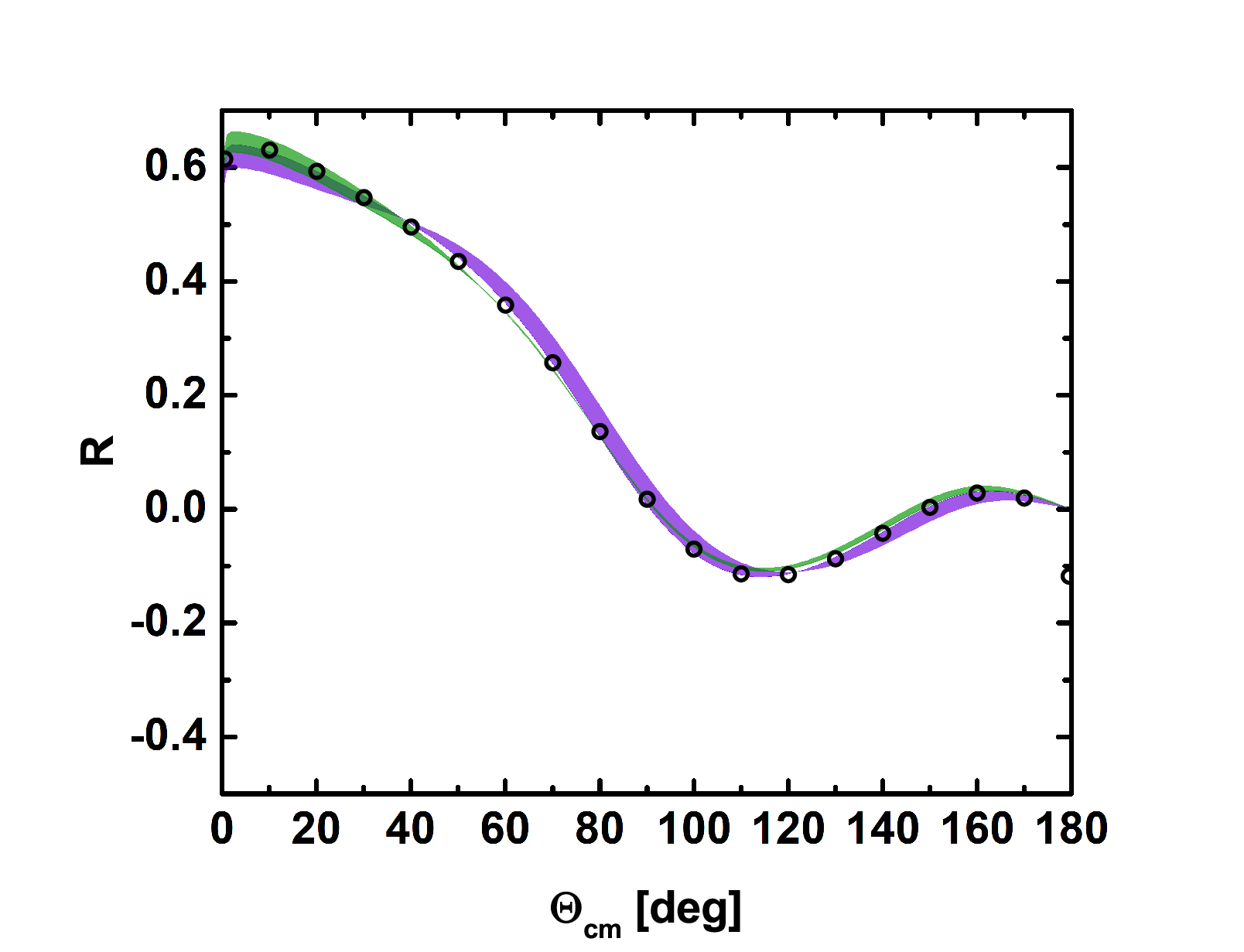}\hspace{-13mm}
\includegraphics[scale=0.24]{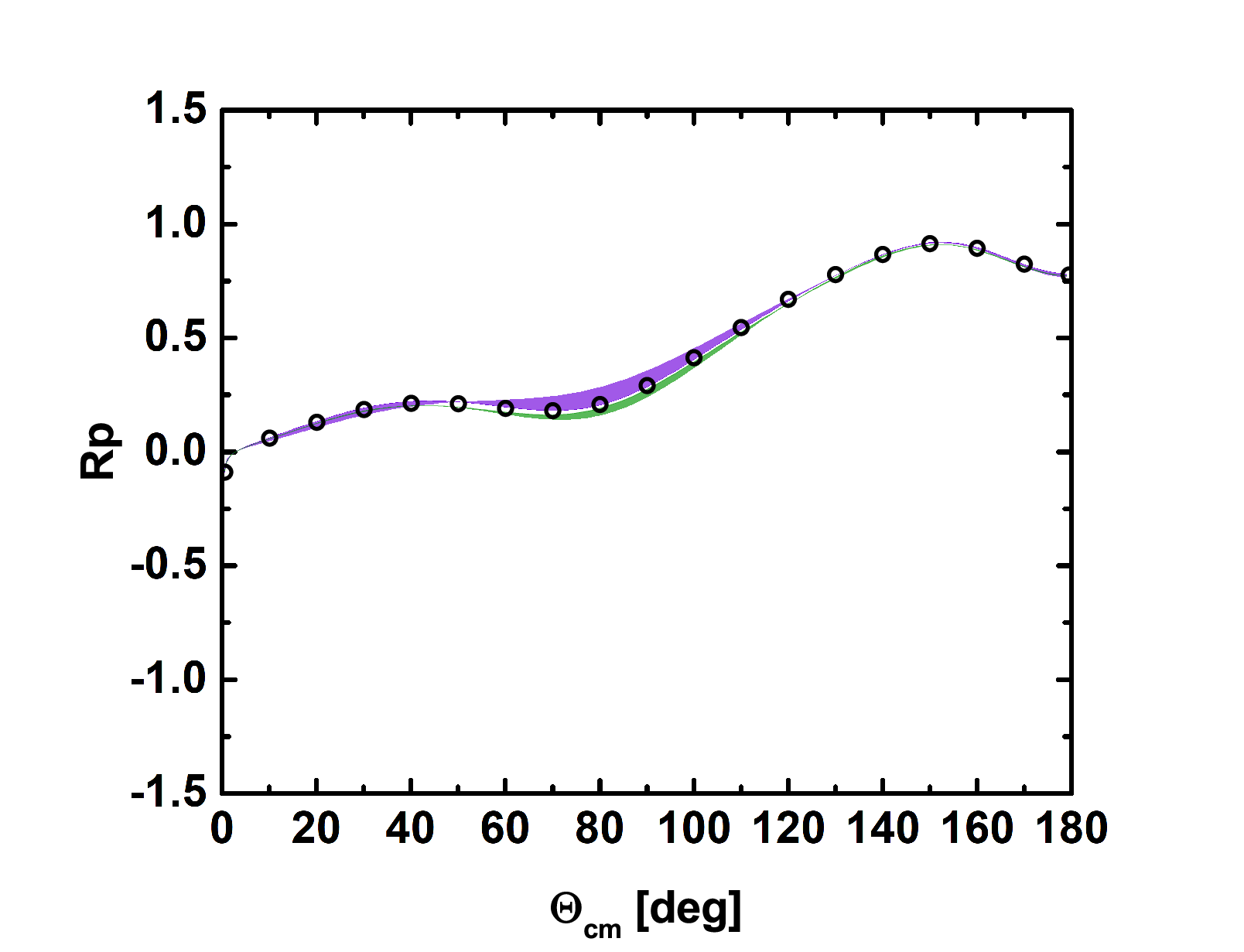}
\caption{ Descriptions of d$\sigma$/d$\Omega$, $A_y$, $A$, $A_p$, $R$, $R_p$ for $T_{\text{lab}}= 100$ MeV . The purple bands are the results of the relativistic chiral nuclear force at NLO and NNLO, the green bands represent the results of the Weinberg chiral nuclear force at NNLO and N$^3$LO, and the circles represent the PWA93 results.}\label{ob100}
\end{figure*}

\begin{figure*}[htpb]
\centering
\includegraphics[scale=0.24]{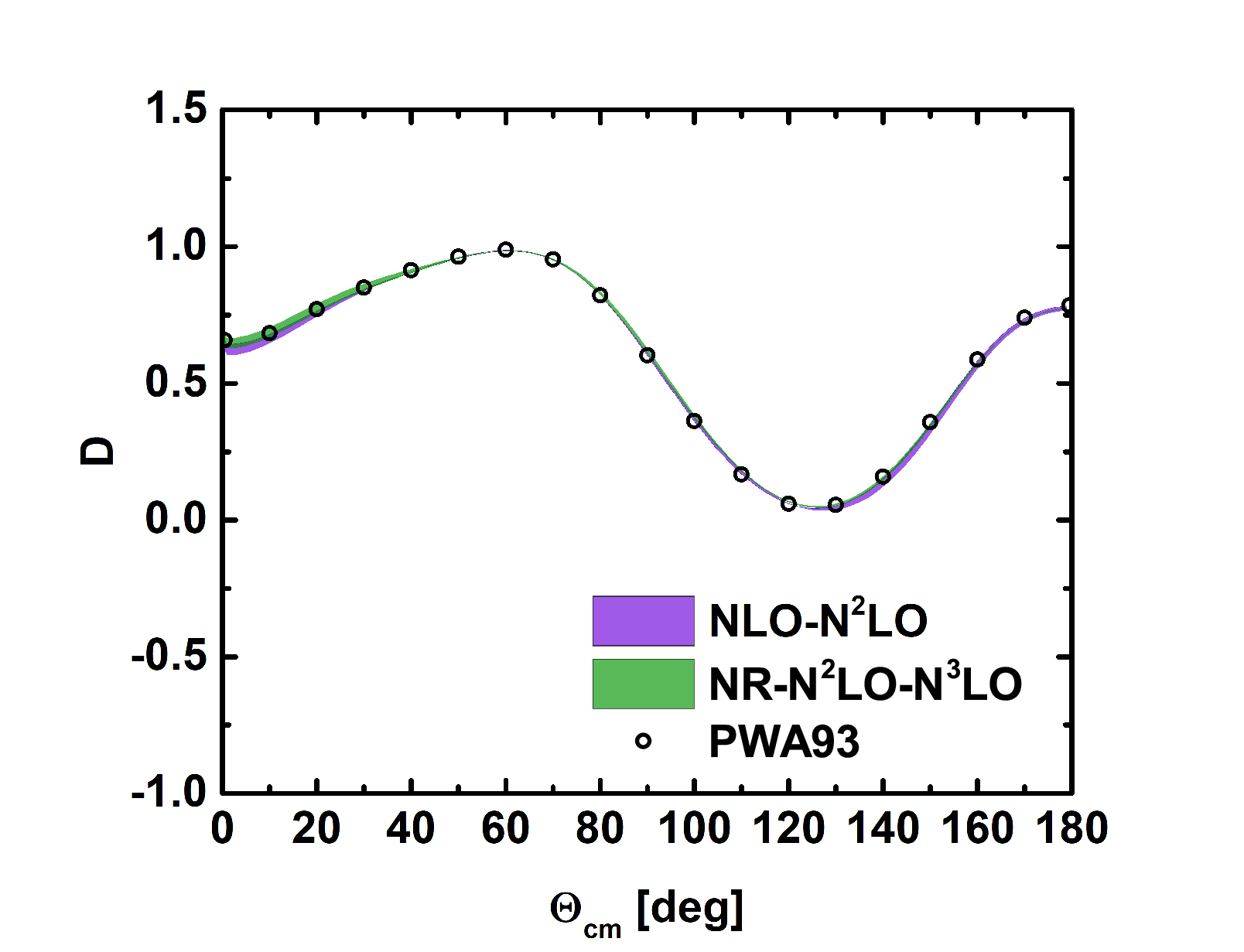}\hspace{-13mm}
\includegraphics[scale=0.24]{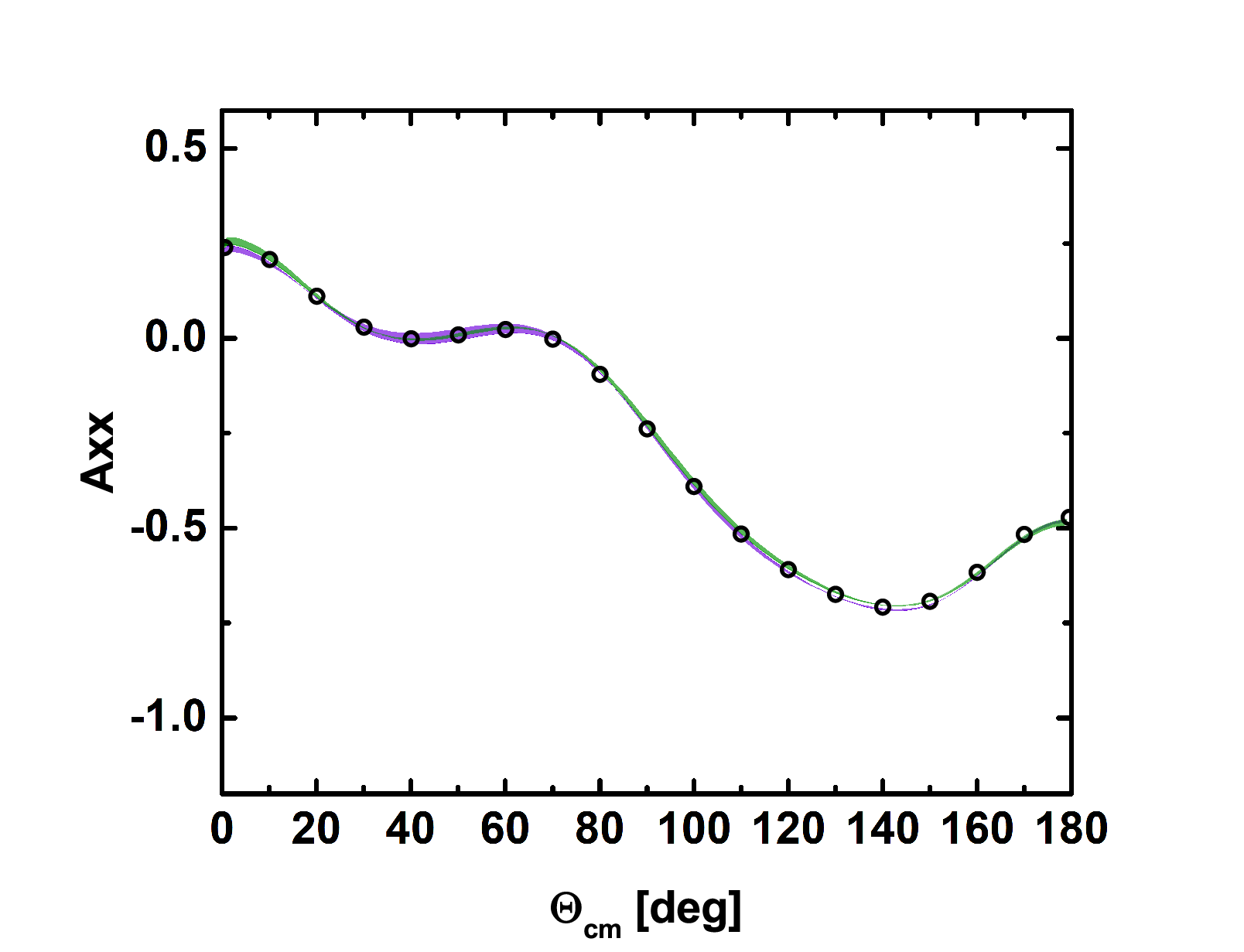}\hspace{-13mm}
\includegraphics[scale=0.24]{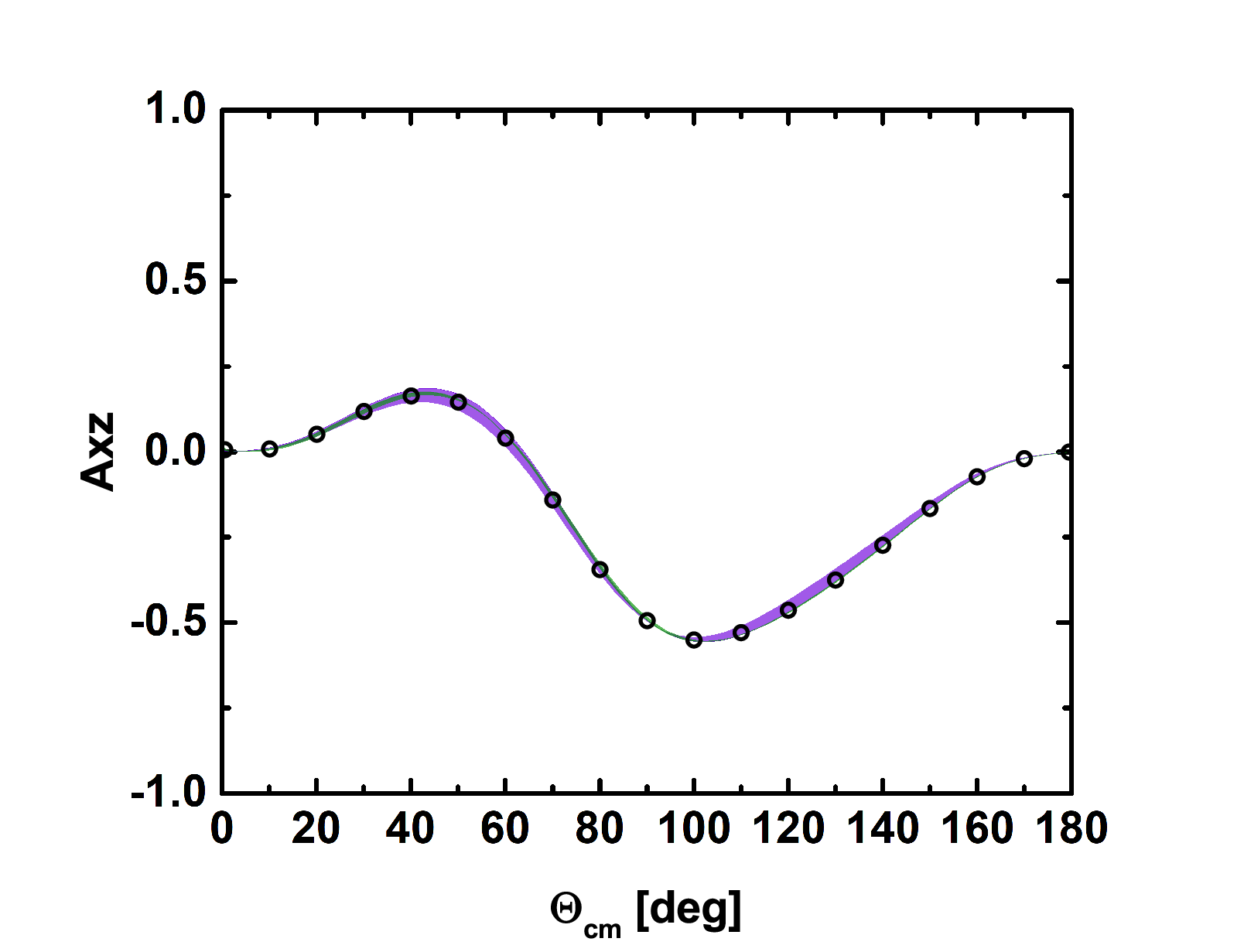}\\
\includegraphics[scale=0.24]{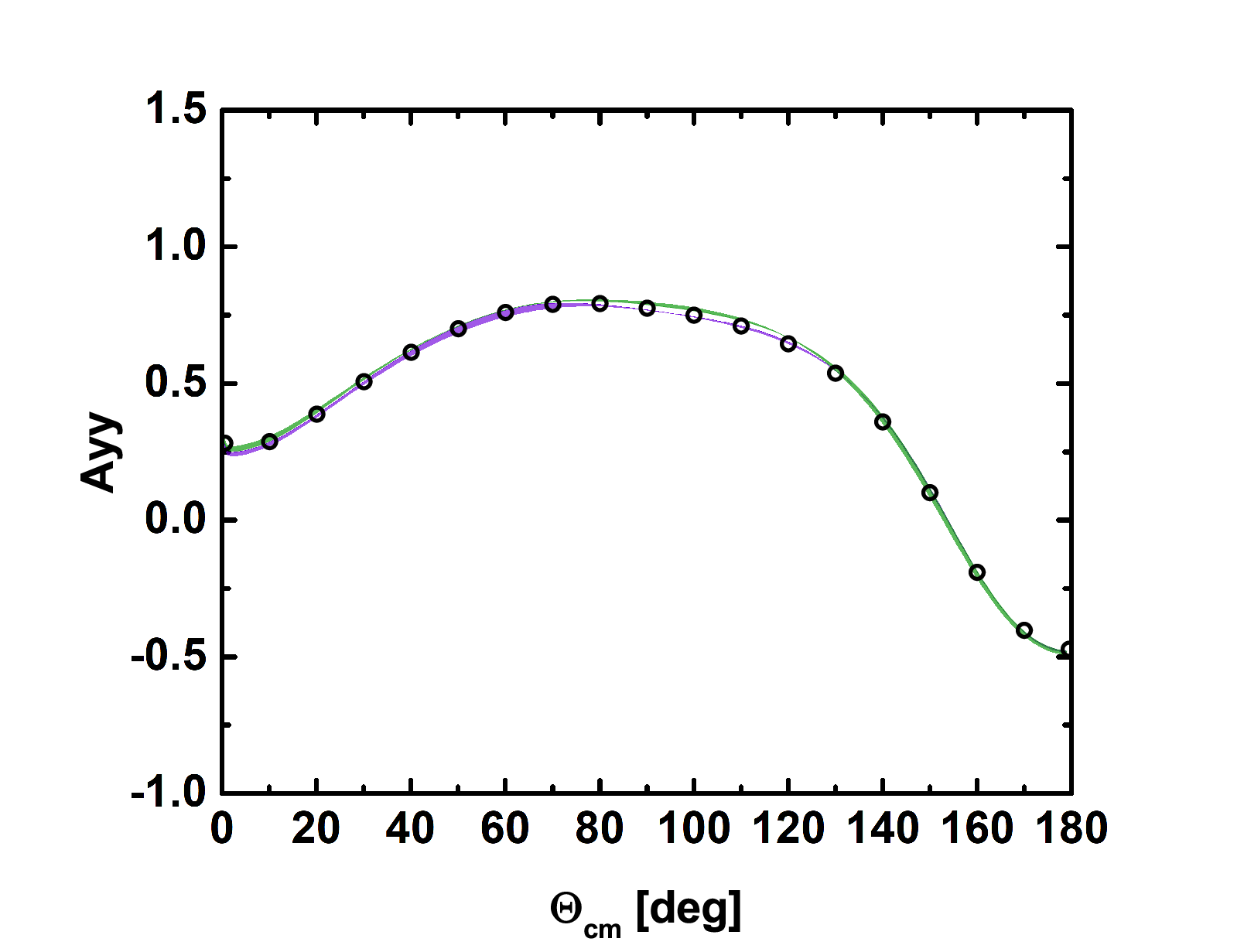}\hspace{-13mm}
\includegraphics[scale=0.24]{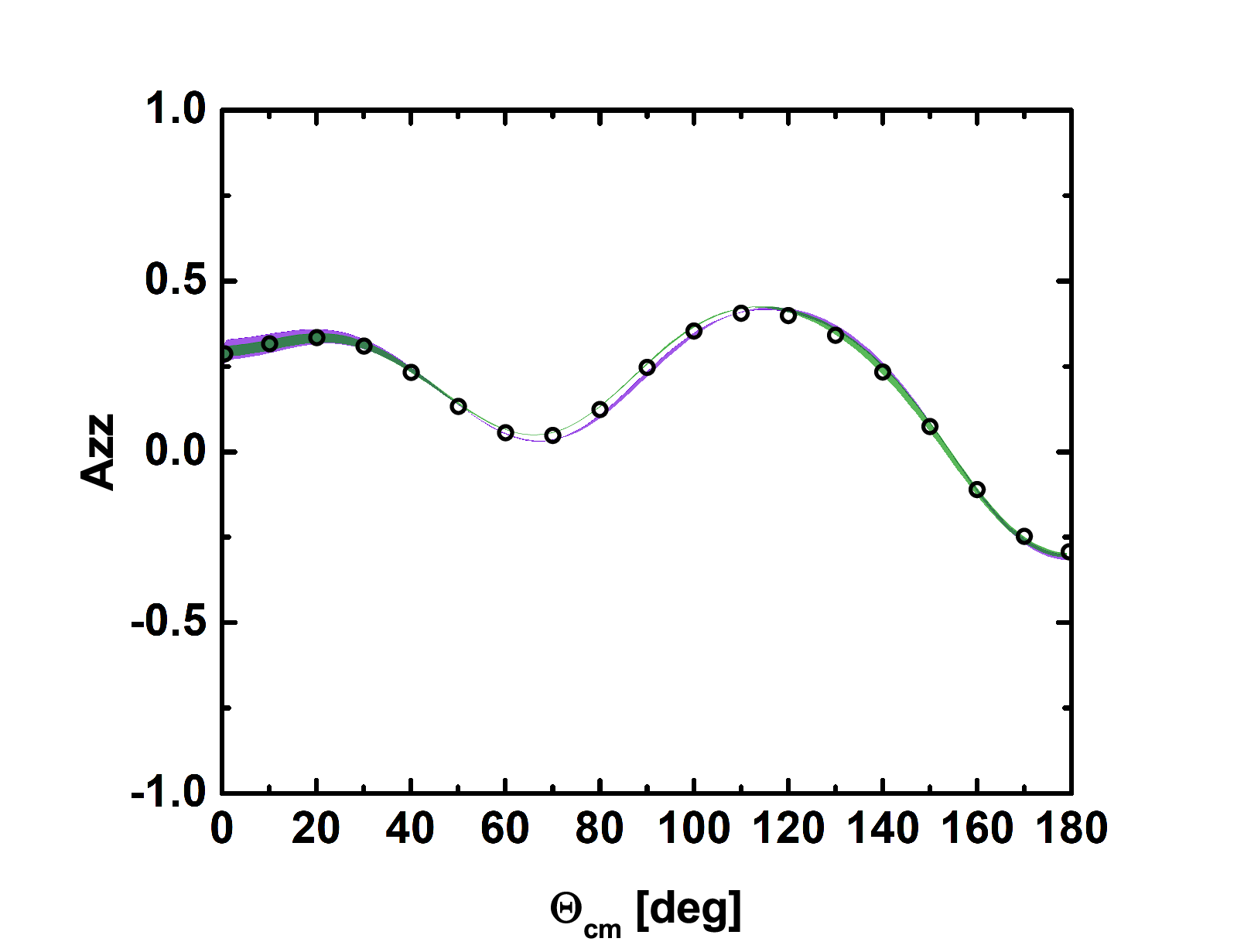}\hspace{-13mm}
\caption{Description of $D$, $A_{xx}$, $A_{xz}$, $A_{yy}$, and $A_{zz}$ for $T_{\text{lab}}= 100$ MeV. The purple bands represent the results of the relativistic chiral nuclear force at NLO and NNLO, the green bands represent the results of the non-relativistic chiral nuclear force at NNLO and N$^3$LO, and the circles represent the PWA93 results.}\label{ob1000}
\end{figure*}

Although the leading-order relativistic chiral nuclear force is as good as the Weinberg next-to-leading-order results in describing the partial waves with total angular momentum $J\leq 1$, its description accuracy for partial waves of higher angular momentum $J >1$ is still insufficient~\cite{Ren:2016jna}. To achieve a high-precision description of nucleon-nucleon scattering data, we have calculated the chiral nuclear force up to the next-to-next-to-leading order within the relativistic framework~\cite{Lu:2021gsb}.

\subsection{relativistic chiral nuclear forces up to NNLO}

Up to the next-next-to-leading order, the complete relativistic chiral nuclear force consists of the following contributions
\begin{equation}\label{NNForce}  V=V_{\mathrm{CT}}^{\mathrm{LO}}+V_{\mathrm{CT}}^{\mathrm{NLO}}+V_{\mathrm{OPE}}+V_{\mathrm{TPE}}^{\mathrm{NLO}}+V_{\mathrm{TPE}}^{\mathrm{NNLO}}-V_{\mathrm{IOPE}},
\end{equation}
which includes the contact term interactions at LO [$\mathcal{O}(p^0)$] and NLO [$\mathcal{O}(p^2)$], the OPE terms, the leading order and next-to-leading order TPE contributions. The last term subtracts the double-counted OPE.

After fitting the data to determine the unknown low-energy constants, the description of the scattering phase shifts by the next-to-next-to-leading-order relativistic chiral nuclear force is shown in Fig.~\ref{fig:EX-uncertainties}. The theoretical uncertainty at the 68$\%$ confidence level is estimated by the Bayesian method~\cite{Furnstahl:2015rha,Melendez:2017phj,Melendez:2019izc}. For comparison, we also present the results obtained by the Weinberg N$^3$LO with different regularization schemes, denoted by ``Idaho" \cite{Entem:2003ft,Machleidt:2011zz} and ``EKM" \cite{Epelbaum:2014efa,Epelbaum:2014sza}, respectively.

We note that within the relativistic framework, the NLO and NNLO descriptions of the phase shifts are in very good agreement with the experimental data up to $T_\mathrm{lab}=200$ MeV and are also similar to the results of the Weinberg N$^3$LO chiral forces. In particular, the relativistic NLO phase shifts almost overlap with the NNLO ones, which better agrees with the experimental data. This shows the relatively fast convergence of the chiral expansion in the relativistic framework. On the other hand, for the $^3F_2$ partial wave, NLO performs better. This is because the sub-leading order TPE contribution of NNLO is relatively large, and the $^3F_2$ partial wave must be sacrificed in fitting all partial waves with $J=2$. When we remove the correlation between the $D$-wave and the $^3P_2$-$^3F_2$ partial wave, i.e., fit the $D$-wave and the $^3P_2$-$^3F_2$ partial waves independently, or slightly adjust the size of the momentum cutoff, the description can be significantly improved. Regarding this issue, we note that in the NR-N$^3$LO-Idaho, to describe better the $^3F_2$ partial wave phase shift, the low-energy constants for the $\pi$N scattering is treated as a semi-free parameter~\cite{Entem:2003ft}.

In Figs.~\ref{ob100} and~\ref{ob1000}, we present the predictions of the high-precision NNLO relativistic chiral nuclear force and the N$^3$LO Weinberg chiral nuclear force for the evolution of the differential scattering cross section ($DSG$), polarization ($P$), depolarization ($D$), transverse rotation parameters ($A$, $A^{\prime}$), longitudinal rotation parameters ($R$, $R^{\prime}$), and spin correlation parameters ($A_{xx}$, $A_{yy}$, $A_{zz}$, $A_{zx}$) with the scattering angle in the center-of-mass system for $T_{\text{lab}}= 100$ MeV. For comparison, we also present the Nijmegen results, where the purple band regions represent the results of the NLO and NNLO relativistic chiral nuclear forces, the green band regions represent the results of the NNLO and N$^3$LO non-relativistic chiral nuclear forces, and the black dots represent the Nijmegen results. The figures show that the high-precision NNLO covariant chiral nuclear force can describe these observables well.

\section{Anti-nucleon nucleon interaction in the relativistic framework}

The revivalistic framework can be naturally extended to study the antinucleon nucleon ($\bar{N} N$) interaction. Recently, the $\bar{N} N$ interaction is in heated discussions partly motivated by the observations of near-threshold $\bar{N}N$ enhancements in charmonium~ decays~\cite{BES:2003aic,BES:2005ega,BESIII:2011aa,BESIII:2016fbr,BESIII:2021xoh,BESIII:2023vvr}, $B$ meson decays~\cite{Belle:2002bro,Belle:2002fay}, and $e^+e^- \rightarrow \bar{p}p$ reactions~\cite{BaBar:2005sdl,BaBar:2005pon}, which provided an opportunity to elucidate the existence of speculated $\bar{N}N$ molecules and stimulated studies of the low-energy $\bar{N}N$ interaction. 

The $\bar{N}N$ interaction, still remaining much less understood compared to the $NN$ interaction because of limited experiment data and sophisticated annihilation processes, is closely connected to the $NN$ interaction in ChEFT in the sense that the intermediate/long-range part of the potential can be obtained by performing $G$-parity transformations to the pion exchange potentials. In contrast, the short-range/annihilation part is described by introducing real/complex contact terms in analogy to the $NN$ interaction with LECs adjusted to data. There are several varieties of chiral $\bar{N}N$ interactions~\cite{Chen:2011yu, Kang:2013uia, Dai:2017ont}. The most accurate chiral $\bar{N}N$ interaction to date was that of the J{\"u}lich group~\cite{Kang:2013uia, Dai:2017ont}. The J{\"u}lich potential has many successful applications in the studies of nucleon electromagnetic form factors~\cite{Haidenbauer:2014kja,Yang:2022qoy}, semileptonic baryonic decays~\cite{Cheng:2017qpv}, near $\bar{p}p$ threshold structures~\cite{Dai:2018tlc,Yang:2022kpm}, and neutron-antineutron oscillations~\cite{Haidenbauer:2019fyd}.

\begin{figure*}[htbp]
\centering
\includegraphics[scale=0.4]{ 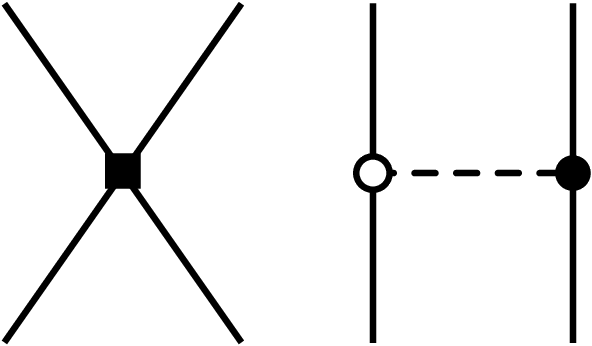}
\caption{Feynman diagrams contributing to the elastic part of the
$\bar{N}N$ interaction at leading order in the covariant power counting. The solid lines denote nucleons/antinucleons, and the dashed line represents the pion. The box denotes the vertex from $\mathcal{L}_{\bar{N}N}^{(0)}$, while the circle/dot shows vertex from $\mathcal{L}_{\pi \bar{N}}^{(1)}$/$\mathcal{L}_{\pi N}^{(1)}$.}
\label{Fig:LO}
\end{figure*}

However, the J{\"u}lich potential based on the heavy baryon chiral EFT shares problems similar to the NN potential, including the slow convergence and renormalization issues. In this section, we briefly introduce the recently developed relativistic $\bar{N} N$ interaction~\cite{Xiao:2024jmu} and the possible near threshold $\bar{N}N$ molecules.

\subsection{The hybrid antinucleon-nucleon potentials}

The $\bar{N}N$ interaction contains real and imaginary parts, which reads 
\begin{align}
    V_{\bar{N}N} = V^{\text{R}} + V^{\text{I}}.
\end{align}
For the real parts of the interaction, 
the underlying covariant power counting is the same as the $NN$ case, which is described in detail in the previous section, where the antinucleon field (spinor $v$) and the nucleon field (spinor $u$) are treated on an equal footing as spin-1/2 fields. The corresponding LO Feynman diagrams are summarized in Fig.~\ref{Fig:LO}, and the relevant Lagrangians are
\begin{align}
\label{eq:lageff}
\mathcal{L}_{\text{eff.}}=\mathcal{L}_{\pi\pi}^{(2)} + \mathcal{L}_{\pi \bar{N}}^{(1)} + \mathcal{L}_{\pi N}^{(1)} + \mathcal{L}_{\bar{N}N}^{(0)},
\end{align}
where the superscript denotes the chiral dimension. The
lowest order $\pi\pi$, $\pi \bar{N}$, $\pi N$, and $\bar{N}N$   Lagrangians read,
\begin{align}
\label{eq:Lag}
\mathcal{L}_{\pi\pi}^{(2)} =& \frac{f_\pi^2}{4} \text{Tr} \left[ \partial_\mu U \partial^\mu U^\dag + \left( U + U^\dag\right)m_\pi^2 \right],\\
\mathcal{L}_{\pi \bar{N}}^{(1)}  =& \mathcal{L}_{\pi N}^{(1)} = \bar{\Psi}\left( i \slashed{D} - m + \frac{g_A}{2} \gamma^\mu \gamma_5 u_\mu \right)\psi,\\
\mathcal{L}_{\bar{N} N}^{(0)} =&  C_S \left( \bar{\psi} \psi \right)\left( \bar{\psi} \psi \right) + C_A \left( \bar{\psi} \gamma_5 \psi \right) \left( \bar{\psi} \gamma_5 \psi \right) \\\nonumber  
  +&  C_V\left( \bar{\psi} \gamma_\mu \psi \right)\left( \bar{\psi} \gamma^\mu \psi \right) \\\nonumber
 +&  C_{AV}\left( \bar{\psi} \gamma_\mu \gamma_5 \psi \right) \left( \bar{\psi} \gamma^\mu \gamma_5 \psi \right) \\\nonumber
 +& C_T\left( \bar{\psi} \sigma_{\mu\nu} \psi \right)\left( \bar{\psi} \sigma^{\mu\nu} \psi \right), 
\end{align}
with the same notation as in Eq.~\eqref{eq:Lag1}.
The Spinor $v\left(\bm{p},s\right)= \gamma_0 C u^*\left( \bm{p},s \right)$ with $C$ representing the charge transformation operator,
\begin{equation}
    C = i \gamma_0 \gamma_2 =\left(
    \begin{matrix}
        0 & i \bm{\sigma}_2\\
        i \bm{\sigma}_2 &0
    \end{matrix}
    \right).
\end{equation}
The Pauli principle does not hold in the $\bar{N}N$ interaction, so the number of low-energy constants is twice that of the $NN$ case. 

Different from the $NN$ interaction, a new feature of the $\bar{N}N$ interaction is the presence of the annihilation process, which leads to an intrinsic difficulty in describing a system that has hundreds of annihilation many-body channels at rest~\cite{Carbonell:2023onq}. Here, we follow the approach of Ref.~\cite{Kang:2013uia} that manifestly fulfills unitarity and considers the contributions to the potential from the annihilation process of the following form,
\begin{align}
\label{eq:Vann}
V=\sum_{X=2\pi,3\pi,...}V_{\bar{N}N \rightarrow X}G_XV_{X \rightarrow \bar{N}N},
\end{align}
where $X$ is the sum over all open annihilation channels, and $G_X$ is the propagator of the
intermediate state $X$. Making use of the identity 
\begin{align}
    \frac{1}{x \pm i \epsilon} = \mathcal{P}\frac{1}{x} \mp i\pi \delta\left(x\right),
\end{align}
The imaginary part of Eq.~\eqref{eq:Vann} is constrained by,
\begin{align}
    \text{Im} V = -\pi \sum_X V_{\bar{N}N \rightarrow X}V_{X \rightarrow \bar{N}N}.
\end{align}
where the $V_{\bar{N}N \rightarrow X}$ and $V_{X \rightarrow \bar{N}N}$ are expanded in the EFT spirit according to the naive dimensional analysis.

\subsection{Phase shifts of anti-nucleon nucleon scattering and possible $N\bar{N}$ molecules}

The partial wave $S$ matrix is related to the on-shell 
$T$ matrix by,
\begin{align}
    S^{SJ}_{L',L}\l( p \r) = \delta_{L',L} - i \frac{ p~ m_N^2 }{8\pi^2 E_p} T^{SJ}_{L',L}\l(p \r).
\end{align}

Phase shifts and mixing angles can be obtained from the matrix $S$ using the idea of “Stapp”~\cite{Stapp:1956mz}. The annihilation process leads to complex phase shifts for the $\bar{N}N$ interaction. We follow the procedure of Ref.~\cite{refId0} to evaluate the phase shifts. For uncoupled channels, the real and imaginary parts of the phase shift $\delta_L$ can be obtained from the on-shell $S$ matrix,
\begin{align}
\nonumber
    \text{Re} \l(\delta_L\r) &= \frac{1}{2} \arctan \frac{\text{Im}\l(S_L\r)}{\text{Re}\l(S_L\r)}, \\
    \text{Im}\l( \delta_L \r)& = -\frac{1}{2}\log\lvert S_L  \rvert.
\end{align}
For coupled channels, the phase shifts $\delta_{L \pm1}$ and mixing angles $\epsilon_J$ are,
\begin{align}
\nonumber
 \text{Re}  \l(\delta_{L \pm 1}\r) &= \frac{1}{2} \arctan \frac{\text{Im} \l( \eta_{L \pm1} \r) }{\text{Re} \l( \eta_{L \pm1} \r) } ,\\  \nonumber
 \text{Im} \l( \delta_{L \pm 1} \r) &= -\frac{1}{2} \log \lvert \eta_{L \pm1} \rvert , \\\nonumber
  \epsilon_J &=   \frac{1}{2} \arctan \l( \frac{i \l( S_{L-1, L-1}+S_{L+1, L+1}\r)}{2\sqrt{S_{L-1, L-1} S_{L+1, L+1}}}\r),
\end{align}
where $\eta_L =  \frac{S_{L,L}}{\cos 2 \epsilon_J}  $.

In the fitting procedure, we perform a simultaneous fit to the $J = 0, 1$ PWA of Ref.~\cite{Zhou:2012ui} at laboratory energies below 125 MeV $\l( p_{\text{lab}} \leq 500  ~ \text{MeV}\r)$  with cutoff values varying in the range $\Lambda=450-600$ MeV, except for the $^{1}S_0$ and $^{3}P_0$ partial waves with $I=1$, where we consider extra data at $p_{\text{lab}}=600$ MeV because of the resonance-like behaviors.

\begin{figure*}[htbp]
\centering
\includegraphics[scale=0.15]{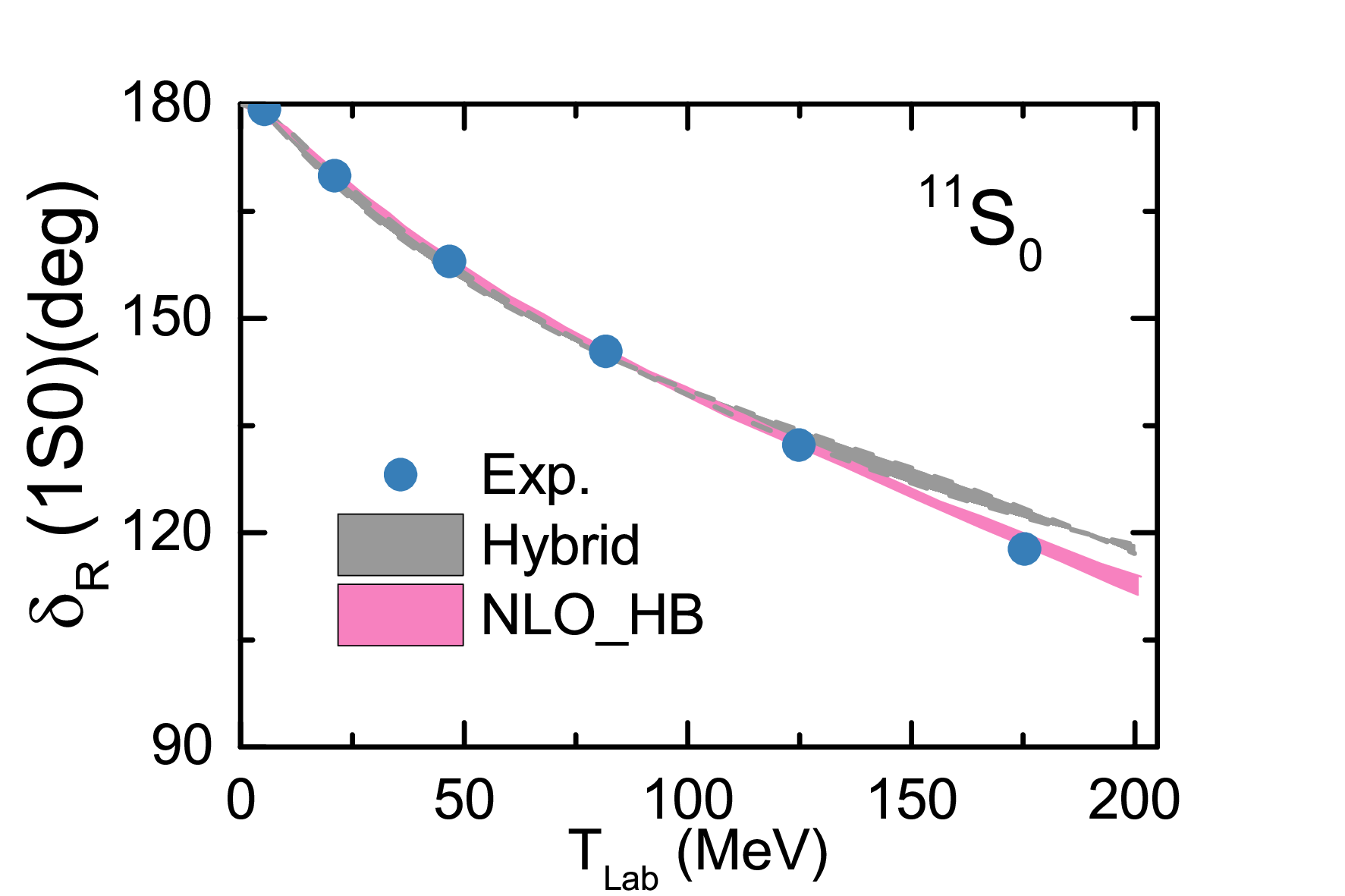}\hspace{-6mm}
\includegraphics[scale=0.15]{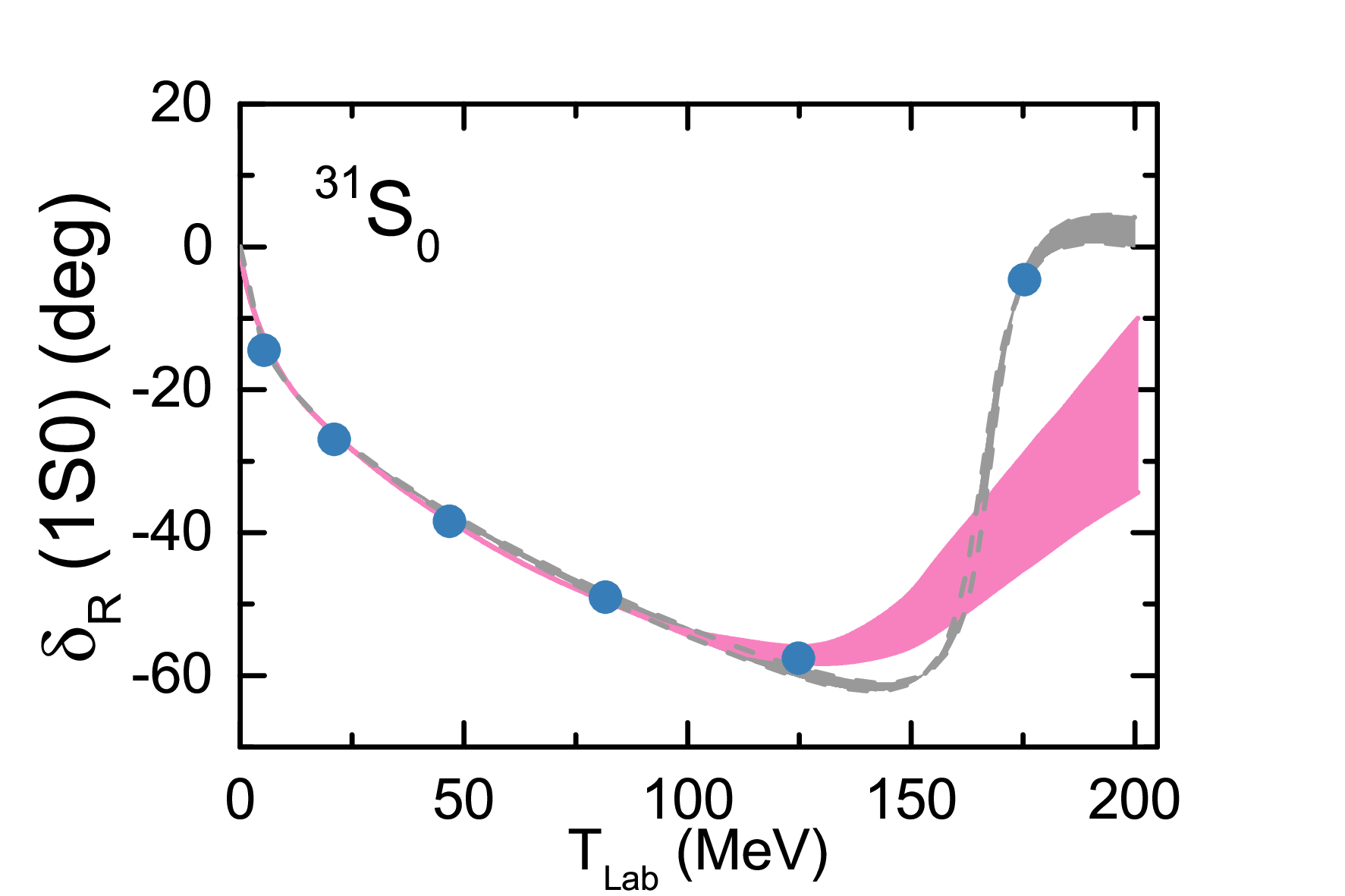}\hspace{-6mm}
\includegraphics[scale=0.15]{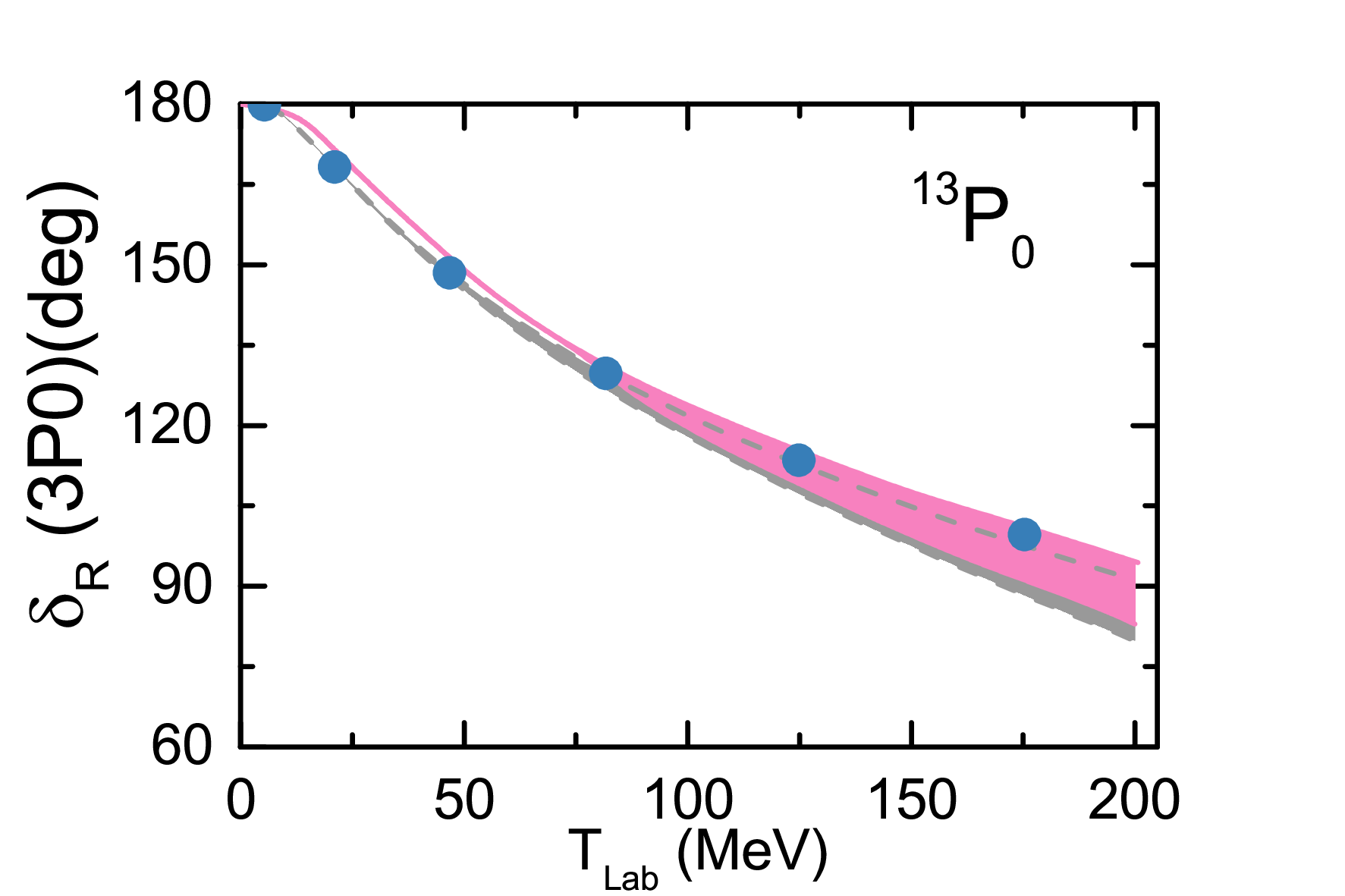}\hspace{-6mm}
\includegraphics[scale=0.15]{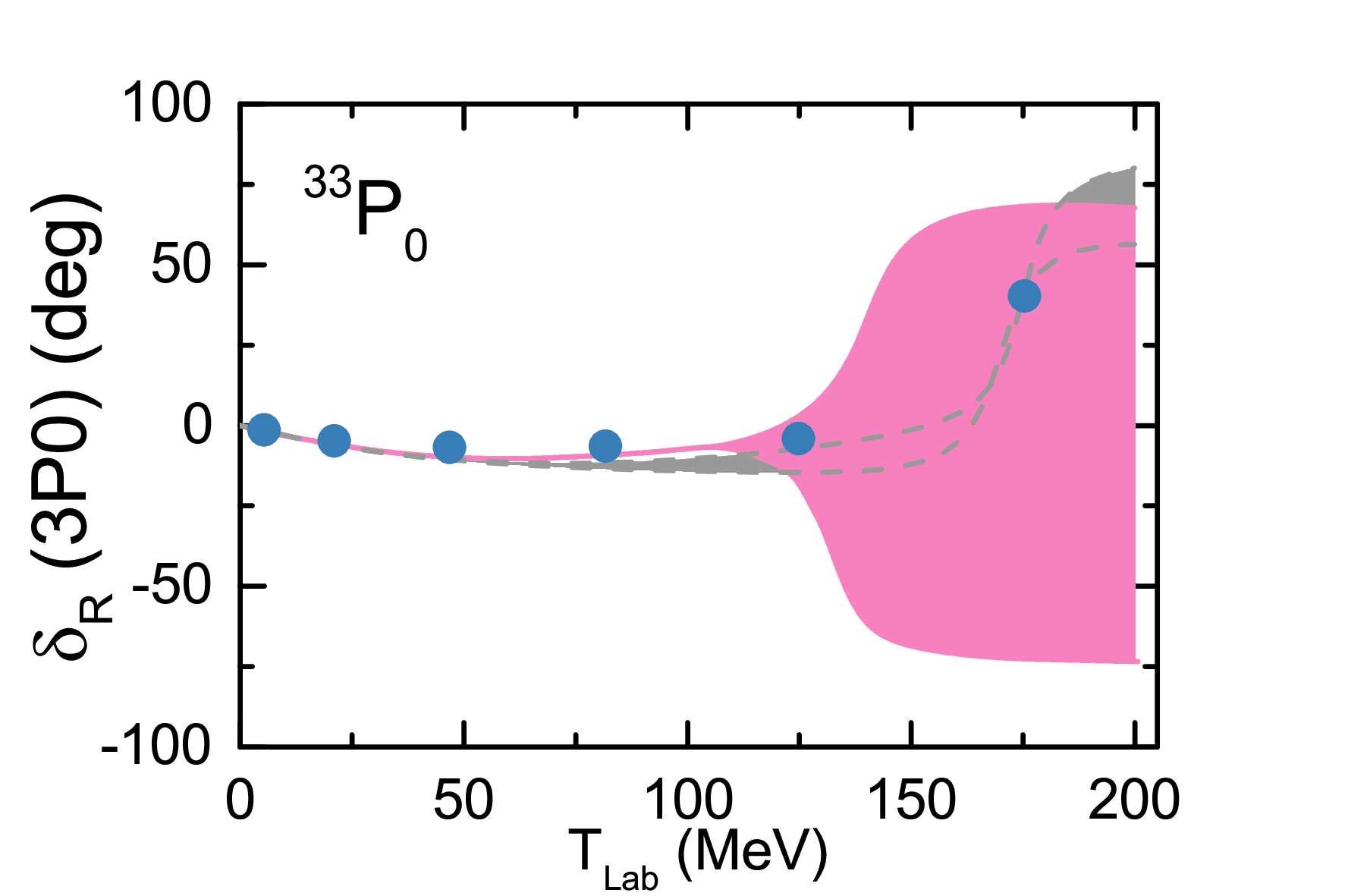}\\
\includegraphics[scale=0.15]{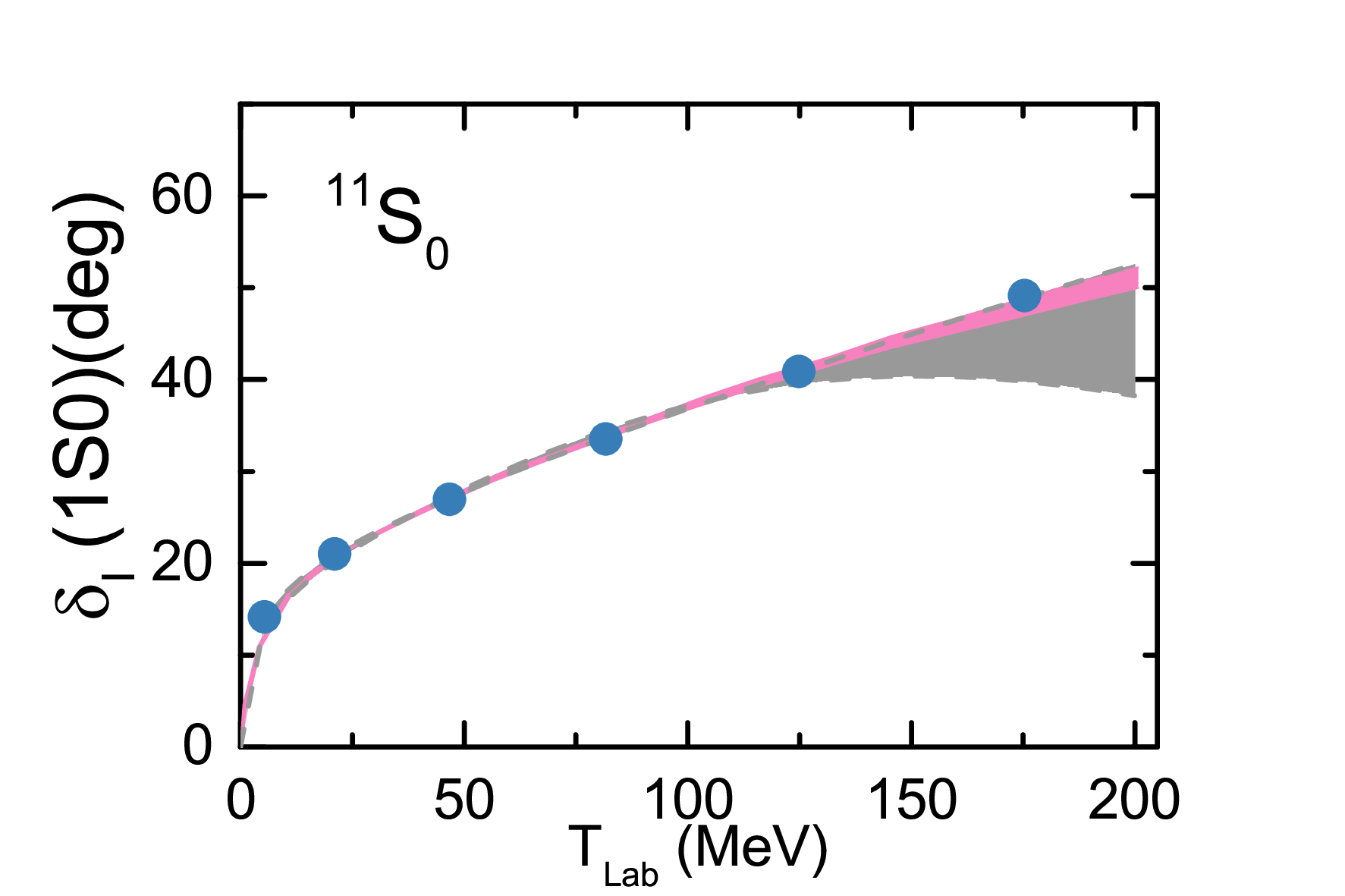}\hspace{-6mm}
\includegraphics[scale=0.15]{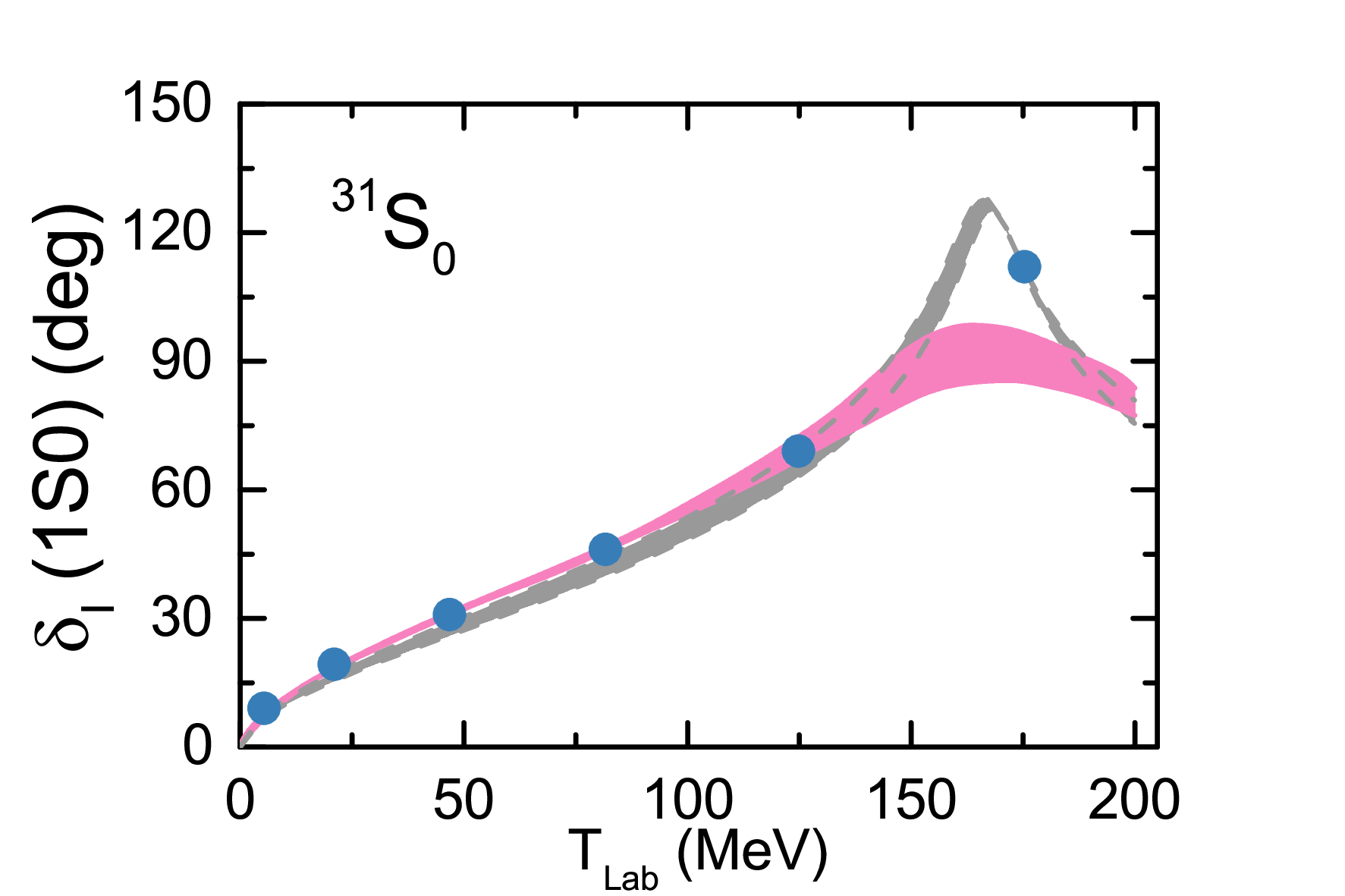}\hspace{-6mm}
\includegraphics[scale=0.15]{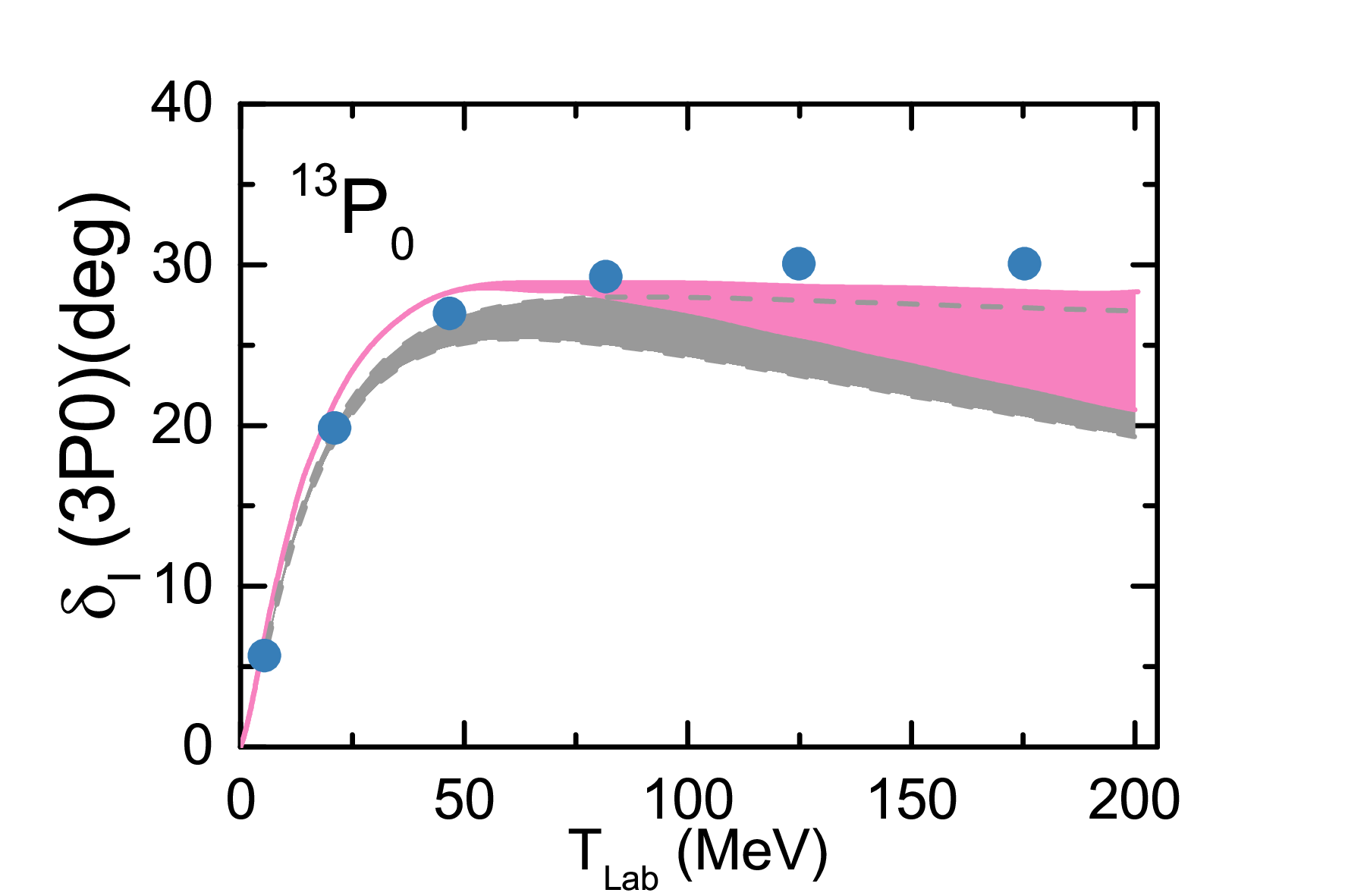}\hspace{-6mm}
\includegraphics[scale=0.15]{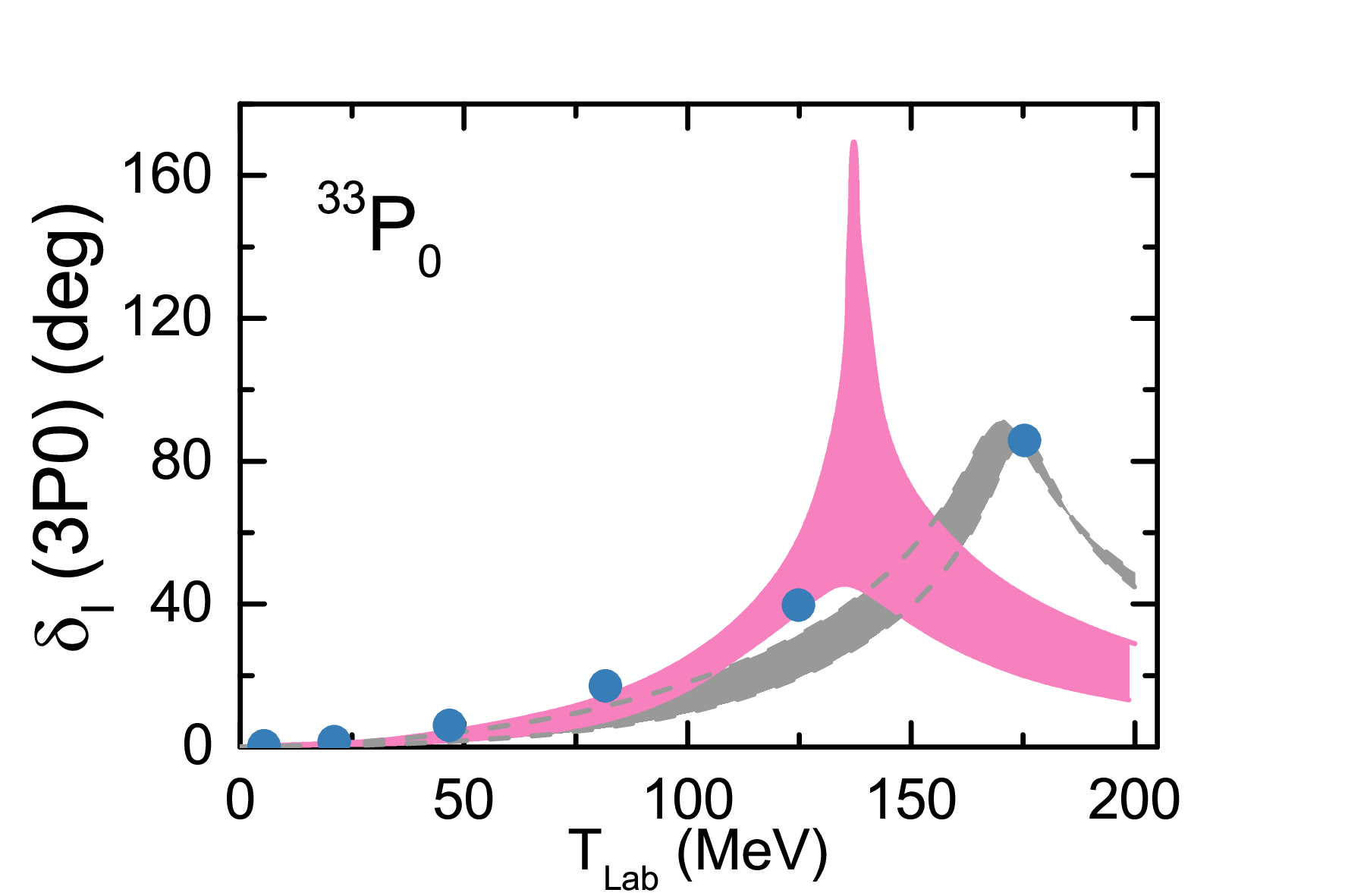}
\caption{Real and imaginary parts of the phase shifts for the $^1S_0$ and $^3P_0$ partial waves. The gray bands show our results with the cutoff in the range
$\Lambda$ = 450–600 MeV. The pink bands show the NLO HB chiral EFT results of Ref.~\cite{Kang:2013uia}. The blue dots refer to the PWA results of Ref.~\cite{Zhou:2012ui}.}
\label{fig:1S03P0}
\end{figure*}

\begin{figure*}[htbp]
\centering
\includegraphics[scale=0.15]{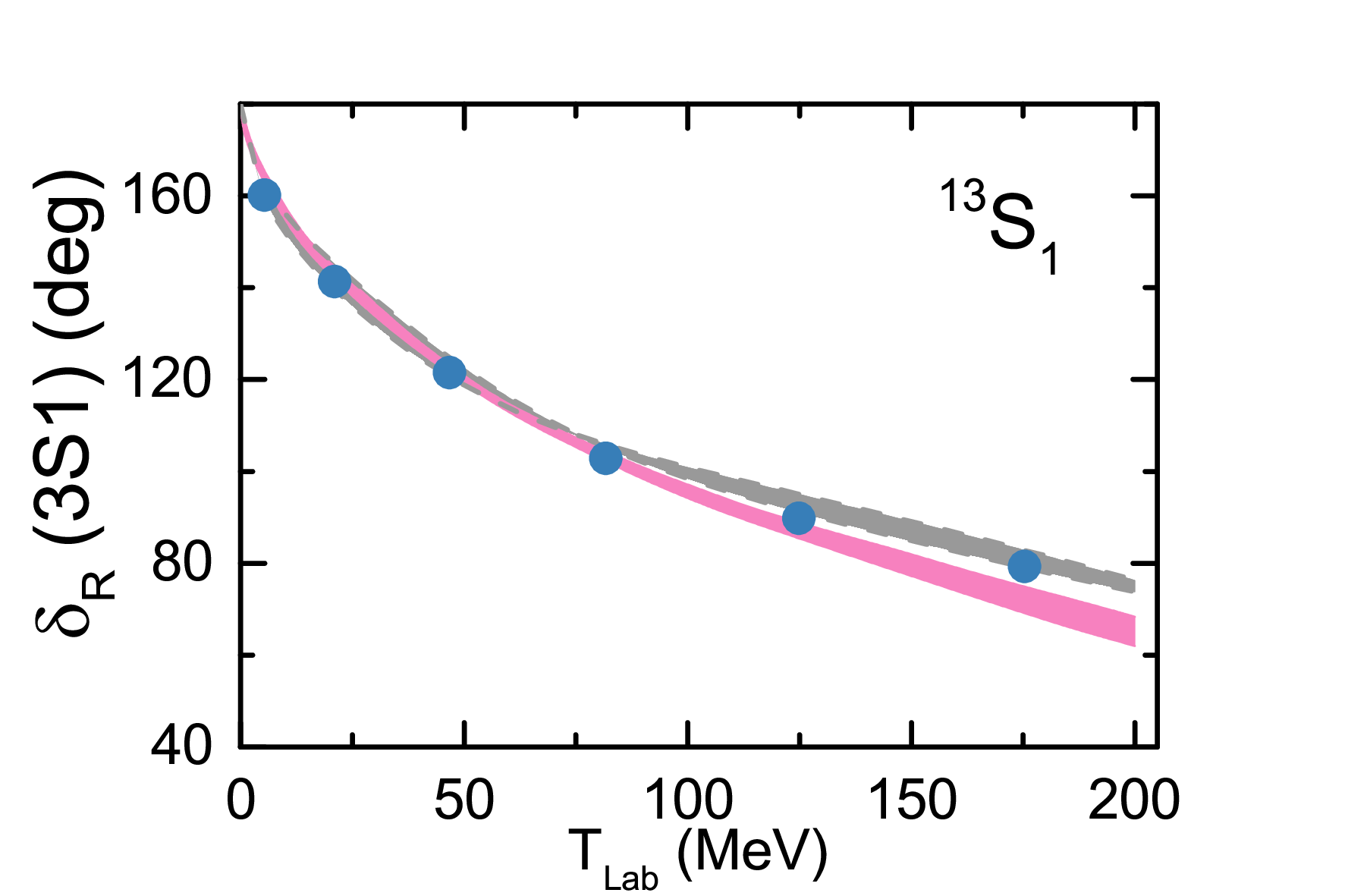}\hspace{-6mm}
\includegraphics[scale=0.15]{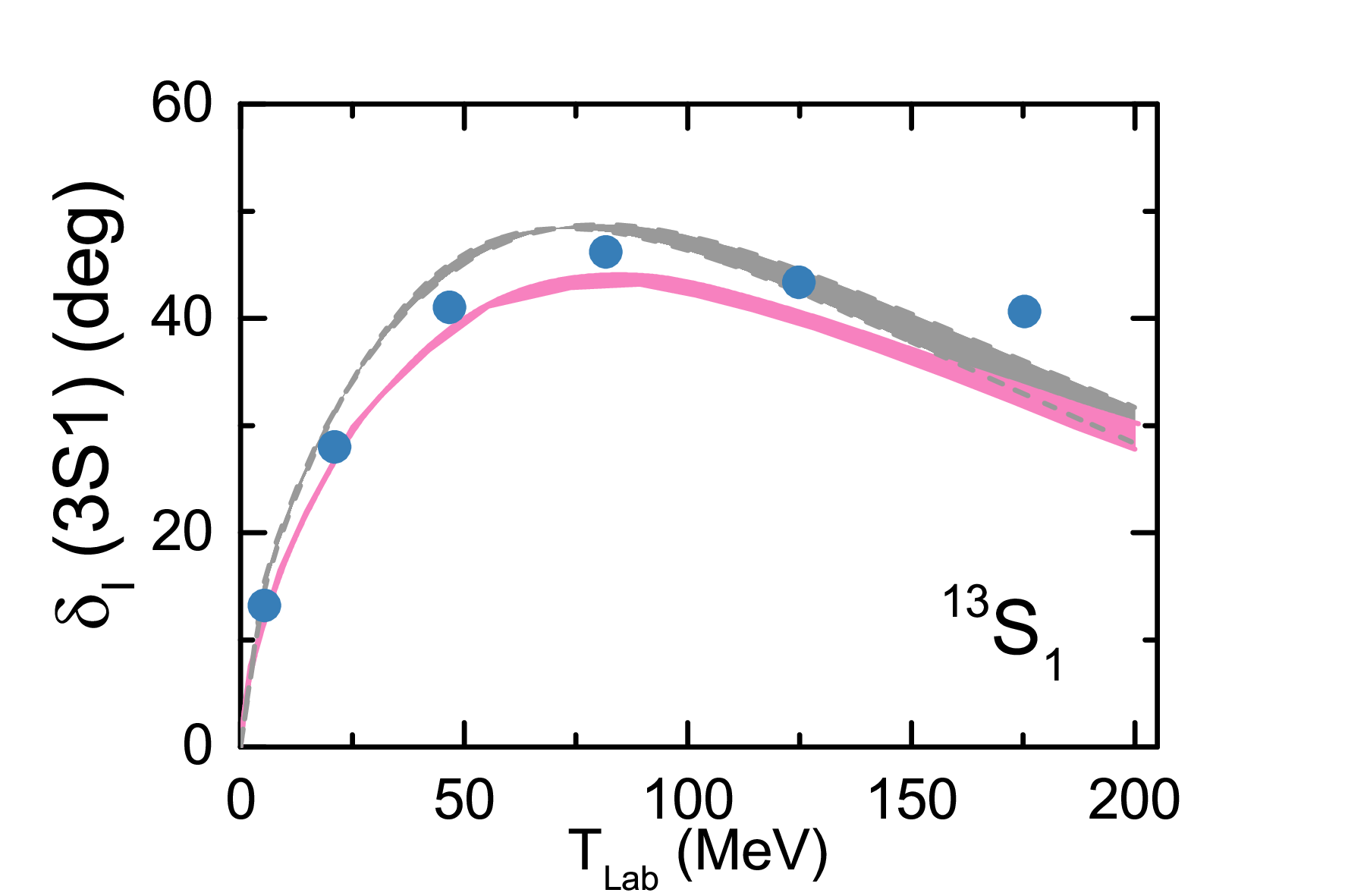}\hspace{-6mm}
\includegraphics[scale=0.15]{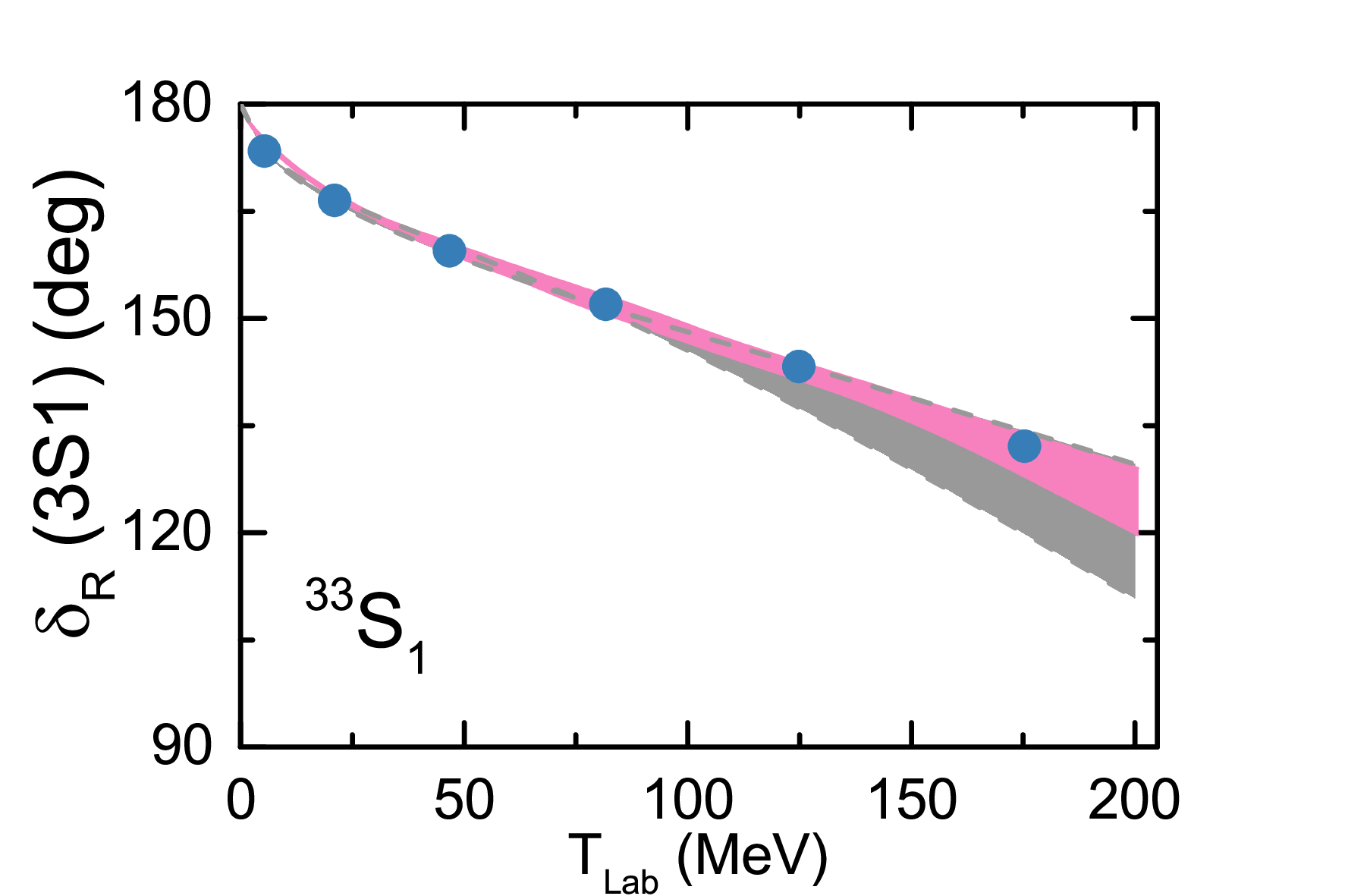}\hspace{-6mm}
\includegraphics[scale=0.15]{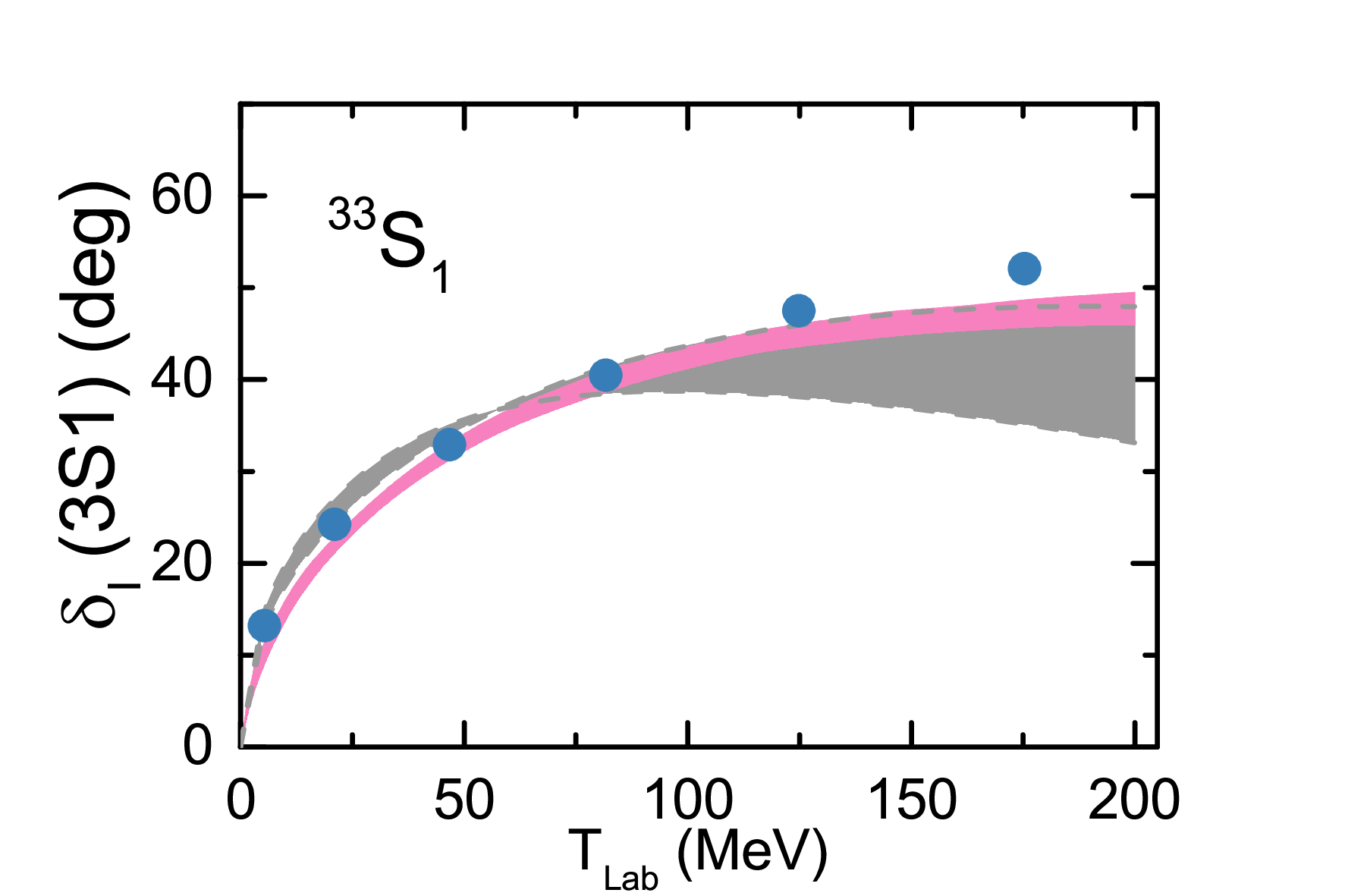}
\caption{Same as Fig.~\ref{fig:1S03P0}, but for the ${^3S_1}$ partial waves with $I=0$ and $I=1$.}
\label{fig:3S1}
\end{figure*}

The phase shifts obtained in our study, the NLO HB results~\cite{Kang:2013uia}, and the $\bar{N}N$ PWA~\cite{Zhou:2012ui} for laboratory energies up to $200$ MeV are shown in Figs.~\ref{fig:1S03P0}-\ref{fig:3S1}. The partial waves are labeled in the spectral notation $^{\l(2I+1\r)\l(2S+1\r)}L_J$, and the bands are generated by varying the cutoff in the range $\Lambda=450-600$ MeV for both our hybrid method and the HB approach. The LO HB phase shifts are not included for comparison because the potential relevant to the annihilation process, in this case, is only non-zero for the $^1S_0$ and $^3S_1$ partial waves. Hence, the descriptions of the phase shifts of other partial waves are very bad. In addition, even for the $^1S_0$ and $^3S_1$ partial waves, the differences between the LO HB phase shifts and the PWA are significant compared with the differences between the NLO HB phase shifts and the PWA results. 

Our results for the $J=0$ partial waves agree with the PWA for the energy region shown here. Compared with the NLO HB results, the overall cutoff dependence of the hybrid phase shifts is weaker, especially for the real parts of the phase shifts of the $I=1$ partial waves. At the same time, one can observe a sizable cutoff dependence in the NLO HB results for
energies above $150$ MeV because of the resonance-like behaviors. Since the number of free parameters for the $J=0$ partial waves in the hybrid method and the NLO HB formalism is identical ($4$ for the $^1S_0$ partial wave and $2$ for the $^3P_0$ partial wave, including annihilation parameters), the relatively weaker cutoff dependence has to do with the relativistic corrections of the scattering equation and the potentials at orders higher than $\mathcal{O}\l(p^2\r)$ (in the conventional Weinberg power counting). An exception exists in the imaginary part of the phase shift of the $^{11}S_0$ partial wave, where the cutoff dependence of the hybrid results is sizable at laboratory energies above $150$ MeV. This is related to the description of the $^{13}P_1$ partial wave, for which the PWA yields a negative phase at low energies, which tends to be positive at higher energies. Improvement might be possible by using an NLO covariant potential for the elastic process.

Next, we turn to the near-threshold $\bar{N}N$ structures. The phase shifts shown in Figs.~\ref{fig:1S03P0}-~\ref{fig:3S1} suggest the existence of bound states in the $^{11}S_0, ^{13}P_0, ^{13}S_1,$ and $^{33}S_1$ channels because their phase shifts are about $180^{\circ}$ at threshold. Therefore, we search for possible $\bar{N}N$ bound states in these channels. The corresponding binding energies obtained with our hybrid potential and the NLO HB potential~\cite{Kang:2013uia} are summarized in Table~\ref{tb:bound_states}. Although these structures have complex $E_B$, the sign of the real part
of $E_B$ is even positive in some cases, according to Refs.~\cite{Kang:2013uia, Badalian:1981xj}, the poles that we found can still be called bound states because they lie on the
physical sheet and move below the threshold when the annihilation potential is switched off.  We find a deeply bound state with $E_B = \l( -102.2 ,-152.5 \r) - \text{i} \l( 79.1 , 199.3 \r)$ MeV in the $^{11}S_0$ channel, whose quantum number is consistent with the pseudoscalar interpretation of $X(1835)$, $X(1840)$, and $X(1880)$ suggested by the BESIII Collaboration~\cite{BES:2005ega,BESIII:2023vvr}, despite that it is located far below the $\bar{N}N$ threshold and our result suffer from relatively large uncertainties. A firm conclusion can only be drawn once reliable theoretical uncertainties can be estimated. We want to mention that the studies employing semi-phenomenological $\bar{N}N$ interactions have found a bound state in the $^{11}S_0$ channel~\cite{Yan:2004xs, Sibirtsev:2004id,Ding:2005ew, Dedonder:2009bk}, although the predicted binding energies are rather different. Therefore, more studies are needed to confirm the nature of this state. Apart from the bound states, the phase shifts exhibit resonance-like structures in the $^{31}S_0$ and $^{33}P_0$ partial waves at energies above $150$ MeV. Thus, we also look for poles in the second Riemann sheet in these two channels. However, we do not find any resonant states.
\begin{table}[h!]
\centering
\caption{$\bar{N}N$ bound states and their binding energies. The uncertainties originate from the variation of the cutoff in the range $\Lambda=450-600$ MeV.}
\label{tb:bound_states}
\resizebox{\linewidth}{!}{
\begin{tabular}{c|c|c}
\hline
\hline
  \multirow{2}{*}{Partial Wave}   & \multicolumn{2}{c}{$E_B$ (MeV)} \\
  \cline{2-3}
  & Hybrid & NLO HB~\cite{Kang:2013uia}\\
     \hline
    $^{11}S_0$ & $\l( -102.2 ,-152.5 \r) - \text{i} \l( 79.1 , 199.3 \r)$ & /~\footnote{\label{note1}It is unclear whether a $\bar{N}N$ bound states can be observed in this channel because the possible structures are only searched for near $\bar{N}N$ threshold. }\\ 
    $^{13}P_0$ & $\l( -1.5 , -2.1 \r) - \text{i} \l( 20.2 , 21.0 \r)$ & $\l( -1.1 , 1.9 \r) - \text{i} \l( 17.8 , 22.4 \r)$ \\
    $^{13}S_1$ & $\l( -7.1 , 28.8 \r) - \text{i} \l( 45.5 , 49.2 \r)$ & $\l( 5.6 , 7.7 \r) - \text{i} \l( 49.2 , 60.5 \r)$ \\
    $^{33}S_1$ & $\l( -17.6 , 7.0 \r) - \text{i} \l( 128.9 , 134.4 \r)$ & /~ \footref{note1} \\
\hline
\hline
\end{tabular}
}
\end{table}

\section{Summary and prospective}

The nucleon-nucleon interaction has been constructed up to the next-to-next-to-leading order (NNLO) in the covariant baryon chiral effective field theory that can describe the Nijmegen partial-wave phase shifts. Within the relativistic framework, the NLO and NNLO results are similar in describing the scattering phase shifts for $T_\mathrm{lab}\le 200$ MeV. However, in the high-energy region, the NNLO results show significant improvement. This shows the good convergence of the covariant chiral expansion. We constructed the relativistic NNLO chiral nuclear force to provide the much-needed microscopic input for relativistic ab initio studies of nuclear matter and nuclear structure and reactions.

A faster convergence implies that the desired accuracy can be achieved at relatively lower orders. Therefore, the relativistic chiral nuclear force has the potential to solve the problem that the Weinberg chiral nuclear force in ab initio calculations must be calculated up to N$^4$LO. However, at the NNLO, the accuracy of the relativistic chiral nuclear force is not yet comparable to that of the N$^4$LO Weinberg chiral nuclear force. Therefore, it is necessary to construct the nuclear force up to N$^3$LO within the relativistic framework. The faster convergence of the relativistic chiral nuclear force hints that at N$^3$LO, the relativistic chiral nuclear force will be comparable to the non-relativistic chiral nuclear force at N$^4$LO.

The Peking group has recently made significant progress in conducting ab initio studies of finite nuclei and nuclear matter based on DBHF~\cite{Shen:2019dls}. For finite nuclei, starting from the relativistic Bonn potential, they solved the DBHF equation using a fully self-consistent Dirac Woods-Saxon basis. Without introducing any free parameters, they performed relativistic ab initio calculations of the ground-state properties of atomic nuclei such as $^4$He, $^{16}$O, and $^{40}$Ca~\cite{Shen:2019dls}. In addition, the DBHF theory for nuclear matter has also undergone substantial refinements. Particularly, the rigorous treatment of the center-of-mass momentum of two nucleons has, to a certain extent, improved the description of the saturation properties of nuclear matter by the DBHF theory~\cite{Tong:2018qwx}, and the restoration of the completeness of the Dirac basis space has resolved the uncertainty problem encountered when extracting the single-particle potential field~\cite{Wang:2021mvg}. These developments imply the possibility of achieving self-consistent unified nuclear structure and reaction calculations within the relativistic framework. However, the current DBHF studies still employed the phenomenological Bonn potential based on one-boson exchanges developed 30 years ago~\cite{Machleidt:1987hj}, which not only lacks important theoretical basis but also does not meet the accuracy requirements of ab initio calculations~\cite{Machleidt:2023jws}. The relativistic chiral nuclear force in the present work will certainly provide a fundamental input for relativistic ab initio studies. The latest work~\cite{Zou:2023quo} showed that with the LO relativistic chiral nuclear force, one can already achieve saturation for symmetric nuclear matter, showcasing the importance of relativistic effects.

The relativistic framework has been extended to study hyperon-nucleon/hyperon interactions~\cite{Li:2016paq,Li:2016mln,Song:2018qqm,Li:2018tbt,Liu:2020uxi,Song:2021yab,Liu:2022nec}. One observes similar patterns as in the nucleon-nucleon sector. With reduced cutoff dependence, one can achieve descriptions of existing experimental and lattice QCD data at LO similar to its non-relativistic counterparts at NLO.  Such studies are currently extended to higher orders and infinite nuclear matter~\cite{Zheng:2025sol}. 

\section{Acknowledgement}
This work is partly supported by the National Key R$\&$D Program of China under Grant No. 2023YFA1606703 and the National Science Foundation of China under Grant No. 12435007. Yang Xiao and Jun-Xu Lu thank the Fundamental Research Funds for Central Universities for the support. Yang Xiao thanks the support from the National Natural Science Foundation of China under Grant No.12347113 and the Chinese Postdoctoral Science Foundation under Grants No.2022M720360. Zhi-Wei Liu acknowledges support from the National Natural Science Foundation of China under Grant No.12405133, No.12347180, China Postdoctoral Science Foundation under Grant No.2023M740189, and the Postdoctoral Fellowship Program of CPSF under Grant No.GZC20233381.

\bibliography{Refs}

\begin{thebibliography}{160}%
\makeatletter
\providecommand \@ifxundefined [1]{%
 \@ifx{#1\undefined}
}%
\providecommand \@ifnum [1]{%
 \ifnum #1\expandafter \@firstoftwo
 \else \expandafter \@secondoftwo
 \fi
}%
\providecommand \@ifx [1]{%
 \ifx #1\expandafter \@firstoftwo
 \else \expandafter \@secondoftwo
 \fi
}%
\providecommand \natexlab [1]{#1}%
\providecommand \enquote  [1]{``#1''}%
\providecommand \bibnamefont  [1]{#1}%
\providecommand \bibfnamefont [1]{#1}%
\providecommand \citenamefont [1]{#1}%
\providecommand \href@noop [0]{\@secondoftwo}%
\providecommand \href [0]{\begingroup \@sanitize@url \@href}%
\providecommand \@href[1]{\@@startlink{#1}\@@href}%
\providecommand \@@href[1]{\endgroup#1\@@endlink}%
\providecommand \@sanitize@url [0]{\catcode `\\12\catcode `\$12\catcode `\&12\catcode `\#12\catcode `\^12\catcode `\_12\catcode `\%12\relax}%
\providecommand \@@startlink[1]{}%
\providecommand \@@endlink[0]{}%
\providecommand \url  [0]{\begingroup\@sanitize@url \@url }%
\providecommand \@url [1]{\endgroup\@href {#1}{\urlprefix }}%
\providecommand \urlprefix  [0]{URL }%
\providecommand \Eprint [0]{\href }%
\providecommand \doibase [0]{http://dx.doi.org/}%
\providecommand \selectlanguage [0]{\@gobble}%
\providecommand \bibinfo  [0]{\@secondoftwo}%
\providecommand \bibfield  [0]{\@secondoftwo}%
\providecommand \translation [1]{[#1]}%
\providecommand \BibitemOpen [0]{}%
\providecommand \bibitemStop [0]{}%
\providecommand \bibitemNoStop [0]{.\EOS\space}%
\providecommand \EOS [0]{\spacefactor3000\relax}%
\providecommand \BibitemShut  [1]{\csname bibitem#1\endcsname}%
\let\auto@bib@innerbib\@empty
\bibitem [{\citenamefont {Yukawa}(1935)}]{Yukawa:1935xg}%
  \BibitemOpen
  \bibfield  {author} {\bibinfo {author} {\bibfnamefont {H.}~\bibnamefont {Yukawa}},\ }\href {\doibase 10.1143/PTPS.1.1} {\bibfield  {journal} {\bibinfo  {journal} {Proc. Phys. Math. Soc. Jap.}\ }\textbf {\bibinfo {volume} {17}},\ \bibinfo {pages} {48} (\bibinfo {year} {1935})}\BibitemShut {NoStop}%
\bibitem [{\citenamefont {De~Tourreil}\ \emph {et~al.}(1975)\citenamefont {De~Tourreil}, \citenamefont {Rouben},\ and\ \citenamefont {Sprung}}]{DeTourreil:1975gz}%
  \BibitemOpen
  \bibfield  {author} {\bibinfo {author} {\bibfnamefont {R.}~\bibnamefont {De~Tourreil}}, \bibinfo {author} {\bibfnamefont {B.}~\bibnamefont {Rouben}}, \ and\ \bibinfo {author} {\bibfnamefont {D.~W.~L.}\ \bibnamefont {Sprung}},\ }\href {\doibase 10.1016/0375-9474(75)90107-4} {\bibfield  {journal} {\bibinfo  {journal} {Nucl. Phys. A}\ }\textbf {\bibinfo {volume} {242}},\ \bibinfo {pages} {445} (\bibinfo {year} {1975})}\BibitemShut {NoStop}%
\bibitem [{\citenamefont {Holinde}\ and\ \citenamefont {Machleidt}(1976)}]{Holinde:1976mkn}%
  \BibitemOpen
  \bibfield  {author} {\bibinfo {author} {\bibfnamefont {K.}~\bibnamefont {Holinde}}\ and\ \bibinfo {author} {\bibfnamefont {R.}~\bibnamefont {Machleidt}},\ }\href {\doibase 10.1016/0375-9474(76)90386-9} {\bibfield  {journal} {\bibinfo  {journal} {Nucl. Phys. A}\ }\textbf {\bibinfo {volume} {256}},\ \bibinfo {pages} {497} (\bibinfo {year} {1976})}\BibitemShut {NoStop}%
\bibitem [{\citenamefont {Nagels}\ \emph {et~al.}(1978)\citenamefont {Nagels}, \citenamefont {Rijken},\ and\ \citenamefont {de~Swart}}]{Nagels:1977ze}%
  \BibitemOpen
  \bibfield  {author} {\bibinfo {author} {\bibfnamefont {M.~M.}\ \bibnamefont {Nagels}}, \bibinfo {author} {\bibfnamefont {T.~A.}\ \bibnamefont {Rijken}}, \ and\ \bibinfo {author} {\bibfnamefont {J.~J.}\ \bibnamefont {de~Swart}},\ }\href {\doibase 10.1103/PhysRevD.17.768} {\bibfield  {journal} {\bibinfo  {journal} {Phys. Rev. D}\ }\textbf {\bibinfo {volume} {17}},\ \bibinfo {pages} {768} (\bibinfo {year} {1978})}\BibitemShut {NoStop}%
\bibitem [{\citenamefont {Oka}\ and\ \citenamefont {Yazaki}(1980)}]{Oka:1980ax}%
  \BibitemOpen
  \bibfield  {author} {\bibinfo {author} {\bibfnamefont {M.}~\bibnamefont {Oka}}\ and\ \bibinfo {author} {\bibfnamefont {K.}~\bibnamefont {Yazaki}},\ }\href {\doibase 10.1016/0370-2693(80)90046-5} {\bibfield  {journal} {\bibinfo  {journal} {Phys. Lett. B}\ }\textbf {\bibinfo {volume} {90}},\ \bibinfo {pages} {41} (\bibinfo {year} {1980})}\BibitemShut {NoStop}%
\bibitem [{\citenamefont {Faessler}\ \emph {et~al.}(1982)\citenamefont {Faessler}, \citenamefont {Fernandez}, \citenamefont {Lubeck},\ and\ \citenamefont {Shimizu}}]{Faessler:1982ik}%
  \BibitemOpen
  \bibfield  {author} {\bibinfo {author} {\bibfnamefont {A.}~\bibnamefont {Faessler}}, \bibinfo {author} {\bibfnamefont {F.}~\bibnamefont {Fernandez}}, \bibinfo {author} {\bibfnamefont {G.}~\bibnamefont {Lubeck}}, \ and\ \bibinfo {author} {\bibfnamefont {K.}~\bibnamefont {Shimizu}},\ }\href {\doibase 10.1016/0370-2693(82)90961-3} {\bibfield  {journal} {\bibinfo  {journal} {Phys. Lett. B}\ }\textbf {\bibinfo {volume} {112}},\ \bibinfo {pages} {201} (\bibinfo {year} {1982})}\BibitemShut {NoStop}%
\bibitem [{\citenamefont {Shimizu}(1989)}]{Shimizu:1989ye}%
  \BibitemOpen
  \bibfield  {author} {\bibinfo {author} {\bibfnamefont {K.}~\bibnamefont {Shimizu}},\ }\href {\doibase 10.1088/0034-4885/52/1/001} {\bibfield  {journal} {\bibinfo  {journal} {Rept. Prog. Phys.}\ }\textbf {\bibinfo {volume} {52}},\ \bibinfo {pages} {1} (\bibinfo {year} {1989})}\BibitemShut {NoStop}%
\bibitem [{\citenamefont {Valcarce}\ \emph {et~al.}(2005)\citenamefont {Valcarce}, \citenamefont {Garcilazo}, \citenamefont {Fernandez},\ and\ \citenamefont {Gonzalez}}]{Valcarce:2005em}%
  \BibitemOpen
  \bibfield  {author} {\bibinfo {author} {\bibfnamefont {A.}~\bibnamefont {Valcarce}}, \bibinfo {author} {\bibfnamefont {H.}~\bibnamefont {Garcilazo}}, \bibinfo {author} {\bibfnamefont {F.}~\bibnamefont {Fernandez}}, \ and\ \bibinfo {author} {\bibfnamefont {P.}~\bibnamefont {Gonzalez}},\ }\href {\doibase 10.1088/0034-4885/68/5/R01} {\bibfield  {journal} {\bibinfo  {journal} {Rept. Prog. Phys.}\ }\textbf {\bibinfo {volume} {68}},\ \bibinfo {pages} {965} (\bibinfo {year} {2005})},\ \Eprint {http://arxiv.org/abs/hep-ph/0502173} {arXiv:hep-ph/0502173} \BibitemShut {NoStop}%
\bibitem [{\citenamefont {Straub}\ \emph {et~al.}(1988)\citenamefont {Straub}, \citenamefont {Zhang}, \citenamefont {Brauer}, \citenamefont {Faessler}, \citenamefont {Khadkikar},\ and\ \citenamefont {Lubeck}}]{Straub:1988gj}%
  \BibitemOpen
  \bibfield  {author} {\bibinfo {author} {\bibfnamefont {U.}~\bibnamefont {Straub}}, \bibinfo {author} {\bibfnamefont {Z.-Y.}\ \bibnamefont {Zhang}}, \bibinfo {author} {\bibfnamefont {K.}~\bibnamefont {Brauer}}, \bibinfo {author} {\bibfnamefont {A.}~\bibnamefont {Faessler}}, \bibinfo {author} {\bibfnamefont {S.~B.}\ \bibnamefont {Khadkikar}}, \ and\ \bibinfo {author} {\bibfnamefont {G.}~\bibnamefont {Lubeck}},\ }\href {\doibase 10.1016/0375-9474(88)90092-9} {\bibfield  {journal} {\bibinfo  {journal} {Nucl. Phys. A}\ }\textbf {\bibinfo {volume} {483}},\ \bibinfo {pages} {686} (\bibinfo {year} {1988})}\BibitemShut {NoStop}%
\bibitem [{\citenamefont {Zhang}\ \emph {et~al.}(1994)\citenamefont {Zhang}, \citenamefont {Faessler}, \citenamefont {Straub},\ and\ \citenamefont {Glozman}}]{Zhang:1994pp}%
  \BibitemOpen
  \bibfield  {author} {\bibinfo {author} {\bibfnamefont {Z.-Y.}\ \bibnamefont {Zhang}}, \bibinfo {author} {\bibfnamefont {A.}~\bibnamefont {Faessler}}, \bibinfo {author} {\bibfnamefont {U.}~\bibnamefont {Straub}}, \ and\ \bibinfo {author} {\bibfnamefont {L.~Y.}\ \bibnamefont {Glozman}},\ }\href {\doibase 10.1016/0375-9474(94)90761-7} {\bibfield  {journal} {\bibinfo  {journal} {Nucl. Phys. A}\ }\textbf {\bibinfo {volume} {578}},\ \bibinfo {pages} {573} (\bibinfo {year} {1994})}\BibitemShut {NoStop}%
\bibitem [{\citenamefont {Zhang}\ \emph {et~al.}(1997)\citenamefont {Zhang}, \citenamefont {Yu}, \citenamefont {Shen}, \citenamefont {Dai}, \citenamefont {Faessler},\ and\ \citenamefont {Straub}}]{Zhang:1997ny}%
  \BibitemOpen
  \bibfield  {author} {\bibinfo {author} {\bibfnamefont {Z.~Y.}\ \bibnamefont {Zhang}}, \bibinfo {author} {\bibfnamefont {Y.~W.}\ \bibnamefont {Yu}}, \bibinfo {author} {\bibfnamefont {P.~N.}\ \bibnamefont {Shen}}, \bibinfo {author} {\bibfnamefont {L.~R.}\ \bibnamefont {Dai}}, \bibinfo {author} {\bibfnamefont {A.}~\bibnamefont {Faessler}}, \ and\ \bibinfo {author} {\bibfnamefont {U.}~\bibnamefont {Straub}},\ }\href {\doibase 10.1016/S0375-9474(97)00033-X} {\bibfield  {journal} {\bibinfo  {journal} {Nucl. Phys. A}\ }\textbf {\bibinfo {volume} {625}},\ \bibinfo {pages} {59} (\bibinfo {year} {1997})}\BibitemShut {NoStop}%
\bibitem [{\citenamefont {Dai}\ \emph {et~al.}(2003)\citenamefont {Dai}, \citenamefont {Zhang}, \citenamefont {Yu},\ and\ \citenamefont {Wang}}]{Dai:2003dz}%
  \BibitemOpen
  \bibfield  {author} {\bibinfo {author} {\bibfnamefont {L.~R.}\ \bibnamefont {Dai}}, \bibinfo {author} {\bibfnamefont {Z.~Y.}\ \bibnamefont {Zhang}}, \bibinfo {author} {\bibfnamefont {Y.~W.}\ \bibnamefont {Yu}}, \ and\ \bibinfo {author} {\bibfnamefont {P.}~\bibnamefont {Wang}},\ }\href {\doibase 10.1016/j.nuclphysa.2003.08.006} {\bibfield  {journal} {\bibinfo  {journal} {Nucl. Phys. A}\ }\textbf {\bibinfo {volume} {727}},\ \bibinfo {pages} {321} (\bibinfo {year} {2003})},\ \Eprint {http://arxiv.org/abs/nucl-th/0404004} {arXiv:nucl-th/0404004} \BibitemShut {NoStop}%
\bibitem [{\citenamefont {Wang}\ \emph {et~al.}(1992)\citenamefont {Wang}, \citenamefont {Wu}, \citenamefont {Teng},\ and\ \citenamefont {Goldman}}]{Wang:1992wi}%
  \BibitemOpen
  \bibfield  {author} {\bibinfo {author} {\bibfnamefont {F.}~\bibnamefont {Wang}}, \bibinfo {author} {\bibfnamefont {G.-h.}\ \bibnamefont {Wu}}, \bibinfo {author} {\bibfnamefont {L.-j.}\ \bibnamefont {Teng}}, \ and\ \bibinfo {author} {\bibfnamefont {J.~T.}\ \bibnamefont {Goldman}},\ }\href {\doibase 10.1103/PhysRevLett.69.2901} {\bibfield  {journal} {\bibinfo  {journal} {Phys. Rev. Lett.}\ }\textbf {\bibinfo {volume} {69}},\ \bibinfo {pages} {2901} (\bibinfo {year} {1992})},\ \Eprint {http://arxiv.org/abs/nucl-th/9210002} {arXiv:nucl-th/9210002} \BibitemShut {NoStop}%
\bibitem [{\citenamefont {Wu}\ \emph {et~al.}(1996)\citenamefont {Wu}, \citenamefont {Teng}, \citenamefont {Ping}, \citenamefont {Wang},\ and\ \citenamefont {Goldman}}]{Wu:1996fm}%
  \BibitemOpen
  \bibfield  {author} {\bibinfo {author} {\bibfnamefont {G.-H.}\ \bibnamefont {Wu}}, \bibinfo {author} {\bibfnamefont {L.-J.}\ \bibnamefont {Teng}}, \bibinfo {author} {\bibfnamefont {J.~L.}\ \bibnamefont {Ping}}, \bibinfo {author} {\bibfnamefont {F.}~\bibnamefont {Wang}}, \ and\ \bibinfo {author} {\bibfnamefont {J.~T.}\ \bibnamefont {Goldman}},\ }\href {\doibase 10.1103/PhysRevC.53.1161} {\bibfield  {journal} {\bibinfo  {journal} {Phys. Rev. C}\ }\textbf {\bibinfo {volume} {53}},\ \bibinfo {pages} {1161} (\bibinfo {year} {1996})}\BibitemShut {NoStop}%
\bibitem [{\citenamefont {Ping}\ \emph {et~al.}(1999)\citenamefont {Ping}, \citenamefont {Wang},\ and\ \citenamefont {Goldman}}]{Ping:1998si}%
  \BibitemOpen
  \bibfield  {author} {\bibinfo {author} {\bibfnamefont {J.-L.}\ \bibnamefont {Ping}}, \bibinfo {author} {\bibfnamefont {F.}~\bibnamefont {Wang}}, \ and\ \bibinfo {author} {\bibfnamefont {J.~T.}\ \bibnamefont {Goldman}},\ }\href {\doibase 10.1016/S0375-9474(99)00321-8} {\bibfield  {journal} {\bibinfo  {journal} {Nucl. Phys. A}\ }\textbf {\bibinfo {volume} {657}},\ \bibinfo {pages} {95} (\bibinfo {year} {1999})},\ \Eprint {http://arxiv.org/abs/nucl-th/9812068} {arXiv:nucl-th/9812068} \BibitemShut {NoStop}%
\bibitem [{\citenamefont {Wu}\ \emph {et~al.}(2000)\citenamefont {Wu}, \citenamefont {Ping}, \citenamefont {Teng}, \citenamefont {Wang},\ and\ \citenamefont {Goldman}}]{Wu:1998wu}%
  \BibitemOpen
  \bibfield  {author} {\bibinfo {author} {\bibfnamefont {G.-h.}\ \bibnamefont {Wu}}, \bibinfo {author} {\bibfnamefont {J.-L.}\ \bibnamefont {Ping}}, \bibinfo {author} {\bibfnamefont {L.-j.}\ \bibnamefont {Teng}}, \bibinfo {author} {\bibfnamefont {F.}~\bibnamefont {Wang}}, \ and\ \bibinfo {author} {\bibfnamefont {J.~T.}\ \bibnamefont {Goldman}},\ }\href {\doibase 10.1016/S0375-9474(00)00141-X} {\bibfield  {journal} {\bibinfo  {journal} {Nucl. Phys. A}\ }\textbf {\bibinfo {volume} {673}},\ \bibinfo {pages} {279} (\bibinfo {year} {2000})},\ \Eprint {http://arxiv.org/abs/nucl-th/9812079} {arXiv:nucl-th/9812079} \BibitemShut {NoStop}%
\bibitem [{\citenamefont {Pang}\ \emph {et~al.}(2002)\citenamefont {Pang}, \citenamefont {Ping}, \citenamefont {Wang},\ and\ \citenamefont {Goldman}}]{Pang:2001xx}%
  \BibitemOpen
  \bibfield  {author} {\bibinfo {author} {\bibfnamefont {H.~R.}\ \bibnamefont {Pang}}, \bibinfo {author} {\bibfnamefont {J.~L.}\ \bibnamefont {Ping}}, \bibinfo {author} {\bibfnamefont {F.}~\bibnamefont {Wang}}, \ and\ \bibinfo {author} {\bibfnamefont {J.~T.}\ \bibnamefont {Goldman}},\ }\href {\doibase 10.1103/PhysRevC.65.014003} {\bibfield  {journal} {\bibinfo  {journal} {Phys. Rev. C}\ }\textbf {\bibinfo {volume} {65}},\ \bibinfo {pages} {014003} (\bibinfo {year} {2002})},\ \Eprint {http://arxiv.org/abs/nucl-th/0106056} {arXiv:nucl-th/0106056} \BibitemShut {NoStop}%
\bibitem [{\citenamefont {Fujiwara}\ \emph {et~al.}(1995)\citenamefont {Fujiwara}, \citenamefont {Nakamoto},\ and\ \citenamefont {Suzuki}}]{Fujiwara:1995td}%
  \BibitemOpen
  \bibfield  {author} {\bibinfo {author} {\bibfnamefont {Y.}~\bibnamefont {Fujiwara}}, \bibinfo {author} {\bibfnamefont {C.}~\bibnamefont {Nakamoto}}, \ and\ \bibinfo {author} {\bibfnamefont {Y.}~\bibnamefont {Suzuki}},\ }\href {\doibase 10.1143/PTP.94.215} {\bibfield  {journal} {\bibinfo  {journal} {Prog. Theor. Phys.}\ }\textbf {\bibinfo {volume} {94}},\ \bibinfo {pages} {215} (\bibinfo {year} {1995})}\BibitemShut {NoStop}%
\bibitem [{\citenamefont {Fujiwara}\ \emph {et~al.}(1996{\natexlab{a}})\citenamefont {Fujiwara}, \citenamefont {Nakamoto},\ and\ \citenamefont {Suzuki}}]{Fujiwara:1995fx}%
  \BibitemOpen
  \bibfield  {author} {\bibinfo {author} {\bibfnamefont {Y.}~\bibnamefont {Fujiwara}}, \bibinfo {author} {\bibfnamefont {C.}~\bibnamefont {Nakamoto}}, \ and\ \bibinfo {author} {\bibfnamefont {Y.}~\bibnamefont {Suzuki}},\ }\href {\doibase 10.1103/PhysRevLett.76.2242} {\bibfield  {journal} {\bibinfo  {journal} {Phys. Rev. Lett.}\ }\textbf {\bibinfo {volume} {76}},\ \bibinfo {pages} {2242} (\bibinfo {year} {1996}{\natexlab{a}})}\BibitemShut {NoStop}%
\bibitem [{\citenamefont {Fujiwara}\ \emph {et~al.}(1996{\natexlab{b}})\citenamefont {Fujiwara}, \citenamefont {Nakamoto},\ and\ \citenamefont {Suzuki}}]{Fujiwara:1996qj}%
  \BibitemOpen
  \bibfield  {author} {\bibinfo {author} {\bibfnamefont {Y.}~\bibnamefont {Fujiwara}}, \bibinfo {author} {\bibfnamefont {C.}~\bibnamefont {Nakamoto}}, \ and\ \bibinfo {author} {\bibfnamefont {Y.}~\bibnamefont {Suzuki}},\ }\href {\doibase 10.1103/PhysRevC.54.2180} {\bibfield  {journal} {\bibinfo  {journal} {Phys. Rev. C}\ }\textbf {\bibinfo {volume} {54}},\ \bibinfo {pages} {2180} (\bibinfo {year} {1996}{\natexlab{b}})}\BibitemShut {NoStop}%
\bibitem [{\citenamefont {Fujita}\ \emph {et~al.}(1998)\citenamefont {Fujita}, \citenamefont {Fujiwara}, \citenamefont {Nakamoto},\ and\ \citenamefont {Suzuki}}]{Fujita:1998sg}%
  \BibitemOpen
  \bibfield  {author} {\bibinfo {author} {\bibfnamefont {T.}~\bibnamefont {Fujita}}, \bibinfo {author} {\bibfnamefont {Y.}~\bibnamefont {Fujiwara}}, \bibinfo {author} {\bibfnamefont {C.}~\bibnamefont {Nakamoto}}, \ and\ \bibinfo {author} {\bibfnamefont {Y.}~\bibnamefont {Suzuki}},\ }\href {\doibase 10.1143/PTP.100.931} {\bibfield  {journal} {\bibinfo  {journal} {Prog. Theor. Phys.}\ }\textbf {\bibinfo {volume} {100}},\ \bibinfo {pages} {931} (\bibinfo {year} {1998})}\BibitemShut {NoStop}%
\bibitem [{\citenamefont {Weinberg}(1990)}]{Weinberg:1990rz}%
  \BibitemOpen
  \bibfield  {author} {\bibinfo {author} {\bibfnamefont {S.}~\bibnamefont {Weinberg}},\ }\href {\doibase 10.1016/0370-2693(90)90938-3} {\bibfield  {journal} {\bibinfo  {journal} {Phys. Lett. B}\ }\textbf {\bibinfo {volume} {251}},\ \bibinfo {pages} {288} (\bibinfo {year} {1990})}\BibitemShut {NoStop}%
\bibitem [{\citenamefont {Weinberg}(1991)}]{Weinberg:1991um}%
  \BibitemOpen
  \bibfield  {author} {\bibinfo {author} {\bibfnamefont {S.}~\bibnamefont {Weinberg}},\ }\href {\doibase 10.1016/0550-3213(91)90231-L} {\bibfield  {journal} {\bibinfo  {journal} {Nucl. Phys. B}\ }\textbf {\bibinfo {volume} {363}},\ \bibinfo {pages} {3} (\bibinfo {year} {1991})}\BibitemShut {NoStop}%
\bibitem [{\citenamefont {Weinberg}(1992)}]{Weinberg:1992yk}%
  \BibitemOpen
  \bibfield  {author} {\bibinfo {author} {\bibfnamefont {S.}~\bibnamefont {Weinberg}},\ }\href {\doibase 10.1016/0370-2693(92)90099-P} {\bibfield  {journal} {\bibinfo  {journal} {Phys. Lett. B}\ }\textbf {\bibinfo {volume} {295}},\ \bibinfo {pages} {114} (\bibinfo {year} {1992})},\ \Eprint {http://arxiv.org/abs/hep-ph/9209257} {arXiv:hep-ph/9209257} \BibitemShut {NoStop}%
\bibitem [{\citenamefont {Epelbaum}\ \emph {et~al.}(2009)\citenamefont {Epelbaum}, \citenamefont {Hammer},\ and\ \citenamefont {Meissner}}]{Epelbaum:2008ga}%
  \BibitemOpen
  \bibfield  {author} {\bibinfo {author} {\bibfnamefont {E.}~\bibnamefont {Epelbaum}}, \bibinfo {author} {\bibfnamefont {H.-W.}\ \bibnamefont {Hammer}}, \ and\ \bibinfo {author} {\bibfnamefont {U.-G.}\ \bibnamefont {Meissner}},\ }\href {\doibase 10.1103/RevModPhys.81.1773} {\bibfield  {journal} {\bibinfo  {journal} {Rev. Mod. Phys.}\ }\textbf {\bibinfo {volume} {81}},\ \bibinfo {pages} {1773} (\bibinfo {year} {2009})},\ \Eprint {http://arxiv.org/abs/0811.1338} {arXiv:0811.1338 [nucl-th]} \BibitemShut {NoStop}%
\bibitem [{\citenamefont {Machleidt}\ and\ \citenamefont {Entem}(2011)}]{Machleidt:2011zz}%
  \BibitemOpen
  \bibfield  {author} {\bibinfo {author} {\bibfnamefont {R.}~\bibnamefont {Machleidt}}\ and\ \bibinfo {author} {\bibfnamefont {D.~R.}\ \bibnamefont {Entem}},\ }\href {\doibase 10.1016/j.physrep.2011.02.001} {\bibfield  {journal} {\bibinfo  {journal} {Phys. Rept.}\ }\textbf {\bibinfo {volume} {503}},\ \bibinfo {pages} {1} (\bibinfo {year} {2011})},\ \Eprint {http://arxiv.org/abs/1105.2919} {arXiv:1105.2919 [nucl-th]} \BibitemShut {NoStop}%
\bibitem [{\citenamefont {Machleidt}\ and\ \citenamefont {Sammarruca}(2024)}]{Machleidt:2024bwl}%
  \BibitemOpen
  \bibfield  {author} {\bibinfo {author} {\bibfnamefont {R.}~\bibnamefont {Machleidt}}\ and\ \bibinfo {author} {\bibfnamefont {F.}~\bibnamefont {Sammarruca}},\ }\href@noop {} {\  (\bibinfo {year} {2024})},\ \Eprint {http://arxiv.org/abs/2402.14032} {arXiv:2402.14032 [nucl-th]} \BibitemShut {NoStop}%
\bibitem [{\citenamefont {Bedaque}\ and\ \citenamefont {van Kolck}(2002)}]{Bedaque:2002mn}%
  \BibitemOpen
  \bibfield  {author} {\bibinfo {author} {\bibfnamefont {P.~F.}\ \bibnamefont {Bedaque}}\ and\ \bibinfo {author} {\bibfnamefont {U.}~\bibnamefont {van Kolck}},\ }\href {\doibase 10.1146/annurev.nucl.52.050102.090637} {\bibfield  {journal} {\bibinfo  {journal} {Ann. Rev. Nucl. Part. Sci.}\ }\textbf {\bibinfo {volume} {52}},\ \bibinfo {pages} {339} (\bibinfo {year} {2002})},\ \Eprint {http://arxiv.org/abs/nucl-th/0203055} {arXiv:nucl-th/0203055} \BibitemShut {NoStop}%
\bibitem [{\citenamefont {Ordonez}\ \emph {et~al.}(1994)\citenamefont {Ordonez}, \citenamefont {Ray},\ and\ \citenamefont {van Kolck}}]{Ordonez:1993tn}%
  \BibitemOpen
  \bibfield  {author} {\bibinfo {author} {\bibfnamefont {C.}~\bibnamefont {Ordonez}}, \bibinfo {author} {\bibfnamefont {L.}~\bibnamefont {Ray}}, \ and\ \bibinfo {author} {\bibfnamefont {U.}~\bibnamefont {van Kolck}},\ }\href {\doibase 10.1103/PhysRevLett.72.1982} {\bibfield  {journal} {\bibinfo  {journal} {Phys. Rev. Lett.}\ }\textbf {\bibinfo {volume} {72}},\ \bibinfo {pages} {1982} (\bibinfo {year} {1994})}\BibitemShut {NoStop}%
\bibitem [{\citenamefont {Ordonez}\ \emph {et~al.}(1996)\citenamefont {Ordonez}, \citenamefont {Ray},\ and\ \citenamefont {van Kolck}}]{Ordonez:1995rz}%
  \BibitemOpen
  \bibfield  {author} {\bibinfo {author} {\bibfnamefont {C.}~\bibnamefont {Ordonez}}, \bibinfo {author} {\bibfnamefont {L.}~\bibnamefont {Ray}}, \ and\ \bibinfo {author} {\bibfnamefont {U.}~\bibnamefont {van Kolck}},\ }\href {\doibase 10.1103/PhysRevC.53.2086} {\bibfield  {journal} {\bibinfo  {journal} {Phys. Rev. C}\ }\textbf {\bibinfo {volume} {53}},\ \bibinfo {pages} {2086} (\bibinfo {year} {1996})},\ \Eprint {http://arxiv.org/abs/hep-ph/9511380} {arXiv:hep-ph/9511380} \BibitemShut {NoStop}%
\bibitem [{\citenamefont {Kaiser}\ \emph {et~al.}(1998)\citenamefont {Kaiser}, \citenamefont {Gerstendorfer},\ and\ \citenamefont {Weise}}]{Kaiser:1998wa}%
  \BibitemOpen
  \bibfield  {author} {\bibinfo {author} {\bibfnamefont {N.}~\bibnamefont {Kaiser}}, \bibinfo {author} {\bibfnamefont {S.}~\bibnamefont {Gerstendorfer}}, \ and\ \bibinfo {author} {\bibfnamefont {W.}~\bibnamefont {Weise}},\ }\href {\doibase 10.1016/S0375-9474(98)00234-6} {\bibfield  {journal} {\bibinfo  {journal} {Nucl. Phys. A}\ }\textbf {\bibinfo {volume} {637}},\ \bibinfo {pages} {395} (\bibinfo {year} {1998})},\ \Eprint {http://arxiv.org/abs/nucl-th/9802071} {arXiv:nucl-th/9802071} \BibitemShut {NoStop}%
\bibitem [{\citenamefont {Krebs}\ \emph {et~al.}(2007)\citenamefont {Krebs}, \citenamefont {Epelbaum},\ and\ \citenamefont {Meissner}}]{Krebs:2007rh}%
  \BibitemOpen
  \bibfield  {author} {\bibinfo {author} {\bibfnamefont {H.}~\bibnamefont {Krebs}}, \bibinfo {author} {\bibfnamefont {E.}~\bibnamefont {Epelbaum}}, \ and\ \bibinfo {author} {\bibfnamefont {U.-G.}\ \bibnamefont {Meissner}},\ }\href {\doibase 10.1140/epja/i2007-10372-y} {\bibfield  {journal} {\bibinfo  {journal} {Eur. Phys. J. A}\ }\textbf {\bibinfo {volume} {32}},\ \bibinfo {pages} {127} (\bibinfo {year} {2007})},\ \Eprint {http://arxiv.org/abs/nucl-th/0703087} {arXiv:nucl-th/0703087} \BibitemShut {NoStop}%
\bibitem [{\citenamefont {Epelbaum}\ \emph {et~al.}(2008{\natexlab{a}})\citenamefont {Epelbaum}, \citenamefont {Krebs},\ and\ \citenamefont {Mei{\ss}ner}}]{Epelbaum:2008td}%
  \BibitemOpen
  \bibfield  {author} {\bibinfo {author} {\bibfnamefont {E.}~\bibnamefont {Epelbaum}}, \bibinfo {author} {\bibfnamefont {H.}~\bibnamefont {Krebs}}, \ and\ \bibinfo {author} {\bibfnamefont {U.-G.}\ \bibnamefont {Mei{\ss}ner}},\ }\href {\doibase 10.1103/PhysRevC.77.034006} {\bibfield  {journal} {\bibinfo  {journal} {Phys. Rev. C}\ }\textbf {\bibinfo {volume} {77}},\ \bibinfo {pages} {034006} (\bibinfo {year} {2008}{\natexlab{a}})},\ \Eprint {http://arxiv.org/abs/0801.1299} {arXiv:0801.1299 [nucl-th]} \BibitemShut {NoStop}%
\bibitem [{\citenamefont {Piarulli}\ \emph {et~al.}(2015)\citenamefont {Piarulli}, \citenamefont {Girlanda}, \citenamefont {Schiavilla}, \citenamefont {Navarro~P\'erez}, \citenamefont {Amaro},\ and\ \citenamefont {Ruiz~Arriola}}]{Piarulli:2014bda}%
  \BibitemOpen
  \bibfield  {author} {\bibinfo {author} {\bibfnamefont {M.}~\bibnamefont {Piarulli}}, \bibinfo {author} {\bibfnamefont {L.}~\bibnamefont {Girlanda}}, \bibinfo {author} {\bibfnamefont {R.}~\bibnamefont {Schiavilla}}, \bibinfo {author} {\bibfnamefont {R.}~\bibnamefont {Navarro~P\'erez}}, \bibinfo {author} {\bibfnamefont {J.~E.}\ \bibnamefont {Amaro}}, \ and\ \bibinfo {author} {\bibfnamefont {E.}~\bibnamefont {Ruiz~Arriola}},\ }\href {\doibase 10.1103/PhysRevC.91.024003} {\bibfield  {journal} {\bibinfo  {journal} {Phys. Rev. C}\ }\textbf {\bibinfo {volume} {91}},\ \bibinfo {pages} {024003} (\bibinfo {year} {2015})},\ \Eprint {http://arxiv.org/abs/1412.6446} {arXiv:1412.6446 [nucl-th]} \BibitemShut {NoStop}%
\bibitem [{\citenamefont {Ekstr\"om}\ \emph {et~al.}(2018)\citenamefont {Ekstr\"om}, \citenamefont {Hagen}, \citenamefont {Morris}, \citenamefont {Papenbrock},\ and\ \citenamefont {Schwartz}}]{Ekstrom:2017koy}%
  \BibitemOpen
  \bibfield  {author} {\bibinfo {author} {\bibfnamefont {A.}~\bibnamefont {Ekstr\"om}}, \bibinfo {author} {\bibfnamefont {G.}~\bibnamefont {Hagen}}, \bibinfo {author} {\bibfnamefont {T.~D.}\ \bibnamefont {Morris}}, \bibinfo {author} {\bibfnamefont {T.}~\bibnamefont {Papenbrock}}, \ and\ \bibinfo {author} {\bibfnamefont {P.~D.}\ \bibnamefont {Schwartz}},\ }\href {\doibase 10.1103/PhysRevC.97.024332} {\bibfield  {journal} {\bibinfo  {journal} {Phys. Rev. C}\ }\textbf {\bibinfo {volume} {97}},\ \bibinfo {pages} {024332} (\bibinfo {year} {2018})},\ \Eprint {http://arxiv.org/abs/1707.09028} {arXiv:1707.09028 [nucl-th]} \BibitemShut {NoStop}%
\bibitem [{\citenamefont {Strohmeier}\ and\ \citenamefont {Kaiser}(2020)}]{Strohmeier:2020dkb}%
  \BibitemOpen
  \bibfield  {author} {\bibinfo {author} {\bibfnamefont {S.}~\bibnamefont {Strohmeier}}\ and\ \bibinfo {author} {\bibfnamefont {N.}~\bibnamefont {Kaiser}},\ }\href {\doibase 10.1016/j.nuclphysa.2020.121980} {\bibfield  {journal} {\bibinfo  {journal} {Nucl. Phys. A}\ }\textbf {\bibinfo {volume} {1002}},\ \bibinfo {pages} {121980} (\bibinfo {year} {2020})},\ \Eprint {http://arxiv.org/abs/2004.12964} {arXiv:2004.12964 [nucl-th]} \BibitemShut {NoStop}%
\bibitem [{\citenamefont {Nosyk}\ \emph {et~al.}(2021)\citenamefont {Nosyk}, \citenamefont {Entem},\ and\ \citenamefont {Machleidt}}]{Nosyk:2021pxb}%
  \BibitemOpen
  \bibfield  {author} {\bibinfo {author} {\bibfnamefont {Y.}~\bibnamefont {Nosyk}}, \bibinfo {author} {\bibfnamefont {D.~R.}\ \bibnamefont {Entem}}, \ and\ \bibinfo {author} {\bibfnamefont {R.}~\bibnamefont {Machleidt}},\ }\href {\doibase 10.1103/PhysRevC.104.054001} {\bibfield  {journal} {\bibinfo  {journal} {Phys. Rev. C}\ }\textbf {\bibinfo {volume} {104}},\ \bibinfo {pages} {054001} (\bibinfo {year} {2021})},\ \Eprint {http://arxiv.org/abs/2107.06452} {arXiv:2107.06452 [nucl-th]} \BibitemShut {NoStop}%
\bibitem [{\citenamefont {van Kolck}(1994)}]{vanKolck:1994yi}%
  \BibitemOpen
  \bibfield  {author} {\bibinfo {author} {\bibfnamefont {U.}~\bibnamefont {van Kolck}},\ }\href {\doibase 10.1103/PhysRevC.49.2932} {\bibfield  {journal} {\bibinfo  {journal} {Phys. Rev. C}\ }\textbf {\bibinfo {volume} {49}},\ \bibinfo {pages} {2932} (\bibinfo {year} {1994})}\BibitemShut {NoStop}%
\bibitem [{\citenamefont {Pandharipande}\ \emph {et~al.}(2005)\citenamefont {Pandharipande}, \citenamefont {Phillips},\ and\ \citenamefont {van Kolck}}]{Pandharipande:2005sx}%
  \BibitemOpen
  \bibfield  {author} {\bibinfo {author} {\bibfnamefont {V.~R.}\ \bibnamefont {Pandharipande}}, \bibinfo {author} {\bibfnamefont {D.~R.}\ \bibnamefont {Phillips}}, \ and\ \bibinfo {author} {\bibfnamefont {U.}~\bibnamefont {van Kolck}},\ }\href {\doibase 10.1103/PhysRevC.71.064002} {\bibfield  {journal} {\bibinfo  {journal} {Phys. Rev. C}\ }\textbf {\bibinfo {volume} {71}},\ \bibinfo {pages} {064002} (\bibinfo {year} {2005})},\ \Eprint {http://arxiv.org/abs/nucl-th/0501061} {arXiv:nucl-th/0501061} \BibitemShut {NoStop}%
\bibitem [{\citenamefont {Epelbaum}\ \emph {et~al.}(2008{\natexlab{b}})\citenamefont {Epelbaum}, \citenamefont {Krebs},\ and\ \citenamefont {Meissner}}]{Epelbaum:2007sq}%
  \BibitemOpen
  \bibfield  {author} {\bibinfo {author} {\bibfnamefont {E.}~\bibnamefont {Epelbaum}}, \bibinfo {author} {\bibfnamefont {H.}~\bibnamefont {Krebs}}, \ and\ \bibinfo {author} {\bibfnamefont {U.-G.}\ \bibnamefont {Meissner}},\ }\href {\doibase 10.1016/j.nuclphysa.2008.02.305} {\bibfield  {journal} {\bibinfo  {journal} {Nucl. Phys. A}\ }\textbf {\bibinfo {volume} {806}},\ \bibinfo {pages} {65} (\bibinfo {year} {2008}{\natexlab{b}})},\ \Eprint {http://arxiv.org/abs/0712.1969} {arXiv:0712.1969 [nucl-th]} \BibitemShut {NoStop}%
\bibitem [{\citenamefont {Kaiser}(2015)}]{Kaiser:2015yca}%
  \BibitemOpen
  \bibfield  {author} {\bibinfo {author} {\bibfnamefont {N.}~\bibnamefont {Kaiser}},\ }\href {\doibase 10.1103/PhysRevC.92.024002} {\bibfield  {journal} {\bibinfo  {journal} {Phys. Rev. C}\ }\textbf {\bibinfo {volume} {92}},\ \bibinfo {pages} {024002} (\bibinfo {year} {2015})},\ \Eprint {http://arxiv.org/abs/1504.05131} {arXiv:1504.05131 [nucl-th]} \BibitemShut {NoStop}%
\bibitem [{\citenamefont {Krebs}\ \emph {et~al.}(2018)\citenamefont {Krebs}, \citenamefont {Gasparyan},\ and\ \citenamefont {Epelbaum}}]{Krebs:2018jkc}%
  \BibitemOpen
  \bibfield  {author} {\bibinfo {author} {\bibfnamefont {H.}~\bibnamefont {Krebs}}, \bibinfo {author} {\bibfnamefont {A.~M.}\ \bibnamefont {Gasparyan}}, \ and\ \bibinfo {author} {\bibfnamefont {E.}~\bibnamefont {Epelbaum}},\ }\href {\doibase 10.1103/PhysRevC.98.014003} {\bibfield  {journal} {\bibinfo  {journal} {Phys. Rev. C}\ }\textbf {\bibinfo {volume} {98}},\ \bibinfo {pages} {014003} (\bibinfo {year} {2018})},\ \Eprint {http://arxiv.org/abs/1803.09613} {arXiv:1803.09613 [nucl-th]} \BibitemShut {NoStop}%
\bibitem [{\citenamefont {Nogga}\ \emph {et~al.}(2005)\citenamefont {Nogga}, \citenamefont {Timmermans},\ and\ \citenamefont {van Kolck}}]{PhysRevC.72.054006}%
  \BibitemOpen
  \bibfield  {author} {\bibinfo {author} {\bibfnamefont {A.}~\bibnamefont {Nogga}}, \bibinfo {author} {\bibfnamefont {R.~G.~E.}\ \bibnamefont {Timmermans}}, \ and\ \bibinfo {author} {\bibfnamefont {U.}~\bibnamefont {van Kolck}},\ }\href {\doibase 10.1103/PhysRevC.72.054006} {\bibfield  {journal} {\bibinfo  {journal} {Phys. Rev. C}\ }\textbf {\bibinfo {volume} {72}},\ \bibinfo {pages} {054006} (\bibinfo {year} {2005})},\ \Eprint {http://arxiv.org/abs/nucl-th/0506005} {arXiv:nucl-th/0506005} \BibitemShut {NoStop}%
\bibitem [{\citenamefont {Birse}(2006)}]{Birse:2005um}%
  \BibitemOpen
  \bibfield  {author} {\bibinfo {author} {\bibfnamefont {M.~C.}\ \bibnamefont {Birse}},\ }\href {\doibase 10.1103/PhysRevC.74.014003} {\bibfield  {journal} {\bibinfo  {journal} {Phys. Rev. C}\ }\textbf {\bibinfo {volume} {74}},\ \bibinfo {pages} {014003} (\bibinfo {year} {2006})},\ \Eprint {http://arxiv.org/abs/nucl-th/0507077} {arXiv:nucl-th/0507077} \BibitemShut {NoStop}%
\bibitem [{\citenamefont {Valderrama}\ and\ \citenamefont {Phillips}(2015)}]{PhysRevLett.114.082502}%
  \BibitemOpen
  \bibfield  {author} {\bibinfo {author} {\bibfnamefont {M.~P.}\ \bibnamefont {Valderrama}}\ and\ \bibinfo {author} {\bibfnamefont {D.~R.}\ \bibnamefont {Phillips}},\ }\href {\doibase 10.1103/PhysRevLett.114.082502} {\bibfield  {journal} {\bibinfo  {journal} {Phys. Rev. Lett.}\ }\textbf {\bibinfo {volume} {114}},\ \bibinfo {pages} {082502} (\bibinfo {year} {2015})}\BibitemShut {NoStop}%
\bibitem [{\citenamefont {Kievsky}\ \emph {et~al.}(2017)\citenamefont {Kievsky}, \citenamefont {Viviani}, \citenamefont {Gattobigio},\ and\ \citenamefont {Girlanda}}]{PhysRevC.95.024001}%
  \BibitemOpen
  \bibfield  {author} {\bibinfo {author} {\bibfnamefont {A.}~\bibnamefont {Kievsky}}, \bibinfo {author} {\bibfnamefont {M.}~\bibnamefont {Viviani}}, \bibinfo {author} {\bibfnamefont {M.}~\bibnamefont {Gattobigio}}, \ and\ \bibinfo {author} {\bibfnamefont {L.}~\bibnamefont {Girlanda}},\ }\href {\doibase 10.1103/PhysRevC.95.024001} {\bibfield  {journal} {\bibinfo  {journal} {Phys. Rev. C}\ }\textbf {\bibinfo {volume} {95}},\ \bibinfo {pages} {024001} (\bibinfo {year} {2017})}\BibitemShut {NoStop}%
\bibitem [{\citenamefont {Ren}\ \emph {et~al.}(2018{\natexlab{a}})\citenamefont {Ren}, \citenamefont {Li}, \citenamefont {Geng}, \citenamefont {Long}, \citenamefont {Ring},\ and\ \citenamefont {Meng}}]{Ren:2016jna}%
  \BibitemOpen
  \bibfield  {author} {\bibinfo {author} {\bibfnamefont {X.-L.}\ \bibnamefont {Ren}}, \bibinfo {author} {\bibfnamefont {K.-W.}\ \bibnamefont {Li}}, \bibinfo {author} {\bibfnamefont {L.-S.}\ \bibnamefont {Geng}}, \bibinfo {author} {\bibfnamefont {B.-W.}\ \bibnamefont {Long}}, \bibinfo {author} {\bibfnamefont {P.}~\bibnamefont {Ring}}, \ and\ \bibinfo {author} {\bibfnamefont {J.}~\bibnamefont {Meng}},\ }\href {\doibase 10.1088/1674-1137/42/1/014103} {\bibfield  {journal} {\bibinfo  {journal} {Chin. Phys. C}\ }\textbf {\bibinfo {volume} {42}},\ \bibinfo {pages} {014103} (\bibinfo {year} {2018}{\natexlab{a}})},\ \Eprint {http://arxiv.org/abs/1611.08475} {arXiv:1611.08475 [nucl-th]} \BibitemShut {NoStop}%
\bibitem [{\citenamefont {Xiao}\ \emph {et~al.}(2019)\citenamefont {Xiao}, \citenamefont {Geng},\ and\ \citenamefont {Ren}}]{Xiao:2018jot}%
  \BibitemOpen
  \bibfield  {author} {\bibinfo {author} {\bibfnamefont {Y.}~\bibnamefont {Xiao}}, \bibinfo {author} {\bibfnamefont {L.-S.}\ \bibnamefont {Geng}}, \ and\ \bibinfo {author} {\bibfnamefont {X.-L.}\ \bibnamefont {Ren}},\ }\href {\doibase 10.1103/PhysRevC.99.024004} {\bibfield  {journal} {\bibinfo  {journal} {Phys. Rev. C}\ }\textbf {\bibinfo {volume} {99}},\ \bibinfo {pages} {024004} (\bibinfo {year} {2019})},\ \Eprint {http://arxiv.org/abs/1812.03005} {arXiv:1812.03005 [nucl-th]} \BibitemShut {NoStop}%
\bibitem [{\citenamefont {Xiao}\ \emph {et~al.}(2020)\citenamefont {Xiao}, \citenamefont {Wang}, \citenamefont {Lu},\ and\ \citenamefont {Geng}}]{Xiao:2020ozd}%
  \BibitemOpen
  \bibfield  {author} {\bibinfo {author} {\bibfnamefont {Y.}~\bibnamefont {Xiao}}, \bibinfo {author} {\bibfnamefont {C.-X.}\ \bibnamefont {Wang}}, \bibinfo {author} {\bibfnamefont {J.-X.}\ \bibnamefont {Lu}}, \ and\ \bibinfo {author} {\bibfnamefont {L.-S.}\ \bibnamefont {Geng}},\ }\href {\doibase 10.1103/PhysRevC.102.054001} {\bibfield  {journal} {\bibinfo  {journal} {Phys. Rev. C}\ }\textbf {\bibinfo {volume} {102}},\ \bibinfo {pages} {054001} (\bibinfo {year} {2020})},\ \Eprint {http://arxiv.org/abs/2007.13675} {arXiv:2007.13675 [nucl-th]} \BibitemShut {NoStop}%
\bibitem [{\citenamefont {Wang}\ \emph {et~al.}(2022)\citenamefont {Wang}, \citenamefont {Lu}, \citenamefont {Xiao},\ and\ \citenamefont {Geng}}]{Wang:2021kos}%
  \BibitemOpen
  \bibfield  {author} {\bibinfo {author} {\bibfnamefont {C.-X.}\ \bibnamefont {Wang}}, \bibinfo {author} {\bibfnamefont {J.-X.}\ \bibnamefont {Lu}}, \bibinfo {author} {\bibfnamefont {Y.}~\bibnamefont {Xiao}}, \ and\ \bibinfo {author} {\bibfnamefont {L.-S.}\ \bibnamefont {Geng}},\ }\href {\doibase 10.1103/PhysRevC.105.014003} {\bibfield  {journal} {\bibinfo  {journal} {Phys. Rev. C}\ }\textbf {\bibinfo {volume} {105}},\ \bibinfo {pages} {014003} (\bibinfo {year} {2022})},\ \Eprint {http://arxiv.org/abs/2110.05278} {arXiv:2110.05278 [nucl-th]} \BibitemShut {NoStop}%
\bibitem [{\citenamefont {Lu}\ \emph {et~al.}(2022)\citenamefont {Lu}, \citenamefont {Wang}, \citenamefont {Xiao}, \citenamefont {Geng}, \citenamefont {Meng},\ and\ \citenamefont {Ring}}]{Lu:2021gsb}%
  \BibitemOpen
  \bibfield  {author} {\bibinfo {author} {\bibfnamefont {J.-X.}\ \bibnamefont {Lu}}, \bibinfo {author} {\bibfnamefont {C.-X.}\ \bibnamefont {Wang}}, \bibinfo {author} {\bibfnamefont {Y.}~\bibnamefont {Xiao}}, \bibinfo {author} {\bibfnamefont {L.-S.}\ \bibnamefont {Geng}}, \bibinfo {author} {\bibfnamefont {J.}~\bibnamefont {Meng}}, \ and\ \bibinfo {author} {\bibfnamefont {P.}~\bibnamefont {Ring}},\ }\href {\doibase 10.1103/PhysRevLett.128.142002} {\bibfield  {journal} {\bibinfo  {journal} {Phys. Rev. Lett.}\ }\textbf {\bibinfo {volume} {128}},\ \bibinfo {pages} {142002} (\bibinfo {year} {2022})},\ \Eprint {http://arxiv.org/abs/2111.07766} {arXiv:2111.07766 [nucl-th]} \BibitemShut {NoStop}%
\bibitem [{\citenamefont {Epelbaum}\ \emph {et~al.}(1998)\citenamefont {Epelbaum}, \citenamefont {Gloeckle},\ and\ \citenamefont {Meissner}}]{Epelbaum:1998ka}%
  \BibitemOpen
  \bibfield  {author} {\bibinfo {author} {\bibfnamefont {E.}~\bibnamefont {Epelbaum}}, \bibinfo {author} {\bibfnamefont {W.}~\bibnamefont {Gloeckle}}, \ and\ \bibinfo {author} {\bibfnamefont {U.-G.}\ \bibnamefont {Meissner}},\ }\href {\doibase 10.1016/S0375-9474(98)00220-6} {\bibfield  {journal} {\bibinfo  {journal} {Nucl. Phys. A}\ }\textbf {\bibinfo {volume} {637}},\ \bibinfo {pages} {107} (\bibinfo {year} {1998})},\ \Eprint {http://arxiv.org/abs/nucl-th/9801064} {arXiv:nucl-th/9801064} \BibitemShut {NoStop}%
\bibitem [{\citenamefont {Epelbaum}\ \emph {et~al.}(2000)\citenamefont {Epelbaum}, \citenamefont {Gloeckle},\ and\ \citenamefont {Meissner}}]{Epelbaum:1999dj}%
  \BibitemOpen
  \bibfield  {author} {\bibinfo {author} {\bibfnamefont {E.}~\bibnamefont {Epelbaum}}, \bibinfo {author} {\bibfnamefont {W.}~\bibnamefont {Gloeckle}}, \ and\ \bibinfo {author} {\bibfnamefont {U.-G.}\ \bibnamefont {Meissner}},\ }\href {\doibase 10.1016/S0375-9474(99)00821-0} {\bibfield  {journal} {\bibinfo  {journal} {Nucl. Phys. A}\ }\textbf {\bibinfo {volume} {671}},\ \bibinfo {pages} {295} (\bibinfo {year} {2000})},\ \Eprint {http://arxiv.org/abs/nucl-th/9910064} {arXiv:nucl-th/9910064} \BibitemShut {NoStop}%
\bibitem [{\citenamefont {Epelbaum}\ \emph {et~al.}(2004{\natexlab{a}})\citenamefont {Epelbaum}, \citenamefont {Gloeckle},\ and\ \citenamefont {Meissner}}]{Epelbaum:2003gr}%
  \BibitemOpen
  \bibfield  {author} {\bibinfo {author} {\bibfnamefont {E.}~\bibnamefont {Epelbaum}}, \bibinfo {author} {\bibfnamefont {W.}~\bibnamefont {Gloeckle}}, \ and\ \bibinfo {author} {\bibfnamefont {U.-G.}\ \bibnamefont {Meissner}},\ }\href {\doibase 10.1140/epja/i2003-10096-0} {\bibfield  {journal} {\bibinfo  {journal} {Eur. Phys. J. A}\ }\textbf {\bibinfo {volume} {19}},\ \bibinfo {pages} {125} (\bibinfo {year} {2004}{\natexlab{a}})},\ \Eprint {http://arxiv.org/abs/nucl-th/0304037} {arXiv:nucl-th/0304037} \BibitemShut {NoStop}%
\bibitem [{\citenamefont {Epelbaum}\ \emph {et~al.}(2004{\natexlab{b}})\citenamefont {Epelbaum}, \citenamefont {Gloeckle},\ and\ \citenamefont {Meissner}}]{Epelbaum:2003xx}%
  \BibitemOpen
  \bibfield  {author} {\bibinfo {author} {\bibfnamefont {E.}~\bibnamefont {Epelbaum}}, \bibinfo {author} {\bibfnamefont {W.}~\bibnamefont {Gloeckle}}, \ and\ \bibinfo {author} {\bibfnamefont {U.-G.}\ \bibnamefont {Meissner}},\ }\href {\doibase 10.1140/epja/i2003-10129-8} {\bibfield  {journal} {\bibinfo  {journal} {Eur. Phys. J. A}\ }\textbf {\bibinfo {volume} {19}},\ \bibinfo {pages} {401} (\bibinfo {year} {2004}{\natexlab{b}})},\ \Eprint {http://arxiv.org/abs/nucl-th/0308010} {arXiv:nucl-th/0308010} \BibitemShut {NoStop}%
\bibitem [{\citenamefont {Entem}\ and\ \citenamefont {Machleidt}(2003)}]{Entem:2003ft}%
  \BibitemOpen
  \bibfield  {author} {\bibinfo {author} {\bibfnamefont {D.~R.}\ \bibnamefont {Entem}}\ and\ \bibinfo {author} {\bibfnamefont {R.}~\bibnamefont {Machleidt}},\ }\href {\doibase 10.1103/PhysRevC.68.041001} {\bibfield  {journal} {\bibinfo  {journal} {Phys. Rev. C}\ }\textbf {\bibinfo {volume} {68}},\ \bibinfo {pages} {041001} (\bibinfo {year} {2003})},\ \Eprint {http://arxiv.org/abs/nucl-th/0304018} {arXiv:nucl-th/0304018} \BibitemShut {NoStop}%
\bibitem [{\citenamefont {Epelbaum}\ \emph {et~al.}(2005)\citenamefont {Epelbaum}, \citenamefont {Glockle},\ and\ \citenamefont {Meissner}}]{Epelbaum:2004fk}%
  \BibitemOpen
  \bibfield  {author} {\bibinfo {author} {\bibfnamefont {E.}~\bibnamefont {Epelbaum}}, \bibinfo {author} {\bibfnamefont {W.}~\bibnamefont {Glockle}}, \ and\ \bibinfo {author} {\bibfnamefont {U.-G.}\ \bibnamefont {Meissner}},\ }\href {\doibase 10.1016/j.nuclphysa.2004.09.107} {\bibfield  {journal} {\bibinfo  {journal} {Nucl. Phys. A}\ }\textbf {\bibinfo {volume} {747}},\ \bibinfo {pages} {362} (\bibinfo {year} {2005})},\ \Eprint {http://arxiv.org/abs/nucl-th/0405048} {arXiv:nucl-th/0405048} \BibitemShut {NoStop}%
\bibitem [{\citenamefont {Epelbaum}\ \emph {et~al.}(2015{\natexlab{a}})\citenamefont {Epelbaum}, \citenamefont {Krebs},\ and\ \citenamefont {Mei\ss{}ner}}]{Epelbaum:2014efa}%
  \BibitemOpen
  \bibfield  {author} {\bibinfo {author} {\bibfnamefont {E.}~\bibnamefont {Epelbaum}}, \bibinfo {author} {\bibfnamefont {H.}~\bibnamefont {Krebs}}, \ and\ \bibinfo {author} {\bibfnamefont {U.~G.}\ \bibnamefont {Mei\ss{}ner}},\ }\href {\doibase 10.1140/epja/i2015-15053-8} {\bibfield  {journal} {\bibinfo  {journal} {Eur. Phys. J. A}\ }\textbf {\bibinfo {volume} {51}},\ \bibinfo {pages} {53} (\bibinfo {year} {2015}{\natexlab{a}})},\ \Eprint {http://arxiv.org/abs/1412.0142} {arXiv:1412.0142 [nucl-th]} \BibitemShut {NoStop}%
\bibitem [{\citenamefont {Saha}\ \emph {et~al.}(2023)\citenamefont {Saha}, \citenamefont {Entem}, \citenamefont {Machleidt},\ and\ \citenamefont {Nosyk}}]{Saha:2022oep}%
  \BibitemOpen
  \bibfield  {author} {\bibinfo {author} {\bibfnamefont {S.~K.}\ \bibnamefont {Saha}}, \bibinfo {author} {\bibfnamefont {D.~R.}\ \bibnamefont {Entem}}, \bibinfo {author} {\bibfnamefont {R.}~\bibnamefont {Machleidt}}, \ and\ \bibinfo {author} {\bibfnamefont {Y.}~\bibnamefont {Nosyk}},\ }\href {\doibase 10.1103/PhysRevC.107.034002} {\bibfield  {journal} {\bibinfo  {journal} {Phys. Rev. C}\ }\textbf {\bibinfo {volume} {107}},\ \bibinfo {pages} {034002} (\bibinfo {year} {2023})},\ \Eprint {http://arxiv.org/abs/2209.13170} {arXiv:2209.13170 [nucl-th]} \BibitemShut {NoStop}%
\bibitem [{\citenamefont {Entem}\ \emph {et~al.}(2015{\natexlab{a}})\citenamefont {Entem}, \citenamefont {Kaiser}, \citenamefont {Machleidt},\ and\ \citenamefont {Nosyk}}]{Entem:2014msa}%
  \BibitemOpen
  \bibfield  {author} {\bibinfo {author} {\bibfnamefont {D.~R.}\ \bibnamefont {Entem}}, \bibinfo {author} {\bibfnamefont {N.}~\bibnamefont {Kaiser}}, \bibinfo {author} {\bibfnamefont {R.}~\bibnamefont {Machleidt}}, \ and\ \bibinfo {author} {\bibfnamefont {Y.}~\bibnamefont {Nosyk}},\ }\href {\doibase 10.1103/PhysRevC.91.014002} {\bibfield  {journal} {\bibinfo  {journal} {Phys. Rev. C}\ }\textbf {\bibinfo {volume} {91}},\ \bibinfo {pages} {014002} (\bibinfo {year} {2015}{\natexlab{a}})},\ \Eprint {http://arxiv.org/abs/1411.5335} {arXiv:1411.5335 [nucl-th]} \BibitemShut {NoStop}%
\bibitem [{\citenamefont {Entem}\ \emph {et~al.}(2017)\citenamefont {Entem}, \citenamefont {Machleidt},\ and\ \citenamefont {Nosyk}}]{Entem:2017gor}%
  \BibitemOpen
  \bibfield  {author} {\bibinfo {author} {\bibfnamefont {D.~R.}\ \bibnamefont {Entem}}, \bibinfo {author} {\bibfnamefont {R.}~\bibnamefont {Machleidt}}, \ and\ \bibinfo {author} {\bibfnamefont {Y.}~\bibnamefont {Nosyk}},\ }\href {\doibase 10.1103/PhysRevC.96.024004} {\bibfield  {journal} {\bibinfo  {journal} {Phys. Rev. C}\ }\textbf {\bibinfo {volume} {96}},\ \bibinfo {pages} {024004} (\bibinfo {year} {2017})},\ \Eprint {http://arxiv.org/abs/1703.05454} {arXiv:1703.05454 [nucl-th]} \BibitemShut {NoStop}%
\bibitem [{\citenamefont {Epelbaum}\ \emph {et~al.}(2015{\natexlab{b}})\citenamefont {Epelbaum}, \citenamefont {Krebs},\ and\ \citenamefont {Mei\ss{}ner}}]{Epelbaum:2014sza}%
  \BibitemOpen
  \bibfield  {author} {\bibinfo {author} {\bibfnamefont {E.}~\bibnamefont {Epelbaum}}, \bibinfo {author} {\bibfnamefont {H.}~\bibnamefont {Krebs}}, \ and\ \bibinfo {author} {\bibfnamefont {U.~G.}\ \bibnamefont {Mei\ss{}ner}},\ }\href {\doibase 10.1103/PhysRevLett.115.122301} {\bibfield  {journal} {\bibinfo  {journal} {Phys. Rev. Lett.}\ }\textbf {\bibinfo {volume} {115}},\ \bibinfo {pages} {122301} (\bibinfo {year} {2015}{\natexlab{b}})},\ \Eprint {http://arxiv.org/abs/1412.4623} {arXiv:1412.4623 [nucl-th]} \BibitemShut {NoStop}%
\bibitem [{\citenamefont {Reinert}\ \emph {et~al.}(2018)\citenamefont {Reinert}, \citenamefont {Krebs},\ and\ \citenamefont {Epelbaum}}]{Reinert:2017usi}%
  \BibitemOpen
  \bibfield  {author} {\bibinfo {author} {\bibfnamefont {P.}~\bibnamefont {Reinert}}, \bibinfo {author} {\bibfnamefont {H.}~\bibnamefont {Krebs}}, \ and\ \bibinfo {author} {\bibfnamefont {E.}~\bibnamefont {Epelbaum}},\ }\href {\doibase 10.1140/epja/i2018-12516-4} {\bibfield  {journal} {\bibinfo  {journal} {Eur. Phys. J. A}\ }\textbf {\bibinfo {volume} {54}},\ \bibinfo {pages} {86} (\bibinfo {year} {2018})},\ \Eprint {http://arxiv.org/abs/1711.08821} {arXiv:1711.08821 [nucl-th]} \BibitemShut {NoStop}%
\bibitem [{\citenamefont {Entem}\ \emph {et~al.}(2015{\natexlab{b}})\citenamefont {Entem}, \citenamefont {Kaiser}, \citenamefont {Machleidt},\ and\ \citenamefont {Nosyk}}]{Entem:2015xwa}%
  \BibitemOpen
  \bibfield  {author} {\bibinfo {author} {\bibfnamefont {D.~R.}\ \bibnamefont {Entem}}, \bibinfo {author} {\bibfnamefont {N.}~\bibnamefont {Kaiser}}, \bibinfo {author} {\bibfnamefont {R.}~\bibnamefont {Machleidt}}, \ and\ \bibinfo {author} {\bibfnamefont {Y.}~\bibnamefont {Nosyk}},\ }\href {\doibase 10.1103/PhysRevC.92.064001} {\bibfield  {journal} {\bibinfo  {journal} {Phys. Rev. C}\ }\textbf {\bibinfo {volume} {92}},\ \bibinfo {pages} {064001} (\bibinfo {year} {2015}{\natexlab{b}})},\ \Eprint {http://arxiv.org/abs/1505.03562} {arXiv:1505.03562 [nucl-th]} \BibitemShut {NoStop}%
\bibitem [{\citenamefont {Wiringa}\ \emph {et~al.}(1995)\citenamefont {Wiringa}, \citenamefont {Stoks},\ and\ \citenamefont {Schiavilla}}]{Wiringa:1994wb}%
  \BibitemOpen
  \bibfield  {author} {\bibinfo {author} {\bibfnamefont {R.~B.}\ \bibnamefont {Wiringa}}, \bibinfo {author} {\bibfnamefont {V.~G.~J.}\ \bibnamefont {Stoks}}, \ and\ \bibinfo {author} {\bibfnamefont {R.}~\bibnamefont {Schiavilla}},\ }\href {\doibase 10.1103/PhysRevC.51.38} {\bibfield  {journal} {\bibinfo  {journal} {Phys. Rev. C}\ }\textbf {\bibinfo {volume} {51}},\ \bibinfo {pages} {38} (\bibinfo {year} {1995})},\ \Eprint {http://arxiv.org/abs/nucl-th/9408016} {arXiv:nucl-th/9408016} \BibitemShut {NoStop}%
\bibitem [{\citenamefont {Stoks}\ \emph {et~al.}(1994)\citenamefont {Stoks}, \citenamefont {Klomp}, \citenamefont {Terheggen},\ and\ \citenamefont {de~Swart}}]{Stoks:1994wp}%
  \BibitemOpen
  \bibfield  {author} {\bibinfo {author} {\bibfnamefont {V.~G.~J.}\ \bibnamefont {Stoks}}, \bibinfo {author} {\bibfnamefont {R.~A.~M.}\ \bibnamefont {Klomp}}, \bibinfo {author} {\bibfnamefont {C.~P.~F.}\ \bibnamefont {Terheggen}}, \ and\ \bibinfo {author} {\bibfnamefont {J.~J.}\ \bibnamefont {de~Swart}},\ }\href {\doibase 10.1103/PhysRevC.49.2950} {\bibfield  {journal} {\bibinfo  {journal} {Phys. Rev. C}\ }\textbf {\bibinfo {volume} {49}},\ \bibinfo {pages} {2950} (\bibinfo {year} {1994})},\ \Eprint {http://arxiv.org/abs/nucl-th/9406039} {arXiv:nucl-th/9406039} \BibitemShut {NoStop}%
\bibitem [{\citenamefont {Machleidt}(2001)}]{Machleidt:2000ge}%
  \BibitemOpen
  \bibfield  {author} {\bibinfo {author} {\bibfnamefont {R.}~\bibnamefont {Machleidt}},\ }\href {\doibase 10.1103/PhysRevC.63.024001} {\bibfield  {journal} {\bibinfo  {journal} {Phys. Rev. C}\ }\textbf {\bibinfo {volume} {63}},\ \bibinfo {pages} {024001} (\bibinfo {year} {2001})},\ \Eprint {http://arxiv.org/abs/nucl-th/0006014} {arXiv:nucl-th/0006014} \BibitemShut {NoStop}%
\bibitem [{\citenamefont {Mei\ss{}ner}(2014)}]{Meissner:2014lgi}%
  \BibitemOpen
  \bibfield  {author} {\bibinfo {author} {\bibfnamefont {U.-G.}\ \bibnamefont {Mei\ss{}ner}},\ }\href {\doibase 10.1080/10619127.2014.972167} {\bibfield  {journal} {\bibinfo  {journal} {Nucl. Phys. News.}\ }\textbf {\bibinfo {volume} {24}},\ \bibinfo {pages} {11} (\bibinfo {year} {2014})},\ \Eprint {http://arxiv.org/abs/1505.06997} {arXiv:1505.06997 [nucl-th]} \BibitemShut {NoStop}%
\bibitem [{\citenamefont {Hammer}\ \emph {et~al.}(2020)\citenamefont {Hammer}, \citenamefont {K\"onig},\ and\ \citenamefont {van Kolck}}]{Hammer:2019poc}%
  \BibitemOpen
  \bibfield  {author} {\bibinfo {author} {\bibfnamefont {H.~W.}\ \bibnamefont {Hammer}}, \bibinfo {author} {\bibfnamefont {S.}~\bibnamefont {K\"onig}}, \ and\ \bibinfo {author} {\bibfnamefont {U.}~\bibnamefont {van Kolck}},\ }\href {\doibase 10.1103/RevModPhys.92.025004} {\bibfield  {journal} {\bibinfo  {journal} {Rev. Mod. Phys.}\ }\textbf {\bibinfo {volume} {92}},\ \bibinfo {pages} {025004} (\bibinfo {year} {2020})},\ \Eprint {http://arxiv.org/abs/1906.12122} {arXiv:1906.12122 [nucl-th]} \BibitemShut {NoStop}%
\bibitem [{\citenamefont {Hebeler}(2021)}]{Hebeler:2020ocj}%
  \BibitemOpen
  \bibfield  {author} {\bibinfo {author} {\bibfnamefont {K.}~\bibnamefont {Hebeler}},\ }\href {\doibase 10.1016/j.physrep.2020.08.009} {\bibfield  {journal} {\bibinfo  {journal} {Phys. Rept.}\ }\textbf {\bibinfo {volume} {890}},\ \bibinfo {pages} {1} (\bibinfo {year} {2021})},\ \Eprint {http://arxiv.org/abs/2002.09548} {arXiv:2002.09548 [nucl-th]} \BibitemShut {NoStop}%
\bibitem [{\citenamefont {Epelbaum}\ \emph {et~al.}(2002)\citenamefont {Epelbaum}, \citenamefont {Nogga}, \citenamefont {Gloeckle}, \citenamefont {Kamada}, \citenamefont {Meissner},\ and\ \citenamefont {Witala}}]{Epelbaum:2002vt}%
  \BibitemOpen
  \bibfield  {author} {\bibinfo {author} {\bibfnamefont {E.}~\bibnamefont {Epelbaum}}, \bibinfo {author} {\bibfnamefont {A.}~\bibnamefont {Nogga}}, \bibinfo {author} {\bibfnamefont {W.}~\bibnamefont {Gloeckle}}, \bibinfo {author} {\bibfnamefont {H.}~\bibnamefont {Kamada}}, \bibinfo {author} {\bibfnamefont {U.~G.}\ \bibnamefont {Meissner}}, \ and\ \bibinfo {author} {\bibfnamefont {H.}~\bibnamefont {Witala}},\ }\href {\doibase 10.1103/PhysRevC.66.064001} {\bibfield  {journal} {\bibinfo  {journal} {Phys. Rev. C}\ }\textbf {\bibinfo {volume} {66}},\ \bibinfo {pages} {064001} (\bibinfo {year} {2002})},\ \Eprint {http://arxiv.org/abs/nucl-th/0208023} {arXiv:nucl-th/0208023} \BibitemShut {NoStop}%
\bibitem [{\citenamefont {Bernard}\ \emph {et~al.}(2008)\citenamefont {Bernard}, \citenamefont {Epelbaum}, \citenamefont {Krebs},\ and\ \citenamefont {Meissner}}]{Bernard:2007sp}%
  \BibitemOpen
  \bibfield  {author} {\bibinfo {author} {\bibfnamefont {V.}~\bibnamefont {Bernard}}, \bibinfo {author} {\bibfnamefont {E.}~\bibnamefont {Epelbaum}}, \bibinfo {author} {\bibfnamefont {H.}~\bibnamefont {Krebs}}, \ and\ \bibinfo {author} {\bibfnamefont {U.-G.}\ \bibnamefont {Meissner}},\ }\href {\doibase 10.1103/PhysRevC.77.064004} {\bibfield  {journal} {\bibinfo  {journal} {Phys. Rev. C}\ }\textbf {\bibinfo {volume} {77}},\ \bibinfo {pages} {064004} (\bibinfo {year} {2008})},\ \Eprint {http://arxiv.org/abs/0712.1967} {arXiv:0712.1967 [nucl-th]} \BibitemShut {NoStop}%
\bibitem [{\citenamefont {Bernard}\ \emph {et~al.}(2011)\citenamefont {Bernard}, \citenamefont {Epelbaum}, \citenamefont {Krebs},\ and\ \citenamefont {Meissner}}]{Bernard:2011zr}%
  \BibitemOpen
  \bibfield  {author} {\bibinfo {author} {\bibfnamefont {V.}~\bibnamefont {Bernard}}, \bibinfo {author} {\bibfnamefont {E.}~\bibnamefont {Epelbaum}}, \bibinfo {author} {\bibfnamefont {H.}~\bibnamefont {Krebs}}, \ and\ \bibinfo {author} {\bibfnamefont {U.~G.}\ \bibnamefont {Meissner}},\ }\href {\doibase 10.1103/PhysRevC.84.054001} {\bibfield  {journal} {\bibinfo  {journal} {Phys. Rev. C}\ }\textbf {\bibinfo {volume} {84}},\ \bibinfo {pages} {054001} (\bibinfo {year} {2011})},\ \Eprint {http://arxiv.org/abs/1108.3816} {arXiv:1108.3816 [nucl-th]} \BibitemShut {NoStop}%
\bibitem [{\citenamefont {Drischler}\ \emph {et~al.}(2019)\citenamefont {Drischler}, \citenamefont {Hebeler},\ and\ \citenamefont {Schwenk}}]{Drischler:2017wtt}%
  \BibitemOpen
  \bibfield  {author} {\bibinfo {author} {\bibfnamefont {C.}~\bibnamefont {Drischler}}, \bibinfo {author} {\bibfnamefont {K.}~\bibnamefont {Hebeler}}, \ and\ \bibinfo {author} {\bibfnamefont {A.}~\bibnamefont {Schwenk}},\ }\href {\doibase 10.1103/PhysRevLett.122.042501} {\bibfield  {journal} {\bibinfo  {journal} {Phys. Rev. Lett.}\ }\textbf {\bibinfo {volume} {122}},\ \bibinfo {pages} {042501} (\bibinfo {year} {2019})},\ \Eprint {http://arxiv.org/abs/1710.08220} {arXiv:1710.08220 [nucl-th]} \BibitemShut {NoStop}%
\bibitem [{\citenamefont {Krebs}\ \emph {et~al.}(2012)\citenamefont {Krebs}, \citenamefont {Gasparyan},\ and\ \citenamefont {Epelbaum}}]{Krebs:2012yv}%
  \BibitemOpen
  \bibfield  {author} {\bibinfo {author} {\bibfnamefont {H.}~\bibnamefont {Krebs}}, \bibinfo {author} {\bibfnamefont {A.}~\bibnamefont {Gasparyan}}, \ and\ \bibinfo {author} {\bibfnamefont {E.}~\bibnamefont {Epelbaum}},\ }\href {\doibase 10.1103/PhysRevC.85.054006} {\bibfield  {journal} {\bibinfo  {journal} {Phys. Rev. C}\ }\textbf {\bibinfo {volume} {85}},\ \bibinfo {pages} {054006} (\bibinfo {year} {2012})},\ \Eprint {http://arxiv.org/abs/1203.0067} {arXiv:1203.0067 [nucl-th]} \BibitemShut {NoStop}%
\bibitem [{\citenamefont {Krebs}\ \emph {et~al.}(2013)\citenamefont {Krebs}, \citenamefont {Gasparyan},\ and\ \citenamefont {Epelbaum}}]{Krebs:2013kha}%
  \BibitemOpen
  \bibfield  {author} {\bibinfo {author} {\bibfnamefont {H.}~\bibnamefont {Krebs}}, \bibinfo {author} {\bibfnamefont {A.}~\bibnamefont {Gasparyan}}, \ and\ \bibinfo {author} {\bibfnamefont {E.}~\bibnamefont {Epelbaum}},\ }\href {\doibase 10.1103/PhysRevC.87.054007} {\bibfield  {journal} {\bibinfo  {journal} {Phys. Rev. C}\ }\textbf {\bibinfo {volume} {87}},\ \bibinfo {pages} {054007} (\bibinfo {year} {2013})},\ \Eprint {http://arxiv.org/abs/1302.2872} {arXiv:1302.2872 [nucl-th]} \BibitemShut {NoStop}%
\bibitem [{\citenamefont {Girlanda}\ \emph {et~al.}(2011)\citenamefont {Girlanda}, \citenamefont {Kievsky},\ and\ \citenamefont {Viviani}}]{Girlanda:2011fh}%
  \BibitemOpen
  \bibfield  {author} {\bibinfo {author} {\bibfnamefont {L.}~\bibnamefont {Girlanda}}, \bibinfo {author} {\bibfnamefont {A.}~\bibnamefont {Kievsky}}, \ and\ \bibinfo {author} {\bibfnamefont {M.}~\bibnamefont {Viviani}},\ }\href {\doibase 10.1103/PhysRevC.84.014001} {\bibfield  {journal} {\bibinfo  {journal} {Phys. Rev. C}\ }\textbf {\bibinfo {volume} {84}},\ \bibinfo {pages} {014001} (\bibinfo {year} {2011})},\ \bibinfo {note} {[Erratum: Phys.Rev.C 102, 019903 (2020)]},\ \Eprint {http://arxiv.org/abs/1102.4799} {arXiv:1102.4799 [nucl-th]} \BibitemShut {NoStop}%
\bibitem [{\citenamefont {Epelbaum}(2006)}]{Epelbaum:2005bjv}%
  \BibitemOpen
  \bibfield  {author} {\bibinfo {author} {\bibfnamefont {E.}~\bibnamefont {Epelbaum}},\ }\href {\doibase 10.1016/j.physletb.2006.06.046} {\bibfield  {journal} {\bibinfo  {journal} {Phys. Lett. B}\ }\textbf {\bibinfo {volume} {639}},\ \bibinfo {pages} {456} (\bibinfo {year} {2006})},\ \Eprint {http://arxiv.org/abs/nucl-th/0511025} {arXiv:nucl-th/0511025} \BibitemShut {NoStop}%
\bibitem [{\citenamefont {Epelbaum}(2007)}]{Epelbaum:2007us}%
  \BibitemOpen
  \bibfield  {author} {\bibinfo {author} {\bibfnamefont {E.}~\bibnamefont {Epelbaum}},\ }\href {\doibase 10.1140/epja/i2007-10496-0} {\bibfield  {journal} {\bibinfo  {journal} {Eur. Phys. J. A}\ }\textbf {\bibinfo {volume} {34}},\ \bibinfo {pages} {197} (\bibinfo {year} {2007})},\ \Eprint {http://arxiv.org/abs/0710.4250} {arXiv:0710.4250 [nucl-th]} \BibitemShut {NoStop}%
\bibitem [{\citenamefont {Rozpedzik}\ \emph {et~al.}(2006)\citenamefont {Rozpedzik}, \citenamefont {Golak}, \citenamefont {Skibinski}, \citenamefont {Witala}, \citenamefont {Glockle}, \citenamefont {Epelbaum}, \citenamefont {Nogga},\ and\ \citenamefont {Kamada}}]{Rozpedzik:2006yi}%
  \BibitemOpen
  \bibfield  {author} {\bibinfo {author} {\bibfnamefont {D.}~\bibnamefont {Rozpedzik}}, \bibinfo {author} {\bibfnamefont {J.}~\bibnamefont {Golak}}, \bibinfo {author} {\bibfnamefont {R.}~\bibnamefont {Skibinski}}, \bibinfo {author} {\bibfnamefont {H.}~\bibnamefont {Witala}}, \bibinfo {author} {\bibfnamefont {W.}~\bibnamefont {Glockle}}, \bibinfo {author} {\bibfnamefont {E.}~\bibnamefont {Epelbaum}}, \bibinfo {author} {\bibfnamefont {A.}~\bibnamefont {Nogga}}, \ and\ \bibinfo {author} {\bibfnamefont {H.}~\bibnamefont {Kamada}},\ }\href@noop {} {\bibfield  {journal} {\bibinfo  {journal} {Acta Phys. Polon. B}\ }\textbf {\bibinfo {volume} {37}},\ \bibinfo {pages} {2889} (\bibinfo {year} {2006})},\ \Eprint {http://arxiv.org/abs/nucl-th/0606017} {arXiv:nucl-th/0606017} \BibitemShut {NoStop}%
\bibitem [{\citenamefont {Epelbaum}\ \emph {et~al.}(2020{\natexlab{a}})\citenamefont {Epelbaum} \emph {et~al.}}]{Epelbaum:2019zqc}%
  \BibitemOpen
  \bibfield  {author} {\bibinfo {author} {\bibfnamefont {E.}~\bibnamefont {Epelbaum}} \emph {et~al.},\ }\href {\doibase 10.1140/epja/s10050-020-00102-2} {\bibfield  {journal} {\bibinfo  {journal} {Eur. Phys. J. A}\ }\textbf {\bibinfo {volume} {56}},\ \bibinfo {pages} {92} (\bibinfo {year} {2020}{\natexlab{a}})},\ \Eprint {http://arxiv.org/abs/1907.03608} {arXiv:1907.03608 [nucl-th]} \BibitemShut {NoStop}%
\bibitem [{\citenamefont {Epelbaum}\ \emph {et~al.}(2018)\citenamefont {Epelbaum}, \citenamefont {Gasparyan}, \citenamefont {Gegelia},\ and\ \citenamefont {Mei\ss{}ner}}]{Epelbaum:2018zli}%
  \BibitemOpen
  \bibfield  {author} {\bibinfo {author} {\bibfnamefont {E.}~\bibnamefont {Epelbaum}}, \bibinfo {author} {\bibfnamefont {A.~M.}\ \bibnamefont {Gasparyan}}, \bibinfo {author} {\bibfnamefont {J.}~\bibnamefont {Gegelia}}, \ and\ \bibinfo {author} {\bibfnamefont {U.-G.}\ \bibnamefont {Mei\ss{}ner}},\ }\href {\doibase 10.1140/epja/i2018-12632-1} {\bibfield  {journal} {\bibinfo  {journal} {Eur. Phys. J. A}\ }\textbf {\bibinfo {volume} {54}},\ \bibinfo {pages} {186} (\bibinfo {year} {2018})},\ \Eprint {http://arxiv.org/abs/1810.02646} {arXiv:1810.02646 [nucl-th]} \BibitemShut {NoStop}%
\bibitem [{\citenamefont {Kaplan}(2020)}]{Kaplan:2019znu}%
  \BibitemOpen
  \bibfield  {author} {\bibinfo {author} {\bibfnamefont {D.~B.}\ \bibnamefont {Kaplan}},\ }\href {\doibase 10.1103/PhysRevC.102.034004} {\bibfield  {journal} {\bibinfo  {journal} {Phys. Rev. C}\ }\textbf {\bibinfo {volume} {102}},\ \bibinfo {pages} {034004} (\bibinfo {year} {2020})},\ \Eprint {http://arxiv.org/abs/1905.07485} {arXiv:1905.07485 [nucl-th]} \BibitemShut {NoStop}%
\bibitem [{\citenamefont {Baru}\ \emph {et~al.}(2019)\citenamefont {Baru}, \citenamefont {Epelbaum}, \citenamefont {Gegelia},\ and\ \citenamefont {Ren}}]{Baru:2019ndr}%
  \BibitemOpen
  \bibfield  {author} {\bibinfo {author} {\bibfnamefont {V.}~\bibnamefont {Baru}}, \bibinfo {author} {\bibfnamefont {E.}~\bibnamefont {Epelbaum}}, \bibinfo {author} {\bibfnamefont {J.}~\bibnamefont {Gegelia}}, \ and\ \bibinfo {author} {\bibfnamefont {X.~L.}\ \bibnamefont {Ren}},\ }\href {\doibase 10.1016/j.physletb.2019.134987} {\bibfield  {journal} {\bibinfo  {journal} {Phys. Lett. B}\ }\textbf {\bibinfo {volume} {798}},\ \bibinfo {pages} {134987} (\bibinfo {year} {2019})},\ \Eprint {http://arxiv.org/abs/1905.02116} {arXiv:1905.02116 [nucl-th]} \BibitemShut {NoStop}%
\bibitem [{\citenamefont {Hebeler}\ \emph {et~al.}(2015)\citenamefont {Hebeler}, \citenamefont {Holt}, \citenamefont {Menendez},\ and\ \citenamefont {Schwenk}}]{Hebeler:2015hla}%
  \BibitemOpen
  \bibfield  {author} {\bibinfo {author} {\bibfnamefont {K.}~\bibnamefont {Hebeler}}, \bibinfo {author} {\bibfnamefont {J.~D.}\ \bibnamefont {Holt}}, \bibinfo {author} {\bibfnamefont {J.}~\bibnamefont {Menendez}}, \ and\ \bibinfo {author} {\bibfnamefont {A.}~\bibnamefont {Schwenk}},\ }\href {\doibase 10.1146/annurev-nucl-102313-025446} {\bibfield  {journal} {\bibinfo  {journal} {Ann. Rev. Nucl. Part. Sci.}\ }\textbf {\bibinfo {volume} {65}},\ \bibinfo {pages} {457} (\bibinfo {year} {2015})},\ \Eprint {http://arxiv.org/abs/1508.06893} {arXiv:1508.06893 [nucl-th]} \BibitemShut {NoStop}%
\bibitem [{\citenamefont {Epelbaum}\ \emph {et~al.}(2020{\natexlab{b}})\citenamefont {Epelbaum}, \citenamefont {Krebs},\ and\ \citenamefont {Reinert}}]{Epelbaum:2019kcf}%
  \BibitemOpen
  \bibfield  {author} {\bibinfo {author} {\bibfnamefont {E.}~\bibnamefont {Epelbaum}}, \bibinfo {author} {\bibfnamefont {H.}~\bibnamefont {Krebs}}, \ and\ \bibinfo {author} {\bibfnamefont {P.}~\bibnamefont {Reinert}},\ }\href {\doibase 10.3389/fphy.2020.00098} {\bibfield  {journal} {\bibinfo  {journal} {Front. in Phys.}\ }\textbf {\bibinfo {volume} {8}},\ \bibinfo {pages} {98} (\bibinfo {year} {2020}{\natexlab{b}})},\ \Eprint {http://arxiv.org/abs/1911.11875} {arXiv:1911.11875 [nucl-th]} \BibitemShut {NoStop}%
\bibitem [{\citenamefont {Epelbaum}\ \emph {et~al.}(2022)\citenamefont {Epelbaum}, \citenamefont {Krebs},\ and\ \citenamefont {Reinert}}]{Epelbaum:2022cyo}%
  \BibitemOpen
  \bibfield  {author} {\bibinfo {author} {\bibfnamefont {E.}~\bibnamefont {Epelbaum}}, \bibinfo {author} {\bibfnamefont {H.}~\bibnamefont {Krebs}}, \ and\ \bibinfo {author} {\bibfnamefont {P.}~\bibnamefont {Reinert}},\ }\enquote {\bibinfo {title} {{Semi-local Nuclear Forces From Chiral EFT: State-of-the-Art and Challenges}},}\ in\ \href {\doibase 10.1007/978-981-15-8818-1_54-1} {\emph {\bibinfo {booktitle} {{Handbook of Nuclear Physics}}}},\ \bibinfo {editor} {edited by\ \bibinfo {editor} {\bibfnamefont {I.}~\bibnamefont {Tanihata}}, \bibinfo {editor} {\bibfnamefont {H.}~\bibnamefont {Toki}}, \ and\ \bibinfo {editor} {\bibfnamefont {T.}~\bibnamefont {Kajino}}}\ (\bibinfo {year} {2022})\ pp.\ \bibinfo {pages} {1--25},\ \Eprint {http://arxiv.org/abs/2206.07072} {arXiv:2206.07072 [nucl-th]} \BibitemShut {NoStop}%
\bibitem [{\citenamefont {Machleidt}(2023)}]{Machleidt:2023jws}%
  \BibitemOpen
  \bibfield  {author} {\bibinfo {author} {\bibfnamefont {R.}~\bibnamefont {Machleidt}},\ }\href {\doibase 10.1007/s00601-023-01857-2} {\bibfield  {journal} {\bibinfo  {journal} {Few Body Syst.}\ }\textbf {\bibinfo {volume} {64}},\ \bibinfo {pages} {77} (\bibinfo {year} {2023})},\ \Eprint {http://arxiv.org/abs/2307.06416} {arXiv:2307.06416 [nucl-th]} \BibitemShut {NoStop}%
\bibitem [{\citenamefont {Ekstr\"om}\ \emph {et~al.}(2015)\citenamefont {Ekstr\"om}, \citenamefont {Jansen}, \citenamefont {Wendt}, \citenamefont {Hagen}, \citenamefont {Papenbrock}, \citenamefont {Carlsson}, \citenamefont {Forss\'en}, \citenamefont {Hjorth-Jensen}, \citenamefont {Navr\'atil},\ and\ \citenamefont {Nazarewicz}}]{Ekstrom:2015rta}%
  \BibitemOpen
  \bibfield  {author} {\bibinfo {author} {\bibfnamefont {A.}~\bibnamefont {Ekstr\"om}}, \bibinfo {author} {\bibfnamefont {G.~R.}\ \bibnamefont {Jansen}}, \bibinfo {author} {\bibfnamefont {K.~A.}\ \bibnamefont {Wendt}}, \bibinfo {author} {\bibfnamefont {G.}~\bibnamefont {Hagen}}, \bibinfo {author} {\bibfnamefont {T.}~\bibnamefont {Papenbrock}}, \bibinfo {author} {\bibfnamefont {B.~D.}\ \bibnamefont {Carlsson}}, \bibinfo {author} {\bibfnamefont {C.}~\bibnamefont {Forss\'en}}, \bibinfo {author} {\bibfnamefont {M.}~\bibnamefont {Hjorth-Jensen}}, \bibinfo {author} {\bibfnamefont {P.}~\bibnamefont {Navr\'atil}}, \ and\ \bibinfo {author} {\bibfnamefont {W.}~\bibnamefont {Nazarewicz}},\ }\href {\doibase 10.1103/PhysRevC.91.051301} {\bibfield  {journal} {\bibinfo  {journal} {Phys. Rev. C}\ }\textbf {\bibinfo {volume} {91}},\ \bibinfo {pages} {051301} (\bibinfo {year} {2015})},\ \Eprint {http://arxiv.org/abs/1502.04682} {arXiv:1502.04682 [nucl-th]} \BibitemShut {NoStop}%
\bibitem [{\citenamefont {Ekstr\"om}\ \emph {et~al.}(2013)\citenamefont {Ekstr\"om} \emph {et~al.}}]{Ekstrom:2013kea}%
  \BibitemOpen
  \bibfield  {author} {\bibinfo {author} {\bibfnamefont {A.}~\bibnamefont {Ekstr\"om}} \emph {et~al.},\ }\href {\doibase 10.1103/PhysRevLett.110.192502} {\bibfield  {journal} {\bibinfo  {journal} {Phys. Rev. Lett.}\ }\textbf {\bibinfo {volume} {110}},\ \bibinfo {pages} {192502} (\bibinfo {year} {2013})},\ \Eprint {http://arxiv.org/abs/1303.4674} {arXiv:1303.4674 [nucl-th]} \BibitemShut {NoStop}%
\bibitem [{\citenamefont {Carlsson}\ \emph {et~al.}(2016)\citenamefont {Carlsson}, \citenamefont {Ekstr\"om}, \citenamefont {Forss\'en}, \citenamefont {Str\"omberg}, \citenamefont {Jansen}, \citenamefont {Lilja}, \citenamefont {Lindby}, \citenamefont {Mattsson},\ and\ \citenamefont {Wendt}}]{Carlsson:2015vda}%
  \BibitemOpen
  \bibfield  {author} {\bibinfo {author} {\bibfnamefont {B.~D.}\ \bibnamefont {Carlsson}}, \bibinfo {author} {\bibfnamefont {A.}~\bibnamefont {Ekstr\"om}}, \bibinfo {author} {\bibfnamefont {C.}~\bibnamefont {Forss\'en}}, \bibinfo {author} {\bibfnamefont {D.~F.}\ \bibnamefont {Str\"omberg}}, \bibinfo {author} {\bibfnamefont {G.~R.}\ \bibnamefont {Jansen}}, \bibinfo {author} {\bibfnamefont {O.}~\bibnamefont {Lilja}}, \bibinfo {author} {\bibfnamefont {M.}~\bibnamefont {Lindby}}, \bibinfo {author} {\bibfnamefont {B.~A.}\ \bibnamefont {Mattsson}}, \ and\ \bibinfo {author} {\bibfnamefont {K.~A.}\ \bibnamefont {Wendt}},\ }\href {\doibase 10.1103/PhysRevX.6.011019} {\bibfield  {journal} {\bibinfo  {journal} {Phys. Rev. X}\ }\textbf {\bibinfo {volume} {6}},\ \bibinfo {pages} {011019} (\bibinfo {year} {2016})},\ \Eprint {http://arxiv.org/abs/1506.02466} {arXiv:1506.02466 [nucl-th]} \BibitemShut {NoStop}%
\bibitem [{\citenamefont {Jurgenson}\ \emph {et~al.}(2009)\citenamefont {Jurgenson}, \citenamefont {Navratil},\ and\ \citenamefont {Furnstahl}}]{Jurgenson:2009qs}%
  \BibitemOpen
  \bibfield  {author} {\bibinfo {author} {\bibfnamefont {E.~D.}\ \bibnamefont {Jurgenson}}, \bibinfo {author} {\bibfnamefont {P.}~\bibnamefont {Navratil}}, \ and\ \bibinfo {author} {\bibfnamefont {R.~J.}\ \bibnamefont {Furnstahl}},\ }\href {\doibase 10.1103/PhysRevLett.103.082501} {\bibfield  {journal} {\bibinfo  {journal} {Phys. Rev. Lett.}\ }\textbf {\bibinfo {volume} {103}},\ \bibinfo {pages} {082501} (\bibinfo {year} {2009})},\ \Eprint {http://arxiv.org/abs/0905.1873} {arXiv:0905.1873 [nucl-th]} \BibitemShut {NoStop}%
\bibitem [{\citenamefont {Roth}\ \emph {et~al.}(2011)\citenamefont {Roth}, \citenamefont {Langhammer}, \citenamefont {Calci}, \citenamefont {Binder},\ and\ \citenamefont {Navratil}}]{Roth:2011ar}%
  \BibitemOpen
  \bibfield  {author} {\bibinfo {author} {\bibfnamefont {R.}~\bibnamefont {Roth}}, \bibinfo {author} {\bibfnamefont {J.}~\bibnamefont {Langhammer}}, \bibinfo {author} {\bibfnamefont {A.}~\bibnamefont {Calci}}, \bibinfo {author} {\bibfnamefont {S.}~\bibnamefont {Binder}}, \ and\ \bibinfo {author} {\bibfnamefont {P.}~\bibnamefont {Navratil}},\ }\href {\doibase 10.1103/PhysRevLett.107.072501} {\bibfield  {journal} {\bibinfo  {journal} {Phys. Rev. Lett.}\ }\textbf {\bibinfo {volume} {107}},\ \bibinfo {pages} {072501} (\bibinfo {year} {2011})},\ \Eprint {http://arxiv.org/abs/1105.3173} {arXiv:1105.3173 [nucl-th]} \BibitemShut {NoStop}%
\bibitem [{\citenamefont {Roth}\ \emph {et~al.}(2012)\citenamefont {Roth}, \citenamefont {Binder}, \citenamefont {Vobig}, \citenamefont {Calci}, \citenamefont {Langhammer},\ and\ \citenamefont {Navratil}}]{Roth:2011vt}%
  \BibitemOpen
  \bibfield  {author} {\bibinfo {author} {\bibfnamefont {R.}~\bibnamefont {Roth}}, \bibinfo {author} {\bibfnamefont {S.}~\bibnamefont {Binder}}, \bibinfo {author} {\bibfnamefont {K.}~\bibnamefont {Vobig}}, \bibinfo {author} {\bibfnamefont {A.}~\bibnamefont {Calci}}, \bibinfo {author} {\bibfnamefont {J.}~\bibnamefont {Langhammer}}, \ and\ \bibinfo {author} {\bibfnamefont {P.}~\bibnamefont {Navratil}},\ }\href {\doibase 10.1103/PhysRevLett.109.052501} {\bibfield  {journal} {\bibinfo  {journal} {Phys. Rev. Lett.}\ }\textbf {\bibinfo {volume} {109}},\ \bibinfo {pages} {052501} (\bibinfo {year} {2012})},\ \Eprint {http://arxiv.org/abs/1112.0287} {arXiv:1112.0287 [nucl-th]} \BibitemShut {NoStop}%
\bibitem [{\citenamefont {Ren}\ \emph {et~al.}(2012)\citenamefont {Ren}, \citenamefont {Geng}, \citenamefont {Martin~Camalich}, \citenamefont {Meng},\ and\ \citenamefont {Toki}}]{Ren:2012aj}%
  \BibitemOpen
  \bibfield  {author} {\bibinfo {author} {\bibfnamefont {X.~L.}\ \bibnamefont {Ren}}, \bibinfo {author} {\bibfnamefont {L.~S.}\ \bibnamefont {Geng}}, \bibinfo {author} {\bibfnamefont {J.}~\bibnamefont {Martin~Camalich}}, \bibinfo {author} {\bibfnamefont {J.}~\bibnamefont {Meng}}, \ and\ \bibinfo {author} {\bibfnamefont {H.}~\bibnamefont {Toki}},\ }\href {\doibase 10.1007/JHEP12(2012)073} {\bibfield  {journal} {\bibinfo  {journal} {JHEP}\ }\textbf {\bibinfo {volume} {12}},\ \bibinfo {pages} {073} (\bibinfo {year} {2012})},\ \Eprint {http://arxiv.org/abs/1209.3641} {arXiv:1209.3641 [nucl-th]} \BibitemShut {NoStop}%
\bibitem [{\citenamefont {Ren}\ \emph {et~al.}(2014{\natexlab{a}})\citenamefont {Ren}, \citenamefont {Geng},\ and\ \citenamefont {Meng}}]{Ren:2013oaa}%
  \BibitemOpen
  \bibfield  {author} {\bibinfo {author} {\bibfnamefont {X.-L.}\ \bibnamefont {Ren}}, \bibinfo {author} {\bibfnamefont {L.-S.}\ \bibnamefont {Geng}}, \ and\ \bibinfo {author} {\bibfnamefont {J.}~\bibnamefont {Meng}},\ }\href {\doibase 10.1103/PhysRevD.89.054034} {\bibfield  {journal} {\bibinfo  {journal} {Phys. Rev. D}\ }\textbf {\bibinfo {volume} {89}},\ \bibinfo {pages} {054034} (\bibinfo {year} {2014}{\natexlab{a}})},\ \Eprint {http://arxiv.org/abs/1307.1896} {arXiv:1307.1896 [nucl-th]} \BibitemShut {NoStop}%
\bibitem [{\citenamefont {Ren}\ \emph {et~al.}(2014{\natexlab{b}})\citenamefont {Ren}, \citenamefont {Geng},\ and\ \citenamefont {Meng}}]{Ren:2013wxa}%
  \BibitemOpen
  \bibfield  {author} {\bibinfo {author} {\bibfnamefont {X.-L.}\ \bibnamefont {Ren}}, \bibinfo {author} {\bibfnamefont {L.-S.}\ \bibnamefont {Geng}}, \ and\ \bibinfo {author} {\bibfnamefont {J.}~\bibnamefont {Meng}},\ }\href {\doibase 10.1140/epjc/s10052-014-2754-1} {\bibfield  {journal} {\bibinfo  {journal} {Eur. Phys. J. C}\ }\textbf {\bibinfo {volume} {74}},\ \bibinfo {pages} {2754} (\bibinfo {year} {2014}{\natexlab{b}})},\ \Eprint {http://arxiv.org/abs/1311.7234} {arXiv:1311.7234 [hep-ph]} \BibitemShut {NoStop}%
\bibitem [{\citenamefont {Ren}\ \emph {et~al.}(2013)\citenamefont {Ren}, \citenamefont {Geng}, \citenamefont {Meng},\ and\ \citenamefont {Toki}}]{Ren:2013dzt}%
  \BibitemOpen
  \bibfield  {author} {\bibinfo {author} {\bibfnamefont {X.-L.}\ \bibnamefont {Ren}}, \bibinfo {author} {\bibfnamefont {L.}~\bibnamefont {Geng}}, \bibinfo {author} {\bibfnamefont {J.}~\bibnamefont {Meng}}, \ and\ \bibinfo {author} {\bibfnamefont {H.}~\bibnamefont {Toki}},\ }\href {\doibase 10.1103/PhysRevD.87.074001} {\bibfield  {journal} {\bibinfo  {journal} {Phys. Rev. D}\ }\textbf {\bibinfo {volume} {87}},\ \bibinfo {pages} {074001} (\bibinfo {year} {2013})},\ \Eprint {http://arxiv.org/abs/1302.1953} {arXiv:1302.1953 [nucl-th]} \BibitemShut {NoStop}%
\bibitem [{\citenamefont {Ren}\ \emph {et~al.}(2017)\citenamefont {Ren}, \citenamefont {Alvarez-Ruso}, \citenamefont {Geng}, \citenamefont {Ledwig}, \citenamefont {Meng},\ and\ \citenamefont {Vicente~Vacas}}]{Ren:2016aeo}%
  \BibitemOpen
  \bibfield  {author} {\bibinfo {author} {\bibfnamefont {X.-L.}\ \bibnamefont {Ren}}, \bibinfo {author} {\bibfnamefont {L.}~\bibnamefont {Alvarez-Ruso}}, \bibinfo {author} {\bibfnamefont {L.-S.}\ \bibnamefont {Geng}}, \bibinfo {author} {\bibfnamefont {T.}~\bibnamefont {Ledwig}}, \bibinfo {author} {\bibfnamefont {J.}~\bibnamefont {Meng}}, \ and\ \bibinfo {author} {\bibfnamefont {M.~J.}\ \bibnamefont {Vicente~Vacas}},\ }\href {\doibase 10.1016/j.physletb.2017.01.024} {\bibfield  {journal} {\bibinfo  {journal} {Phys. Lett. B}\ }\textbf {\bibinfo {volume} {766}},\ \bibinfo {pages} {325} (\bibinfo {year} {2017})},\ \Eprint {http://arxiv.org/abs/1606.03820} {arXiv:1606.03820 [nucl-th]} \BibitemShut {NoStop}%
\bibitem [{\citenamefont {Ren}\ \emph {et~al.}(2015)\citenamefont {Ren}, \citenamefont {Geng},\ and\ \citenamefont {Meng}}]{Ren:2014vea}%
  \BibitemOpen
  \bibfield  {author} {\bibinfo {author} {\bibfnamefont {X.-L.}\ \bibnamefont {Ren}}, \bibinfo {author} {\bibfnamefont {L.-S.}\ \bibnamefont {Geng}}, \ and\ \bibinfo {author} {\bibfnamefont {J.}~\bibnamefont {Meng}},\ }\href {\doibase 10.1103/PhysRevD.91.051502} {\bibfield  {journal} {\bibinfo  {journal} {Phys. Rev. D}\ }\textbf {\bibinfo {volume} {91}},\ \bibinfo {pages} {051502} (\bibinfo {year} {2015})},\ \Eprint {http://arxiv.org/abs/1404.4799} {arXiv:1404.4799 [hep-ph]} \BibitemShut {NoStop}%
\bibitem [{\citenamefont {Ren}\ \emph {et~al.}(2018{\natexlab{b}})\citenamefont {Ren}, \citenamefont {Ling},\ and\ \citenamefont {Geng}}]{Ren:2017fbv}%
  \BibitemOpen
  \bibfield  {author} {\bibinfo {author} {\bibfnamefont {X.-L.}\ \bibnamefont {Ren}}, \bibinfo {author} {\bibfnamefont {X.-Z.}\ \bibnamefont {Ling}}, \ and\ \bibinfo {author} {\bibfnamefont {L.-S.}\ \bibnamefont {Geng}},\ }\href {\doibase 10.1016/j.physletb.2018.05.063} {\bibfield  {journal} {\bibinfo  {journal} {Phys. Lett. B}\ }\textbf {\bibinfo {volume} {783}},\ \bibinfo {pages} {7} (\bibinfo {year} {2018}{\natexlab{b}})},\ \Eprint {http://arxiv.org/abs/1710.07164} {arXiv:1710.07164 [hep-ph]} \BibitemShut {NoStop}%
\bibitem [{\citenamefont {Geng}\ \emph {et~al.}(2008)\citenamefont {Geng}, \citenamefont {Martin~Camalich}, \citenamefont {Alvarez-Ruso},\ and\ \citenamefont {Vicente~Vacas}}]{Geng:2008mf}%
  \BibitemOpen
  \bibfield  {author} {\bibinfo {author} {\bibfnamefont {L.~S.}\ \bibnamefont {Geng}}, \bibinfo {author} {\bibfnamefont {J.}~\bibnamefont {Martin~Camalich}}, \bibinfo {author} {\bibfnamefont {L.}~\bibnamefont {Alvarez-Ruso}}, \ and\ \bibinfo {author} {\bibfnamefont {M.~J.}\ \bibnamefont {Vicente~Vacas}},\ }\href {\doibase 10.1103/PhysRevLett.101.222002} {\bibfield  {journal} {\bibinfo  {journal} {Phys. Rev. Lett.}\ }\textbf {\bibinfo {volume} {101}},\ \bibinfo {pages} {222002} (\bibinfo {year} {2008})},\ \Eprint {http://arxiv.org/abs/0805.1419} {arXiv:0805.1419 [hep-ph]} \BibitemShut {NoStop}%
\bibitem [{\citenamefont {Liu}\ \emph {et~al.}(2018)\citenamefont {Liu}, \citenamefont {Xiao},\ and\ \citenamefont {Geng}}]{Liu:2018euh}%
  \BibitemOpen
  \bibfield  {author} {\bibinfo {author} {\bibfnamefont {M.-Z.}\ \bibnamefont {Liu}}, \bibinfo {author} {\bibfnamefont {Y.}~\bibnamefont {Xiao}}, \ and\ \bibinfo {author} {\bibfnamefont {L.-S.}\ \bibnamefont {Geng}},\ }\href {\doibase 10.1103/PhysRevD.98.014040} {\bibfield  {journal} {\bibinfo  {journal} {Phys. Rev. D}\ }\textbf {\bibinfo {volume} {98}},\ \bibinfo {pages} {014040} (\bibinfo {year} {2018})},\ \Eprint {http://arxiv.org/abs/1807.00912} {arXiv:1807.00912 [hep-ph]} \BibitemShut {NoStop}%
\bibitem [{\citenamefont {Shi}\ \emph {et~al.}(2019)\citenamefont {Shi}, \citenamefont {Xiao},\ and\ \citenamefont {Geng}}]{Shi:2018rhk}%
  \BibitemOpen
  \bibfield  {author} {\bibinfo {author} {\bibfnamefont {R.-X.}\ \bibnamefont {Shi}}, \bibinfo {author} {\bibfnamefont {Y.}~\bibnamefont {Xiao}}, \ and\ \bibinfo {author} {\bibfnamefont {L.-S.}\ \bibnamefont {Geng}},\ }\href {\doibase 10.1103/PhysRevD.100.054019} {\bibfield  {journal} {\bibinfo  {journal} {Phys. Rev. D}\ }\textbf {\bibinfo {volume} {100}},\ \bibinfo {pages} {054019} (\bibinfo {year} {2019})},\ \Eprint {http://arxiv.org/abs/1812.07833} {arXiv:1812.07833 [hep-ph]} \BibitemShut {NoStop}%
\bibitem [{\citenamefont {Xiao}\ \emph {et~al.}(2018)\citenamefont {Xiao}, \citenamefont {Ren}, \citenamefont {Lu}, \citenamefont {Geng},\ and\ \citenamefont {Mei\ss{}ner}}]{Xiao:2018rvd}%
  \BibitemOpen
  \bibfield  {author} {\bibinfo {author} {\bibfnamefont {Y.}~\bibnamefont {Xiao}}, \bibinfo {author} {\bibfnamefont {X.-L.}\ \bibnamefont {Ren}}, \bibinfo {author} {\bibfnamefont {J.-X.}\ \bibnamefont {Lu}}, \bibinfo {author} {\bibfnamefont {L.-S.}\ \bibnamefont {Geng}}, \ and\ \bibinfo {author} {\bibfnamefont {U.-G.}\ \bibnamefont {Mei\ss{}ner}},\ }\href {\doibase 10.1140/epjc/s10052-018-5960-4} {\bibfield  {journal} {\bibinfo  {journal} {Eur. Phys. J. C}\ }\textbf {\bibinfo {volume} {78}},\ \bibinfo {pages} {489} (\bibinfo {year} {2018})},\ \Eprint {http://arxiv.org/abs/1803.04251} {arXiv:1803.04251 [hep-ph]} \BibitemShut {NoStop}%
\bibitem [{\citenamefont {Siemens}\ \emph {et~al.}(2016)\citenamefont {Siemens}, \citenamefont {Bernard}, \citenamefont {Epelbaum}, \citenamefont {Gasparyan}, \citenamefont {Krebs},\ and\ \citenamefont {Mei\ss{}ner}}]{Siemens:2016hdi}%
  \BibitemOpen
  \bibfield  {author} {\bibinfo {author} {\bibfnamefont {D.}~\bibnamefont {Siemens}}, \bibinfo {author} {\bibfnamefont {V.}~\bibnamefont {Bernard}}, \bibinfo {author} {\bibfnamefont {E.}~\bibnamefont {Epelbaum}}, \bibinfo {author} {\bibfnamefont {A.}~\bibnamefont {Gasparyan}}, \bibinfo {author} {\bibfnamefont {H.}~\bibnamefont {Krebs}}, \ and\ \bibinfo {author} {\bibfnamefont {U.-G.}\ \bibnamefont {Mei\ss{}ner}},\ }\href {\doibase 10.1103/PhysRevC.94.014620} {\bibfield  {journal} {\bibinfo  {journal} {Phys. Rev. C}\ }\textbf {\bibinfo {volume} {94}},\ \bibinfo {pages} {014620} (\bibinfo {year} {2016})},\ \Eprint {http://arxiv.org/abs/1602.02640} {arXiv:1602.02640 [nucl-th]} \BibitemShut {NoStop}%
\bibitem [{\citenamefont {Shi}\ \emph {et~al.}(2022)\citenamefont {Shi}, \citenamefont {Li}, \citenamefont {Lu},\ and\ \citenamefont {Geng}}]{Shi:2022dhw}%
  \BibitemOpen
  \bibfield  {author} {\bibinfo {author} {\bibfnamefont {R.-X.}\ \bibnamefont {Shi}}, \bibinfo {author} {\bibfnamefont {S.-Y.}\ \bibnamefont {Li}}, \bibinfo {author} {\bibfnamefont {J.-X.}\ \bibnamefont {Lu}}, \ and\ \bibinfo {author} {\bibfnamefont {L.-S.}\ \bibnamefont {Geng}},\ }\href {\doibase 10.1016/j.scib.2022.10.026} {\bibfield  {journal} {\bibinfo  {journal} {Sci. Bull.}\ }\textbf {\bibinfo {volume} {67}},\ \bibinfo {pages} {2298} (\bibinfo {year} {2022})},\ \Eprint {http://arxiv.org/abs/2206.11773} {arXiv:2206.11773 [hep-ph]} \BibitemShut {NoStop}%
\bibitem [{\citenamefont {Shen}\ \emph {et~al.}(2019)\citenamefont {Shen}, \citenamefont {Liang}, \citenamefont {Long}, \citenamefont {Meng},\ and\ \citenamefont {Ring}}]{Shen:2019dls}%
  \BibitemOpen
  \bibfield  {author} {\bibinfo {author} {\bibfnamefont {S.}~\bibnamefont {Shen}}, \bibinfo {author} {\bibfnamefont {H.}~\bibnamefont {Liang}}, \bibinfo {author} {\bibfnamefont {W.~H.}\ \bibnamefont {Long}}, \bibinfo {author} {\bibfnamefont {J.}~\bibnamefont {Meng}}, \ and\ \bibinfo {author} {\bibfnamefont {P.}~\bibnamefont {Ring}},\ }\href {\doibase 10.1016/j.ppnp.2019.103713} {\bibfield  {journal} {\bibinfo  {journal} {Prog. Part. Nucl. Phys.}\ }\textbf {\bibinfo {volume} {109}},\ \bibinfo {pages} {103713} (\bibinfo {year} {2019})},\ \Eprint {http://arxiv.org/abs/1904.04977} {arXiv:1904.04977 [nucl-th]} \BibitemShut {NoStop}%
\bibitem [{\citenamefont {Chen}\ \emph {et~al.}(2013)\citenamefont {Chen}, \citenamefont {Yao},\ and\ \citenamefont {Zheng}}]{Chen:2012nx}%
  \BibitemOpen
  \bibfield  {author} {\bibinfo {author} {\bibfnamefont {Y.-H.}\ \bibnamefont {Chen}}, \bibinfo {author} {\bibfnamefont {D.-L.}\ \bibnamefont {Yao}}, \ and\ \bibinfo {author} {\bibfnamefont {H.~Q.}\ \bibnamefont {Zheng}},\ }\href {\doibase 10.1103/PhysRevD.87.054019} {\bibfield  {journal} {\bibinfo  {journal} {Phys. Rev. D}\ }\textbf {\bibinfo {volume} {87}},\ \bibinfo {pages} {054019} (\bibinfo {year} {2013})},\ \Eprint {http://arxiv.org/abs/1212.1893} {arXiv:1212.1893 [hep-ph]} \BibitemShut {NoStop}%
\bibitem [{\citenamefont {Blankenbecler}\ and\ \citenamefont {Sugar}(1966)}]{Blankenbecler:1965gx}%
  \BibitemOpen
  \bibfield  {author} {\bibinfo {author} {\bibfnamefont {R.}~\bibnamefont {Blankenbecler}}\ and\ \bibinfo {author} {\bibfnamefont {R.}~\bibnamefont {Sugar}},\ }\href {\doibase 10.1103/PhysRev.142.1051} {\bibfield  {journal} {\bibinfo  {journal} {Phys. Rev.}\ }\textbf {\bibinfo {volume} {142}},\ \bibinfo {pages} {1051} (\bibinfo {year} {1966})}\BibitemShut {NoStop}%
\bibitem [{\citenamefont {Stapp}\ \emph {et~al.}(1957)\citenamefont {Stapp}, \citenamefont {Ypsilantis},\ and\ \citenamefont {Metropolis}}]{Stapp:1956mz}%
  \BibitemOpen
  \bibfield  {author} {\bibinfo {author} {\bibfnamefont {H.~P.}\ \bibnamefont {Stapp}}, \bibinfo {author} {\bibfnamefont {T.~J.}\ \bibnamefont {Ypsilantis}}, \ and\ \bibinfo {author} {\bibfnamefont {N.}~\bibnamefont {Metropolis}},\ }\href {\doibase 10.1103/PhysRev.105.302} {\bibfield  {journal} {\bibinfo  {journal} {Phys. Rev.}\ }\textbf {\bibinfo {volume} {105}},\ \bibinfo {pages} {302} (\bibinfo {year} {1957})}\BibitemShut {NoStop}%
\bibitem [{\citenamefont {Stoks}\ \emph {et~al.}(1993)\citenamefont {Stoks}, \citenamefont {Klomp}, \citenamefont {Rentmeester},\ and\ \citenamefont {de~Swart}}]{Stoks:1993tb}%
  \BibitemOpen
  \bibfield  {author} {\bibinfo {author} {\bibfnamefont {V.~G.~J.}\ \bibnamefont {Stoks}}, \bibinfo {author} {\bibfnamefont {R.~A.~M.}\ \bibnamefont {Klomp}}, \bibinfo {author} {\bibfnamefont {M.~C.~M.}\ \bibnamefont {Rentmeester}}, \ and\ \bibinfo {author} {\bibfnamefont {J.~J.}\ \bibnamefont {de~Swart}},\ }\href {\doibase 10.1103/PhysRevC.48.792} {\bibfield  {journal} {\bibinfo  {journal} {Phys. Rev. C}\ }\textbf {\bibinfo {volume} {48}},\ \bibinfo {pages} {792} (\bibinfo {year} {1993})}\BibitemShut {NoStop}%
\bibitem [{\citenamefont {Ren}\ \emph {et~al.}(2021)\citenamefont {Ren}, \citenamefont {Wang}, \citenamefont {Li}, \citenamefont {Geng},\ and\ \citenamefont {Meng}}]{Ren:2017yvw}%
  \BibitemOpen
  \bibfield  {author} {\bibinfo {author} {\bibfnamefont {X.-L.}\ \bibnamefont {Ren}}, \bibinfo {author} {\bibfnamefont {C.-X.}\ \bibnamefont {Wang}}, \bibinfo {author} {\bibfnamefont {K.-W.}\ \bibnamefont {Li}}, \bibinfo {author} {\bibfnamefont {L.-S.}\ \bibnamefont {Geng}}, \ and\ \bibinfo {author} {\bibfnamefont {J.}~\bibnamefont {Meng}},\ }\href {\doibase 10.1088/0256-307X/38/6/062101} {\bibfield  {journal} {\bibinfo  {journal} {Chin. Phys. Lett.}\ }\textbf {\bibinfo {volume} {38}},\ \bibinfo {pages} {062101} (\bibinfo {year} {2021})},\ \Eprint {http://arxiv.org/abs/1712.10083} {arXiv:1712.10083 [nucl-th]} \BibitemShut {NoStop}%
\bibitem [{\citenamefont {Wang}\ \emph {et~al.}(2021{\natexlab{a}})\citenamefont {Wang}, \citenamefont {Geng},\ and\ \citenamefont {Long}}]{Wang:2020myr}%
  \BibitemOpen
  \bibfield  {author} {\bibinfo {author} {\bibfnamefont {C.-X.}\ \bibnamefont {Wang}}, \bibinfo {author} {\bibfnamefont {L.-S.}\ \bibnamefont {Geng}}, \ and\ \bibinfo {author} {\bibfnamefont {B.}~\bibnamefont {Long}},\ }\href {\doibase 10.1088/1674-1137/abe368} {\bibfield  {journal} {\bibinfo  {journal} {Chin. Phys. C}\ }\textbf {\bibinfo {volume} {45}},\ \bibinfo {pages} {054101} (\bibinfo {year} {2021}{\natexlab{a}})},\ \Eprint {http://arxiv.org/abs/2001.08483} {arXiv:2001.08483 [nucl-th]} \BibitemShut {NoStop}%
\bibitem [{\citenamefont {Bai}\ \emph {et~al.}(2020)\citenamefont {Bai}, \citenamefont {Wang}, \citenamefont {Xiao},\ and\ \citenamefont {Geng}}]{Bai:2020yml}%
  \BibitemOpen
  \bibfield  {author} {\bibinfo {author} {\bibfnamefont {Q.-Q.}\ \bibnamefont {Bai}}, \bibinfo {author} {\bibfnamefont {C.-X.}\ \bibnamefont {Wang}}, \bibinfo {author} {\bibfnamefont {Y.}~\bibnamefont {Xiao}}, \ and\ \bibinfo {author} {\bibfnamefont {L.-S.}\ \bibnamefont {Geng}},\ }\href {\doibase 10.1016/j.physletb.2020.135745} {\bibfield  {journal} {\bibinfo  {journal} {Phys. Lett.}\ }\textbf {\bibinfo {volume} {B}},\ \bibinfo {pages} {135745} (\bibinfo {year} {2020})},\ \Eprint {http://arxiv.org/abs/2007.01638} {arXiv:2007.01638 [nucl-th]} \BibitemShut {NoStop}%
\bibitem [{\citenamefont {Bai}\ \emph {et~al.}(2022)\citenamefont {Bai}, \citenamefont {Wang}, \citenamefont {Xiao}, \citenamefont {Lu},\ and\ \citenamefont {Geng}}]{Bai:2021uim}%
  \BibitemOpen
  \bibfield  {author} {\bibinfo {author} {\bibfnamefont {Q.-Q.}\ \bibnamefont {Bai}}, \bibinfo {author} {\bibfnamefont {C.-X.}\ \bibnamefont {Wang}}, \bibinfo {author} {\bibfnamefont {Y.}~\bibnamefont {Xiao}}, \bibinfo {author} {\bibfnamefont {J.-X.}\ \bibnamefont {Lu}}, \ and\ \bibinfo {author} {\bibfnamefont {L.-S.}\ \bibnamefont {Geng}},\ }\href {\doibase 10.1016/j.physletb.2022.137347} {\bibfield  {journal} {\bibinfo  {journal} {Phys. Lett. B}\ }\textbf {\bibinfo {volume} {833}},\ \bibinfo {pages} {137347} (\bibinfo {year} {2022})},\ \Eprint {http://arxiv.org/abs/2105.06113} {arXiv:2105.06113 [hep-ph]} \BibitemShut {NoStop}%
\bibitem [{\citenamefont {Inoue}\ \emph {et~al.}(2012)\citenamefont {Inoue}, \citenamefont {Aoki}, \citenamefont {Doi}, \citenamefont {Hatsuda}, \citenamefont {Ikeda}, \citenamefont {Ishii}, \citenamefont {Murano}, \citenamefont {Nemura},\ and\ \citenamefont {Sasaki}}]{Inoue:2011ai}%
  \BibitemOpen
  \bibfield  {author} {\bibinfo {author} {\bibfnamefont {T.}~\bibnamefont {Inoue}}, \bibinfo {author} {\bibfnamefont {S.}~\bibnamefont {Aoki}}, \bibinfo {author} {\bibfnamefont {T.}~\bibnamefont {Doi}}, \bibinfo {author} {\bibfnamefont {T.}~\bibnamefont {Hatsuda}}, \bibinfo {author} {\bibfnamefont {Y.}~\bibnamefont {Ikeda}}, \bibinfo {author} {\bibfnamefont {N.}~\bibnamefont {Ishii}}, \bibinfo {author} {\bibfnamefont {K.}~\bibnamefont {Murano}}, \bibinfo {author} {\bibfnamefont {H.}~\bibnamefont {Nemura}}, \ and\ \bibinfo {author} {\bibfnamefont {K.}~\bibnamefont {Sasaki}} (\bibinfo {collaboration} {HAL QCD}),\ }\href {\doibase 10.1016/j.nuclphysa.2012.02.008} {\bibfield  {journal} {\bibinfo  {journal} {Nucl. Phys. A}\ }\textbf {\bibinfo {volume} {881}},\ \bibinfo {pages} {28} (\bibinfo {year} {2012})},\ \Eprint {http://arxiv.org/abs/1112.5926} {arXiv:1112.5926 [hep-lat]} \BibitemShut {NoStop}%
\bibitem [{\citenamefont {Furnstahl}\ \emph {et~al.}(2015)\citenamefont {Furnstahl}, \citenamefont {Klco}, \citenamefont {Phillips},\ and\ \citenamefont {Wesolowski}}]{Furnstahl:2015rha}%
  \BibitemOpen
  \bibfield  {author} {\bibinfo {author} {\bibfnamefont {R.~J.}\ \bibnamefont {Furnstahl}}, \bibinfo {author} {\bibfnamefont {N.}~\bibnamefont {Klco}}, \bibinfo {author} {\bibfnamefont {D.~R.}\ \bibnamefont {Phillips}}, \ and\ \bibinfo {author} {\bibfnamefont {S.}~\bibnamefont {Wesolowski}},\ }\href {\doibase 10.1103/PhysRevC.92.024005} {\bibfield  {journal} {\bibinfo  {journal} {Phys. Rev. C}\ }\textbf {\bibinfo {volume} {92}},\ \bibinfo {pages} {024005} (\bibinfo {year} {2015})},\ \Eprint {http://arxiv.org/abs/1506.01343} {arXiv:1506.01343 [nucl-th]} \BibitemShut {NoStop}%
\bibitem [{\citenamefont {Melendez}\ \emph {et~al.}(2017)\citenamefont {Melendez}, \citenamefont {Wesolowski},\ and\ \citenamefont {Furnstahl}}]{Melendez:2017phj}%
  \BibitemOpen
  \bibfield  {author} {\bibinfo {author} {\bibfnamefont {J.~A.}\ \bibnamefont {Melendez}}, \bibinfo {author} {\bibfnamefont {S.}~\bibnamefont {Wesolowski}}, \ and\ \bibinfo {author} {\bibfnamefont {R.~J.}\ \bibnamefont {Furnstahl}},\ }\href {\doibase 10.1103/PhysRevC.96.024003} {\bibfield  {journal} {\bibinfo  {journal} {Phys. Rev. C}\ }\textbf {\bibinfo {volume} {96}},\ \bibinfo {pages} {024003} (\bibinfo {year} {2017})},\ \Eprint {http://arxiv.org/abs/1704.03308} {arXiv:1704.03308 [nucl-th]} \BibitemShut {NoStop}%
\bibitem [{\citenamefont {Melendez}\ \emph {et~al.}(2019)\citenamefont {Melendez}, \citenamefont {Furnstahl}, \citenamefont {Phillips}, \citenamefont {Pratola},\ and\ \citenamefont {Wesolowski}}]{Melendez:2019izc}%
  \BibitemOpen
  \bibfield  {author} {\bibinfo {author} {\bibfnamefont {J.~A.}\ \bibnamefont {Melendez}}, \bibinfo {author} {\bibfnamefont {R.~J.}\ \bibnamefont {Furnstahl}}, \bibinfo {author} {\bibfnamefont {D.~R.}\ \bibnamefont {Phillips}}, \bibinfo {author} {\bibfnamefont {M.~T.}\ \bibnamefont {Pratola}}, \ and\ \bibinfo {author} {\bibfnamefont {S.}~\bibnamefont {Wesolowski}},\ }\href {\doibase 10.1103/PhysRevC.100.044001} {\bibfield  {journal} {\bibinfo  {journal} {Phys. Rev. C}\ }\textbf {\bibinfo {volume} {100}},\ \bibinfo {pages} {044001} (\bibinfo {year} {2019})},\ \Eprint {http://arxiv.org/abs/1904.10581} {arXiv:1904.10581 [nucl-th]} \BibitemShut {NoStop}%
\bibitem [{\citenamefont {Bai}\ \emph {et~al.}(2003)\citenamefont {Bai} \emph {et~al.}}]{BES:2003aic}%
  \BibitemOpen
  \bibfield  {author} {\bibinfo {author} {\bibfnamefont {J.~Z.}\ \bibnamefont {Bai}} \emph {et~al.} (\bibinfo {collaboration} {BES}),\ }\href {\doibase 10.1103/PhysRevLett.91.022001} {\bibfield  {journal} {\bibinfo  {journal} {Phys. Rev. Lett.}\ }\textbf {\bibinfo {volume} {91}},\ \bibinfo {pages} {022001} (\bibinfo {year} {2003})},\ \Eprint {http://arxiv.org/abs/hep-ex/0303006} {arXiv:hep-ex/0303006} \BibitemShut {NoStop}%
\bibitem [{\citenamefont {Ablikim}\ \emph {et~al.}(2005)\citenamefont {Ablikim} \emph {et~al.}}]{BES:2005ega}%
  \BibitemOpen
  \bibfield  {author} {\bibinfo {author} {\bibfnamefont {M.}~\bibnamefont {Ablikim}} \emph {et~al.} (\bibinfo {collaboration} {BES}),\ }\href {\doibase 10.1103/PhysRevLett.95.262001} {\bibfield  {journal} {\bibinfo  {journal} {Phys. Rev. Lett.}\ }\textbf {\bibinfo {volume} {95}},\ \bibinfo {pages} {262001} (\bibinfo {year} {2005})},\ \Eprint {http://arxiv.org/abs/hep-ex/0508025} {arXiv:hep-ex/0508025} \BibitemShut {NoStop}%
\bibitem [{\citenamefont {Ablikim}\ \emph {et~al.}(2012)\citenamefont {Ablikim} \emph {et~al.}}]{BESIII:2011aa}%
  \BibitemOpen
  \bibfield  {author} {\bibinfo {author} {\bibfnamefont {M.}~\bibnamefont {Ablikim}} \emph {et~al.} (\bibinfo {collaboration} {BESIII}),\ }\href {\doibase 10.1103/PhysRevLett.108.112003} {\bibfield  {journal} {\bibinfo  {journal} {Phys. Rev. Lett.}\ }\textbf {\bibinfo {volume} {108}},\ \bibinfo {pages} {112003} (\bibinfo {year} {2012})},\ \Eprint {http://arxiv.org/abs/1112.0942} {arXiv:1112.0942 [hep-ex]} \BibitemShut {NoStop}%
\bibitem [{\citenamefont {Ablikim}\ \emph {et~al.}(2016)\citenamefont {Ablikim} \emph {et~al.}}]{BESIII:2016fbr}%
  \BibitemOpen
  \bibfield  {author} {\bibinfo {author} {\bibfnamefont {M.}~\bibnamefont {Ablikim}} \emph {et~al.} (\bibinfo {collaboration} {BESIII}),\ }\href {\doibase 10.1103/PhysRevLett.117.042002} {\bibfield  {journal} {\bibinfo  {journal} {Phys. Rev. Lett.}\ }\textbf {\bibinfo {volume} {117}},\ \bibinfo {pages} {042002} (\bibinfo {year} {2016})},\ \Eprint {http://arxiv.org/abs/1603.09653} {arXiv:1603.09653 [hep-ex]} \BibitemShut {NoStop}%
\bibitem [{\citenamefont {Ablikim}\ \emph {et~al.}(2022)\citenamefont {Ablikim} \emph {et~al.}}]{BESIII:2021xoh}%
  \BibitemOpen
  \bibfield  {author} {\bibinfo {author} {\bibfnamefont {M.}~\bibnamefont {Ablikim}} \emph {et~al.} (\bibinfo {collaboration} {BESIII}),\ }\href {\doibase 10.1103/PhysRevLett.129.022002} {\bibfield  {journal} {\bibinfo  {journal} {Phys. Rev. Lett.}\ }\textbf {\bibinfo {volume} {129}},\ \bibinfo {pages} {022002} (\bibinfo {year} {2022})},\ \Eprint {http://arxiv.org/abs/2112.14369} {arXiv:2112.14369 [hep-ex]} \BibitemShut {NoStop}%
\bibitem [{\citenamefont {Ablikim}\ \emph {et~al.}(2024)\citenamefont {Ablikim} \emph {et~al.}}]{BESIII:2023vvr}%
  \BibitemOpen
  \bibfield  {author} {\bibinfo {author} {\bibfnamefont {M.}~\bibnamefont {Ablikim}} \emph {et~al.} (\bibinfo {collaboration} {BESIII}),\ }\href {\doibase 10.1103/PhysRevLett.132.151901} {\bibfield  {journal} {\bibinfo  {journal} {Phys. Rev. Lett.}\ }\textbf {\bibinfo {volume} {132}},\ \bibinfo {pages} {151901} (\bibinfo {year} {2024})},\ \Eprint {http://arxiv.org/abs/2310.17937} {arXiv:2310.17937 [hep-ex]} \BibitemShut {NoStop}%
\bibitem [{\citenamefont {Abe}\ \emph {et~al.}(2002{\natexlab{a}})\citenamefont {Abe} \emph {et~al.}}]{Belle:2002bro}%
  \BibitemOpen
  \bibfield  {author} {\bibinfo {author} {\bibfnamefont {K.}~\bibnamefont {Abe}} \emph {et~al.} (\bibinfo {collaboration} {Belle}),\ }\href {\doibase 10.1103/PhysRevLett.88.181803} {\bibfield  {journal} {\bibinfo  {journal} {Phys. Rev. Lett.}\ }\textbf {\bibinfo {volume} {88}},\ \bibinfo {pages} {181803} (\bibinfo {year} {2002}{\natexlab{a}})},\ \Eprint {http://arxiv.org/abs/hep-ex/0202017} {arXiv:hep-ex/0202017} \BibitemShut {NoStop}%
\bibitem [{\citenamefont {Abe}\ \emph {et~al.}(2002{\natexlab{b}})\citenamefont {Abe} \emph {et~al.}}]{Belle:2002fay}%
  \BibitemOpen
  \bibfield  {author} {\bibinfo {author} {\bibfnamefont {K.}~\bibnamefont {Abe}} \emph {et~al.} (\bibinfo {collaboration} {Belle}),\ }\href {\doibase 10.1103/PhysRevLett.89.151802} {\bibfield  {journal} {\bibinfo  {journal} {Phys. Rev. Lett.}\ }\textbf {\bibinfo {volume} {89}},\ \bibinfo {pages} {151802} (\bibinfo {year} {2002}{\natexlab{b}})},\ \Eprint {http://arxiv.org/abs/hep-ex/0205083} {arXiv:hep-ex/0205083} \BibitemShut {NoStop}%
\bibitem [{\citenamefont {Aubert}\ \emph {et~al.}(2005)\citenamefont {Aubert} \emph {et~al.}}]{BaBar:2005sdl}%
  \BibitemOpen
  \bibfield  {author} {\bibinfo {author} {\bibfnamefont {B.}~\bibnamefont {Aubert}} \emph {et~al.} (\bibinfo {collaboration} {BaBar}),\ }\href {\doibase 10.1103/PhysRevD.72.051101} {\bibfield  {journal} {\bibinfo  {journal} {Phys. Rev. D}\ }\textbf {\bibinfo {volume} {72}},\ \bibinfo {pages} {051101} (\bibinfo {year} {2005})},\ \Eprint {http://arxiv.org/abs/hep-ex/0507012} {arXiv:hep-ex/0507012} \BibitemShut {NoStop}%
\bibitem [{\citenamefont {Aubert}\ \emph {et~al.}(2006)\citenamefont {Aubert} \emph {et~al.}}]{BaBar:2005pon}%
  \BibitemOpen
  \bibfield  {author} {\bibinfo {author} {\bibfnamefont {B.}~\bibnamefont {Aubert}} \emph {et~al.} (\bibinfo {collaboration} {BaBar}),\ }\href {\doibase 10.1103/PhysRevD.73.012005} {\bibfield  {journal} {\bibinfo  {journal} {Phys. Rev. D}\ }\textbf {\bibinfo {volume} {73}},\ \bibinfo {pages} {012005} (\bibinfo {year} {2006})},\ \Eprint {http://arxiv.org/abs/hep-ex/0512023} {arXiv:hep-ex/0512023} \BibitemShut {NoStop}%
\bibitem [{\citenamefont {Chen}\ and\ \citenamefont {Ma}(2011)}]{Chen:2011yu}%
  \BibitemOpen
  \bibfield  {author} {\bibinfo {author} {\bibfnamefont {G.~Y.}\ \bibnamefont {Chen}}\ and\ \bibinfo {author} {\bibfnamefont {J.~P.}\ \bibnamefont {Ma}},\ }\href {\doibase 10.1103/PhysRevD.83.094029} {\bibfield  {journal} {\bibinfo  {journal} {Phys. Rev. D}\ }\textbf {\bibinfo {volume} {83}},\ \bibinfo {pages} {094029} (\bibinfo {year} {2011})},\ \Eprint {http://arxiv.org/abs/1101.4071} {arXiv:1101.4071 [hep-ph]} \BibitemShut {NoStop}%
\bibitem [{\citenamefont {Kang}\ \emph {et~al.}(2014)\citenamefont {Kang}, \citenamefont {Haidenbauer},\ and\ \citenamefont {Mei\ss{}ner}}]{Kang:2013uia}%
  \BibitemOpen
  \bibfield  {author} {\bibinfo {author} {\bibfnamefont {X.-W.}\ \bibnamefont {Kang}}, \bibinfo {author} {\bibfnamefont {J.}~\bibnamefont {Haidenbauer}}, \ and\ \bibinfo {author} {\bibfnamefont {U.-G.}\ \bibnamefont {Mei\ss{}ner}},\ }\href {\doibase 10.1007/JHEP02(2014)113} {\bibfield  {journal} {\bibinfo  {journal} {JHEP}\ }\textbf {\bibinfo {volume} {02}},\ \bibinfo {pages} {113} (\bibinfo {year} {2014})},\ \Eprint {http://arxiv.org/abs/1311.1658} {arXiv:1311.1658 [hep-ph]} \BibitemShut {NoStop}%
\bibitem [{\citenamefont {Dai}\ \emph {et~al.}(2017)\citenamefont {Dai}, \citenamefont {Haidenbauer},\ and\ \citenamefont {Mei\ss{}ner}}]{Dai:2017ont}%
  \BibitemOpen
  \bibfield  {author} {\bibinfo {author} {\bibfnamefont {L.-Y.}\ \bibnamefont {Dai}}, \bibinfo {author} {\bibfnamefont {J.}~\bibnamefont {Haidenbauer}}, \ and\ \bibinfo {author} {\bibfnamefont {U.-G.}\ \bibnamefont {Mei\ss{}ner}},\ }\href {\doibase 10.1007/JHEP07(2017)078} {\bibfield  {journal} {\bibinfo  {journal} {JHEP}\ }\textbf {\bibinfo {volume} {07}},\ \bibinfo {pages} {078} (\bibinfo {year} {2017})},\ \Eprint {http://arxiv.org/abs/1702.02065} {arXiv:1702.02065 [nucl-th]} \BibitemShut {NoStop}%
\bibitem [{\citenamefont {Haidenbauer}\ \emph {et~al.}(2014)\citenamefont {Haidenbauer}, \citenamefont {Kang},\ and\ \citenamefont {Mei\ss{}ner}}]{Haidenbauer:2014kja}%
  \BibitemOpen
  \bibfield  {author} {\bibinfo {author} {\bibfnamefont {J.}~\bibnamefont {Haidenbauer}}, \bibinfo {author} {\bibfnamefont {X.~W.}\ \bibnamefont {Kang}}, \ and\ \bibinfo {author} {\bibfnamefont {U.~G.}\ \bibnamefont {Mei\ss{}ner}},\ }\href {\doibase 10.1016/j.nuclphysa.2014.06.007} {\bibfield  {journal} {\bibinfo  {journal} {Nucl. Phys. A}\ }\textbf {\bibinfo {volume} {929}},\ \bibinfo {pages} {102} (\bibinfo {year} {2014})},\ \Eprint {http://arxiv.org/abs/1405.1628} {arXiv:1405.1628 [nucl-th]} \BibitemShut {NoStop}%
\bibitem [{\citenamefont {Yang}\ \emph {et~al.}(2023{\natexlab{a}})\citenamefont {Yang}, \citenamefont {Guo}, \citenamefont {Dai}, \citenamefont {Haidenbauer}, \citenamefont {Kang},\ and\ \citenamefont {Mei\ss{}ner}}]{Yang:2022qoy}%
  \BibitemOpen
  \bibfield  {author} {\bibinfo {author} {\bibfnamefont {Q.-H.}\ \bibnamefont {Yang}}, \bibinfo {author} {\bibfnamefont {D.}~\bibnamefont {Guo}}, \bibinfo {author} {\bibfnamefont {L.-Y.}\ \bibnamefont {Dai}}, \bibinfo {author} {\bibfnamefont {J.}~\bibnamefont {Haidenbauer}}, \bibinfo {author} {\bibfnamefont {X.-W.}\ \bibnamefont {Kang}}, \ and\ \bibinfo {author} {\bibfnamefont {U.-G.}\ \bibnamefont {Mei\ss{}ner}},\ }\href {\doibase 10.1016/j.scib.2023.09.036} {\bibfield  {journal} {\bibinfo  {journal} {Sci. Bull.}\ }\textbf {\bibinfo {volume} {68}},\ \bibinfo {pages} {2729} (\bibinfo {year} {2023}{\natexlab{a}})},\ \Eprint {http://arxiv.org/abs/2206.01494} {arXiv:2206.01494 [nucl-th]} \BibitemShut {NoStop}%
\bibitem [{\citenamefont {Cheng}\ and\ \citenamefont {Kang}(2018)}]{Cheng:2017qpv}%
  \BibitemOpen
  \bibfield  {author} {\bibinfo {author} {\bibfnamefont {H.-Y.}\ \bibnamefont {Cheng}}\ and\ \bibinfo {author} {\bibfnamefont {X.-W.}\ \bibnamefont {Kang}},\ }\href {\doibase 10.1016/j.physletb.2018.02.060} {\bibfield  {journal} {\bibinfo  {journal} {Phys. Lett. B}\ }\textbf {\bibinfo {volume} {780}},\ \bibinfo {pages} {100} (\bibinfo {year} {2018})},\ \Eprint {http://arxiv.org/abs/1712.00566} {arXiv:1712.00566 [hep-ph]} \BibitemShut {NoStop}%
\bibitem [{\citenamefont {Dai}\ \emph {et~al.}(2018)\citenamefont {Dai}, \citenamefont {Haidenbauer},\ and\ \citenamefont {Mei\ss{}ner}}]{Dai:2018tlc}%
  \BibitemOpen
  \bibfield  {author} {\bibinfo {author} {\bibfnamefont {L.-Y.}\ \bibnamefont {Dai}}, \bibinfo {author} {\bibfnamefont {J.}~\bibnamefont {Haidenbauer}}, \ and\ \bibinfo {author} {\bibfnamefont {U.-G.}\ \bibnamefont {Mei\ss{}ner}},\ }\href {\doibase 10.1103/PhysRevD.98.014005} {\bibfield  {journal} {\bibinfo  {journal} {Phys. Rev. D}\ }\textbf {\bibinfo {volume} {98}},\ \bibinfo {pages} {014005} (\bibinfo {year} {2018})},\ \Eprint {http://arxiv.org/abs/1804.07077} {arXiv:1804.07077 [hep-ph]} \BibitemShut {NoStop}%
\bibitem [{\citenamefont {Yang}\ \emph {et~al.}(2023{\natexlab{b}})\citenamefont {Yang}, \citenamefont {Guo},\ and\ \citenamefont {Dai}}]{Yang:2022kpm}%
  \BibitemOpen
  \bibfield  {author} {\bibinfo {author} {\bibfnamefont {Q.-H.}\ \bibnamefont {Yang}}, \bibinfo {author} {\bibfnamefont {D.}~\bibnamefont {Guo}}, \ and\ \bibinfo {author} {\bibfnamefont {L.-Y.}\ \bibnamefont {Dai}},\ }\href {\doibase 10.1103/PhysRevD.107.034030} {\bibfield  {journal} {\bibinfo  {journal} {Phys. Rev. D}\ }\textbf {\bibinfo {volume} {107}},\ \bibinfo {pages} {034030} (\bibinfo {year} {2023}{\natexlab{b}})},\ \Eprint {http://arxiv.org/abs/2209.10101} {arXiv:2209.10101 [hep-ph]} \BibitemShut {NoStop}%
\bibitem [{\citenamefont {Haidenbauer}\ and\ \citenamefont {Mei\ss{}ner}(2020)}]{Haidenbauer:2019fyd}%
  \BibitemOpen
  \bibfield  {author} {\bibinfo {author} {\bibfnamefont {J.}~\bibnamefont {Haidenbauer}}\ and\ \bibinfo {author} {\bibfnamefont {U.-G.}\ \bibnamefont {Mei\ss{}ner}},\ }\href {\doibase 10.1088/1674-1137/44/3/033101} {\bibfield  {journal} {\bibinfo  {journal} {Chin. Phys. C}\ }\textbf {\bibinfo {volume} {44}},\ \bibinfo {pages} {033101} (\bibinfo {year} {2020})},\ \Eprint {http://arxiv.org/abs/1910.14423} {arXiv:1910.14423 [hep-ph]} \BibitemShut {NoStop}%
\bibitem [{\citenamefont {Xiao}\ \emph {et~al.}(2024)\citenamefont {Xiao}, \citenamefont {Lu},\ and\ \citenamefont {Geng}}]{Xiao:2024jmu}%
  \BibitemOpen
  \bibfield  {author} {\bibinfo {author} {\bibfnamefont {Y.}~\bibnamefont {Xiao}}, \bibinfo {author} {\bibfnamefont {J.-X.}\ \bibnamefont {Lu}}, \ and\ \bibinfo {author} {\bibfnamefont {L.-S.}\ \bibnamefont {Geng}},\ }\href@noop {} {\  (\bibinfo {year} {2024})},\ \Eprint {http://arxiv.org/abs/2406.01292} {arXiv:2406.01292 [nucl-th]} \BibitemShut {NoStop}%
\bibitem [{\citenamefont {Carbonell}\ \emph {et~al.}(2023)\citenamefont {Carbonell}, \citenamefont {Hupin},\ and\ \citenamefont {Wycech}}]{Carbonell:2023onq}%
  \BibitemOpen
  \bibfield  {author} {\bibinfo {author} {\bibfnamefont {J.}~\bibnamefont {Carbonell}}, \bibinfo {author} {\bibfnamefont {G.}~\bibnamefont {Hupin}}, \ and\ \bibinfo {author} {\bibfnamefont {S.}~\bibnamefont {Wycech}},\ }\href {\doibase 10.1140/epja/s10050-023-01161-x} {\bibfield  {journal} {\bibinfo  {journal} {Eur. Phys. J. A}\ }\textbf {\bibinfo {volume} {59}},\ \bibinfo {pages} {259} (\bibinfo {year} {2023})},\ \Eprint {http://arxiv.org/abs/2309.14831} {arXiv:2309.14831 [nucl-th]} \BibitemShut {NoStop}%
\bibitem [{\citenamefont {{Bystricky, J.}}\ \emph {et~al.}(1987)\citenamefont {{Bystricky, J.}}, \citenamefont {{Lechanoine-Leluc, C.}},\ and\ \citenamefont {{Lehar, F.}}}]{refId0}%
  \BibitemOpen
  \bibfield  {author} {\bibinfo {author} {\bibnamefont {{Bystricky, J.}}}, \bibinfo {author} {\bibnamefont {{Lechanoine-Leluc, C.}}}, \ and\ \bibinfo {author} {\bibnamefont {{Lehar, F.}}},\ }\href {\doibase 10.1051/jphys:01987004802019900} {\bibfield  {journal} {\bibinfo  {journal} {J. Phys. France}\ }\textbf {\bibinfo {volume} {48}},\ \bibinfo {pages} {199} (\bibinfo {year} {1987})}\BibitemShut {NoStop}%
\bibitem [{\citenamefont {Zhou}\ and\ \citenamefont {Timmermans}(2012)}]{Zhou:2012ui}%
  \BibitemOpen
  \bibfield  {author} {\bibinfo {author} {\bibfnamefont {D.}~\bibnamefont {Zhou}}\ and\ \bibinfo {author} {\bibfnamefont {R.~G.~E.}\ \bibnamefont {Timmermans}},\ }\href {\doibase 10.1103/PhysRevC.86.044003} {\bibfield  {journal} {\bibinfo  {journal} {Phys. Rev. C}\ }\textbf {\bibinfo {volume} {86}},\ \bibinfo {pages} {044003} (\bibinfo {year} {2012})},\ \Eprint {http://arxiv.org/abs/1210.7074} {arXiv:1210.7074 [hep-ph]} \BibitemShut {NoStop}%
\bibitem [{\citenamefont {Badalian}\ \emph {et~al.}(1982)\citenamefont {Badalian}, \citenamefont {Kok}, \citenamefont {Polikarpov},\ and\ \citenamefont {Simonov}}]{Badalian:1981xj}%
  \BibitemOpen
  \bibfield  {author} {\bibinfo {author} {\bibfnamefont {A.~M.}\ \bibnamefont {Badalian}}, \bibinfo {author} {\bibfnamefont {L.~P.}\ \bibnamefont {Kok}}, \bibinfo {author} {\bibfnamefont {M.~I.}\ \bibnamefont {Polikarpov}}, \ and\ \bibinfo {author} {\bibfnamefont {Y.~A.}\ \bibnamefont {Simonov}},\ }\href {\doibase 10.1016/0370-1573(82)90014-X} {\bibfield  {journal} {\bibinfo  {journal} {Phys. Rept.}\ }\textbf {\bibinfo {volume} {82}},\ \bibinfo {pages} {31} (\bibinfo {year} {1982})}\BibitemShut {NoStop}%
\bibitem [{\citenamefont {Yan}\ \emph {et~al.}(2005)\citenamefont {Yan}, \citenamefont {Li}, \citenamefont {Wu},\ and\ \citenamefont {Ma}}]{Yan:2004xs}%
  \BibitemOpen
  \bibfield  {author} {\bibinfo {author} {\bibfnamefont {M.-L.}\ \bibnamefont {Yan}}, \bibinfo {author} {\bibfnamefont {S.}~\bibnamefont {Li}}, \bibinfo {author} {\bibfnamefont {B.}~\bibnamefont {Wu}}, \ and\ \bibinfo {author} {\bibfnamefont {B.-Q.}\ \bibnamefont {Ma}},\ }\href {\doibase 10.1103/PhysRevD.72.034027} {\bibfield  {journal} {\bibinfo  {journal} {Phys. Rev. D}\ }\textbf {\bibinfo {volume} {72}},\ \bibinfo {pages} {034027} (\bibinfo {year} {2005})},\ \Eprint {http://arxiv.org/abs/hep-ph/0405087} {arXiv:hep-ph/0405087} \BibitemShut {NoStop}%
\bibitem [{\citenamefont {Sibirtsev}\ \emph {et~al.}(2005)\citenamefont {Sibirtsev}, \citenamefont {Haidenbauer}, \citenamefont {Krewald}, \citenamefont {Meissner},\ and\ \citenamefont {Thomas}}]{Sibirtsev:2004id}%
  \BibitemOpen
  \bibfield  {author} {\bibinfo {author} {\bibfnamefont {A.}~\bibnamefont {Sibirtsev}}, \bibinfo {author} {\bibfnamefont {J.}~\bibnamefont {Haidenbauer}}, \bibinfo {author} {\bibfnamefont {S.}~\bibnamefont {Krewald}}, \bibinfo {author} {\bibfnamefont {U.-G.}\ \bibnamefont {Meissner}}, \ and\ \bibinfo {author} {\bibfnamefont {A.~W.}\ \bibnamefont {Thomas}},\ }\href {\doibase 10.1103/PhysRevD.71.054010} {\bibfield  {journal} {\bibinfo  {journal} {Phys. Rev. D}\ }\textbf {\bibinfo {volume} {71}},\ \bibinfo {pages} {054010} (\bibinfo {year} {2005})},\ \Eprint {http://arxiv.org/abs/hep-ph/0411386} {arXiv:hep-ph/0411386} \BibitemShut {NoStop}%
\bibitem [{\citenamefont {Ding}\ and\ \citenamefont {Yan}(2005)}]{Ding:2005ew}%
  \BibitemOpen
  \bibfield  {author} {\bibinfo {author} {\bibfnamefont {G.-J.}\ \bibnamefont {Ding}}\ and\ \bibinfo {author} {\bibfnamefont {M.-L.}\ \bibnamefont {Yan}},\ }\href {\doibase 10.1103/PhysRevC.72.015208} {\bibfield  {journal} {\bibinfo  {journal} {Phys. Rev. C}\ }\textbf {\bibinfo {volume} {72}},\ \bibinfo {pages} {015208} (\bibinfo {year} {2005})},\ \Eprint {http://arxiv.org/abs/hep-ph/0502127} {arXiv:hep-ph/0502127} \BibitemShut {NoStop}%
\bibitem [{\citenamefont {Dedonder}\ \emph {et~al.}(2009)\citenamefont {Dedonder}, \citenamefont {Loiseau}, \citenamefont {El-Bennich},\ and\ \citenamefont {Wycech}}]{Dedonder:2009bk}%
  \BibitemOpen
  \bibfield  {author} {\bibinfo {author} {\bibfnamefont {J.~P.}\ \bibnamefont {Dedonder}}, \bibinfo {author} {\bibfnamefont {B.}~\bibnamefont {Loiseau}}, \bibinfo {author} {\bibfnamefont {B.}~\bibnamefont {El-Bennich}}, \ and\ \bibinfo {author} {\bibfnamefont {S.}~\bibnamefont {Wycech}},\ }\href {\doibase 10.1103/PhysRevC.80.045207} {\bibfield  {journal} {\bibinfo  {journal} {Phys. Rev. C}\ }\textbf {\bibinfo {volume} {80}},\ \bibinfo {pages} {045207} (\bibinfo {year} {2009})},\ \Eprint {http://arxiv.org/abs/0904.2163} {arXiv:0904.2163 [nucl-th]} \BibitemShut {NoStop}%
\bibitem [{\citenamefont {Tong}\ \emph {et~al.}(2018)\citenamefont {Tong}, \citenamefont {Ren}, \citenamefont {Ring}, \citenamefont {Shen}, \citenamefont {Wang},\ and\ \citenamefont {Meng}}]{Tong:2018qwx}%
  \BibitemOpen
  \bibfield  {author} {\bibinfo {author} {\bibfnamefont {H.}~\bibnamefont {Tong}}, \bibinfo {author} {\bibfnamefont {X.-L.}\ \bibnamefont {Ren}}, \bibinfo {author} {\bibfnamefont {P.}~\bibnamefont {Ring}}, \bibinfo {author} {\bibfnamefont {S.-H.}\ \bibnamefont {Shen}}, \bibinfo {author} {\bibfnamefont {S.-B.}\ \bibnamefont {Wang}}, \ and\ \bibinfo {author} {\bibfnamefont {J.}~\bibnamefont {Meng}},\ }\href {\doibase 10.1103/PhysRevC.98.054302} {\bibfield  {journal} {\bibinfo  {journal} {Phys. Rev. C}\ }\textbf {\bibinfo {volume} {98}},\ \bibinfo {pages} {054302} (\bibinfo {year} {2018})},\ \Eprint {http://arxiv.org/abs/1808.09138} {arXiv:1808.09138 [nucl-th]} \BibitemShut {NoStop}%
\bibitem [{\citenamefont {Wang}\ \emph {et~al.}(2021{\natexlab{b}})\citenamefont {Wang}, \citenamefont {Zhao}, \citenamefont {Ring},\ and\ \citenamefont {Meng}}]{Wang:2021mvg}%
  \BibitemOpen
  \bibfield  {author} {\bibinfo {author} {\bibfnamefont {S.}~\bibnamefont {Wang}}, \bibinfo {author} {\bibfnamefont {Q.}~\bibnamefont {Zhao}}, \bibinfo {author} {\bibfnamefont {P.}~\bibnamefont {Ring}}, \ and\ \bibinfo {author} {\bibfnamefont {J.}~\bibnamefont {Meng}},\ }\href {\doibase 10.1103/PhysRevC.103.054319} {\bibfield  {journal} {\bibinfo  {journal} {Phys. Rev. C}\ }\textbf {\bibinfo {volume} {103}},\ \bibinfo {pages} {054319} (\bibinfo {year} {2021}{\natexlab{b}})},\ \Eprint {http://arxiv.org/abs/2103.12960} {arXiv:2103.12960 [nucl-th]} \BibitemShut {NoStop}%
\bibitem [{\citenamefont {Machleidt}\ \emph {et~al.}(1987)\citenamefont {Machleidt}, \citenamefont {Holinde},\ and\ \citenamefont {Elster}}]{Machleidt:1987hj}%
  \BibitemOpen
  \bibfield  {author} {\bibinfo {author} {\bibfnamefont {R.}~\bibnamefont {Machleidt}}, \bibinfo {author} {\bibfnamefont {K.}~\bibnamefont {Holinde}}, \ and\ \bibinfo {author} {\bibfnamefont {C.}~\bibnamefont {Elster}},\ }\href {\doibase 10.1016/S0370-1573(87)80002-9} {\bibfield  {journal} {\bibinfo  {journal} {Phys. Rept.}\ }\textbf {\bibinfo {volume} {149}},\ \bibinfo {pages} {1} (\bibinfo {year} {1987})}\BibitemShut {NoStop}%
\bibitem [{\citenamefont {Zou}\ \emph {et~al.}(2024)\citenamefont {Zou}, \citenamefont {Lu}, \citenamefont {Zhao}, \citenamefont {Geng},\ and\ \citenamefont {Meng}}]{Zou:2023quo}%
  \BibitemOpen
  \bibfield  {author} {\bibinfo {author} {\bibfnamefont {W.-J.}\ \bibnamefont {Zou}}, \bibinfo {author} {\bibfnamefont {J.-X.}\ \bibnamefont {Lu}}, \bibinfo {author} {\bibfnamefont {P.-W.}\ \bibnamefont {Zhao}}, \bibinfo {author} {\bibfnamefont {L.-S.}\ \bibnamefont {Geng}}, \ and\ \bibinfo {author} {\bibfnamefont {J.}~\bibnamefont {Meng}},\ }\href {\doibase 10.1016/j.physletb.2024.138732} {\bibfield  {journal} {\bibinfo  {journal} {Phys. Lett. B}\ }\textbf {\bibinfo {volume} {854}},\ \bibinfo {pages} {138732} (\bibinfo {year} {2024})},\ \Eprint {http://arxiv.org/abs/2312.15672} {arXiv:2312.15672 [nucl-th]} \BibitemShut {NoStop}%
\bibitem [{\citenamefont {Li}\ \emph {et~al.}(2016)\citenamefont {Li}, \citenamefont {Ren}, \citenamefont {Geng},\ and\ \citenamefont {Long}}]{Li:2016paq}%
  \BibitemOpen
  \bibfield  {author} {\bibinfo {author} {\bibfnamefont {K.-W.}\ \bibnamefont {Li}}, \bibinfo {author} {\bibfnamefont {X.-L.}\ \bibnamefont {Ren}}, \bibinfo {author} {\bibfnamefont {L.-S.}\ \bibnamefont {Geng}}, \ and\ \bibinfo {author} {\bibfnamefont {B.}~\bibnamefont {Long}},\ }\href {\doibase 10.1103/PhysRevD.94.014029} {\bibfield  {journal} {\bibinfo  {journal} {Phys. Rev. D}\ }\textbf {\bibinfo {volume} {94}},\ \bibinfo {pages} {014029} (\bibinfo {year} {2016})},\ \Eprint {http://arxiv.org/abs/1603.07802} {arXiv:1603.07802 [hep-ph]} \BibitemShut {NoStop}%
\bibitem [{\citenamefont {Li}\ \emph {et~al.}(2018{\natexlab{a}})\citenamefont {Li}, \citenamefont {Ren}, \citenamefont {Geng},\ and\ \citenamefont {Long}}]{Li:2016mln}%
  \BibitemOpen
  \bibfield  {author} {\bibinfo {author} {\bibfnamefont {K.-W.}\ \bibnamefont {Li}}, \bibinfo {author} {\bibfnamefont {X.-L.}\ \bibnamefont {Ren}}, \bibinfo {author} {\bibfnamefont {L.-S.}\ \bibnamefont {Geng}}, \ and\ \bibinfo {author} {\bibfnamefont {B.-W.}\ \bibnamefont {Long}},\ }\href {\doibase 10.1088/1674-1137/42/1/014105} {\bibfield  {journal} {\bibinfo  {journal} {Chin. Phys. C}\ }\textbf {\bibinfo {volume} {42}},\ \bibinfo {pages} {014105} (\bibinfo {year} {2018}{\natexlab{a}})},\ \Eprint {http://arxiv.org/abs/1612.08482} {arXiv:1612.08482 [nucl-th]} \BibitemShut {NoStop}%
\bibitem [{\citenamefont {Song}\ \emph {et~al.}(2018)\citenamefont {Song}, \citenamefont {Li},\ and\ \citenamefont {Geng}}]{Song:2018qqm}%
  \BibitemOpen
  \bibfield  {author} {\bibinfo {author} {\bibfnamefont {J.}~\bibnamefont {Song}}, \bibinfo {author} {\bibfnamefont {K.-W.}\ \bibnamefont {Li}}, \ and\ \bibinfo {author} {\bibfnamefont {L.-S.}\ \bibnamefont {Geng}},\ }\href {\doibase 10.1103/PhysRevC.97.065201} {\bibfield  {journal} {\bibinfo  {journal} {Phys. Rev. C}\ }\textbf {\bibinfo {volume} {97}},\ \bibinfo {pages} {065201} (\bibinfo {year} {2018})},\ \Eprint {http://arxiv.org/abs/1802.04433} {arXiv:1802.04433 [nucl-th]} \BibitemShut {NoStop}%
\bibitem [{\citenamefont {Li}\ \emph {et~al.}(2018{\natexlab{b}})\citenamefont {Li}, \citenamefont {Hyodo},\ and\ \citenamefont {Geng}}]{Li:2018tbt}%
  \BibitemOpen
  \bibfield  {author} {\bibinfo {author} {\bibfnamefont {K.-W.}\ \bibnamefont {Li}}, \bibinfo {author} {\bibfnamefont {T.}~\bibnamefont {Hyodo}}, \ and\ \bibinfo {author} {\bibfnamefont {L.-S.}\ \bibnamefont {Geng}},\ }\href {\doibase 10.1103/PhysRevC.98.065203} {\bibfield  {journal} {\bibinfo  {journal} {Phys. Rev. C}\ }\textbf {\bibinfo {volume} {98}},\ \bibinfo {pages} {065203} (\bibinfo {year} {2018}{\natexlab{b}})},\ \Eprint {http://arxiv.org/abs/1809.03199} {arXiv:1809.03199 [nucl-th]} \BibitemShut {NoStop}%
\bibitem [{\citenamefont {Liu}\ \emph {et~al.}(2021)\citenamefont {Liu}, \citenamefont {Song}, \citenamefont {Li},\ and\ \citenamefont {Geng}}]{Liu:2020uxi}%
  \BibitemOpen
  \bibfield  {author} {\bibinfo {author} {\bibfnamefont {Z.-W.}\ \bibnamefont {Liu}}, \bibinfo {author} {\bibfnamefont {J.}~\bibnamefont {Song}}, \bibinfo {author} {\bibfnamefont {K.-W.}\ \bibnamefont {Li}}, \ and\ \bibinfo {author} {\bibfnamefont {L.-S.}\ \bibnamefont {Geng}},\ }\href {\doibase 10.1103/PhysRevC.103.025201} {\bibfield  {journal} {\bibinfo  {journal} {Phys. Rev. C}\ }\textbf {\bibinfo {volume} {103}},\ \bibinfo {pages} {025201} (\bibinfo {year} {2021})},\ \Eprint {http://arxiv.org/abs/2011.05510} {arXiv:2011.05510 [nucl-th]} \BibitemShut {NoStop}%
\bibitem [{\citenamefont {Song}\ \emph {et~al.}(2022)\citenamefont {Song}, \citenamefont {Liu}, \citenamefont {Li},\ and\ \citenamefont {Geng}}]{Song:2021yab}%
  \BibitemOpen
  \bibfield  {author} {\bibinfo {author} {\bibfnamefont {J.}~\bibnamefont {Song}}, \bibinfo {author} {\bibfnamefont {Z.-W.}\ \bibnamefont {Liu}}, \bibinfo {author} {\bibfnamefont {K.-W.}\ \bibnamefont {Li}}, \ and\ \bibinfo {author} {\bibfnamefont {L.-S.}\ \bibnamefont {Geng}},\ }\href {\doibase 10.1103/PhysRevC.105.035203} {\bibfield  {journal} {\bibinfo  {journal} {Phys. Rev. C}\ }\textbf {\bibinfo {volume} {105}},\ \bibinfo {pages} {035203} (\bibinfo {year} {2022})},\ \Eprint {http://arxiv.org/abs/2107.04742} {arXiv:2107.04742 [nucl-th]} \BibitemShut {NoStop}%
\bibitem [{\citenamefont {Liu}\ \emph {et~al.}(2023)\citenamefont {Liu}, \citenamefont {Li},\ and\ \citenamefont {Geng}}]{Liu:2022nec}%
  \BibitemOpen
  \bibfield  {author} {\bibinfo {author} {\bibfnamefont {Z.-W.}\ \bibnamefont {Liu}}, \bibinfo {author} {\bibfnamefont {K.-W.}\ \bibnamefont {Li}}, \ and\ \bibinfo {author} {\bibfnamefont {L.-S.}\ \bibnamefont {Geng}},\ }\href {\doibase 10.1088/1674-1137/ac988a} {\bibfield  {journal} {\bibinfo  {journal} {Chin. Phys. C}\ }\textbf {\bibinfo {volume} {47}},\ \bibinfo {pages} {024108} (\bibinfo {year} {2023})},\ \Eprint {http://arxiv.org/abs/2201.04997} {arXiv:2201.04997 [hep-ph]} \BibitemShut {NoStop}%
\bibitem [{\citenamefont {Zheng}\ \emph {et~al.}(2025)\citenamefont {Zheng}, \citenamefont {Liu}, \citenamefont {Geng}, \citenamefont {Hu},\ and\ \citenamefont {Wang}}]{Zheng:2025sol}%
  \BibitemOpen
  \bibfield  {author} {\bibinfo {author} {\bibfnamefont {R.-Y.}\ \bibnamefont {Zheng}}, \bibinfo {author} {\bibfnamefont {Z.-W.}\ \bibnamefont {Liu}}, \bibinfo {author} {\bibfnamefont {L.-S.}\ \bibnamefont {Geng}}, \bibinfo {author} {\bibfnamefont {J.-N.}\ \bibnamefont {Hu}}, \ and\ \bibinfo {author} {\bibfnamefont {S.}~\bibnamefont {Wang}},\ }\href@noop {} {\  (\bibinfo {year} {2025})},\ \Eprint {http://arxiv.org/abs/2501.02826} {arXiv:2501.02826 [nucl-th]} \BibitemShut {NoStop}%
\end{thebibliography}%

\end{document}